\documentclass[11pt,letterpaper]{article}
\pdfoutput=1
\usepackage{jinstpub}
\usepackage{color}
\usepackage[table]{xcolor}
\usepackage{graphicx}

\usepackage{verbatim}
\usepackage{amsmath}
\usepackage{amssymb}
\usepackage{subfig}
\usepackage{url}
\usepackage{bbold}
\usepackage{slashed}

\usepackage{multirow}
\usepackage{threeparttable}
\usepackage{paralist}

\DeclareRobustCommand{\Sec}[1]{Sec.~\ref{#1}}
\DeclareRobustCommand{\Secs}[2]{Secs.~\ref{#1} and \ref{#2}}
\DeclareRobustCommand{\App}[1]{App.~\ref{#1}}
\DeclareRobustCommand{\Tab}[1]{Table~\ref{#1}}

\DeclareRobustCommand{\Fig}[1]{Fig.~\ref{#1}}

\DeclareRobustCommand{\Figss}[3]{Figs.~\ref{#1}, \ref{#2}, and \ref{#3}}
\DeclareRobustCommand{\Eq}[1]{Eq.~(\ref{#1})}

\DeclareRobustCommand{\Ref}[1]{Ref.~\cite{#1}}
\DeclareRobustCommand{\Refs}[1]{Refs.~\cite{#1}}

\newcommand{\be}{\begin{equation}}
\newcommand{\ee}{\end{equation}}

\newcommand{\tev}{\mathrm{TeV}}

\newcommand{\pt}{p_{\mathrm{T}}}

\newcommand{\at}{\makeatletter @\makeatother}

\begin{document}
\title{The importance of calorimetry for \\ highly-boosted jet substructure}

\author[a]{Evan Coleman,}
\author[b]{Marat Freytsis,}
\author[c]{Andreas Hinzmann,}
\author[a]{Meenakshi Narain,}
\author[d]{\\ Jesse Thaler,}
\author[e]{Nhan Tran,}
\author[e]{Caterina Vernieri}

\affiliation[a]{Brown University, Providence, RI 02912, USA}
\affiliation[b]{Institure of Theoretical Science, University of Oregon, Eugene, OR 97403, USA}
\affiliation[c]{University of Hamburg, Hamburg, Germany}
\affiliation[d]{Center for Theoretical Physics, Massachusetts Institute of Technology, Cambridge, MA 02139, USA}
\affiliation[e]{Fermi National Accelerator Laboratory, Batavia, IL 60510, USA}

\emailAdd{evan\_a\_coleman@brown.edu}
\emailAdd{andreas.hinzmann@cern.ch}
\emailAdd{freytsis@uoregon.edu}
\emailAdd{meenakshi\_narain@brown.edu}
\emailAdd{jthaler@mit.edu}
\emailAdd{ntran@fnal.gov}
\emailAdd{cvernier@fnal.gov}

\abstract{
Jet substructure techniques are playing an essential role in exploring the TeV scale at the Large Hadron Collider (LHC), since they facilitate the efficient reconstruction and identification of highly-boosted objects.
Both for the LHC and for future colliders, there is a growing interest in using jet substructure methods based only on charged-particle information.
The reason is that silicon-based tracking detectors offer excellent granularity and precise vertexing, which can improve the angular resolution on highly-collimated jets and mitigate the impact of pileup.
In this paper, we assess how much jet substructure performance degrades by using track-only information, and we demonstrate physics contexts in which calorimetry is most beneficial.
Specifically, we consider five different hadronic final states---$W$ bosons, $Z$ bosons, top quarks, light quarks, gluons---and test the pairwise discrimination power with a multi-variate combination of substructure observables.
In the idealized case of perfect reconstruction, we quantify the loss in discrimination performance when using just charged particles compared to using all detected particles.
We also consider the intermediate case of using charged particles plus photons, which provides valuable information about neutral pions.
In the more realistic case of a segmented calorimeter, we assess the potential performance gains from improving calorimeter granularity and resolution, comparing a CMS-like detector to more ambitious future detector concepts.
Broadly speaking, we find large performance gains from neutral-particle information and from improved calorimetry in cases where jet mass resolution drives the discrimination power, whereas the gains are more modest if an absolute mass scale calibration is not required.
}

\preprint{
\begin{flushright}
FERMILAB-PUB-17-347-E \\ MIT-CTP {4935}
\end{flushright}
}
\maketitle

\section{Introduction}
\label{sec:introduction}

Current multipurpose detectors for hadron collider experiments~\cite{CMS, Aad:2008zzm} are performing very well at reconstructing final states at unprecedented energies while operating in a challenging environment of high instantaneous luminosities.
This is vital for probing rare and highly-energetic processes that are potentially sensitive to new physics.
Exploring the multi-TeV regime in search of new phenomena related to the $W$, $Z$, and Higgs bosons, as well as to the top quark, represents one of the most important goals of the Large Hadron Collider (LHC) and future higher-energy colliders.
The success of current and future detectors requires advanced techniques to reconstruct highly-boosted ($\pt \gg m$) heavy Standard Model (SM) objects. 
As objects become more energetic, their decay products become increasingly collimated, making it difficult to detect individual particles and resolve event structure with proper resolution.
The tools provided by recent developments in jet substructure \cite{Seymour:1991cb,Seymour:1993mx,Butterworth:2002tt,Butterworth:2007ke,Butterworth:2008iy} are particularly important in exploring these physical signatures at extreme kinematics~\cite{Abdesselam:2010pt,Altheimer:2012mn,Altheimer:2013yza,Adams:2015hiv,Larkoski:2017jix}.

The main features of a multipurpose detector are a strong magnetic field, a high-quality tracking system, a layered muon detection system, a high-resolution and highly-granular electromagnetic calorimeter (ECAL), and a hermetic hadronic calorimeter (HCAL).
For jet substructure, in some cases, track-only observables have been exploited that only use information provided by charged particles.
Compared to calorimetric objects, tracks have the advantage of being more robust to pileup (multiple collisions per bunch crossing), and they typically exhibit better angular and low-to-intermediate momentum resolution as well as lower momentum thresholds.
Track-based observables have been used for example in LHC studies and measurements~\cite{Aad:2014gea,Aad:2015cua,ATLAS-CONF-2016-055,Aad:2016oit,ATLAS:2016vmy,ATLAS:2016wzt,Krohn:2012fg,Waalewijn:2012sv,Chang:2013iba,Chang:2013rca,Schaetzel:2013vka}, and they have also been discussed in the context of future high-energy colliders~\cite{Larkoski:2015yqa,Spannowsky:2015eba,Bressler:2015uma}.

This paper presents a two-fold exploration of the important impact that neutral-particle information and calorimeter sub-detectors can have on jet substructure reconstruction.
Both of these studies use hadronic object identification as a benchmark, assessing the pairwise discrimination power between five different jet categories:  boosted hadronic $W$ bosons ($W$), boosted hadronic $Z$ bosons ($Z$), boosted top quarks ($t$), light quarks ($q$), and gluons ($g$).
First, we quantify the truth-level discrimination power that is lost when neutral-particle information is reduced or ignored.
Then, we perform a detector-level study with three different detector simulations, where calorimeter granularity is steadily increased compared to a CMS-like configuration.
The goal of both analyses is to understand the technological demands on future detectors if we wish to fully exploit jet substructure information and maximize the physics potential of colliders beyond the LHC.

The rest of this paper is organized as follows.
In \Sec{sec:setup}, we discuss the general strategy of the study, including the event generator samples, the choice of jet substructure observables, and the setup to build multivariate jet discriminants.
We present our truth-level study of neutral-particle information in \Sec{sec:results}, and we discuss the impact on neutral object reconstruction of different calorimeter scenarios in \Sec{sec:ImprovedDetector}.
We conclude in \Sec{sec:conclusions} with a summary of our findings, and we discuss the implications for current physics analyses and future detector designs.

\section{General study strategy}
\label{sec:setup}

In jet substructure, one studies the products of hadronic fragmentation and decays, consisting primarily of charged and neutral hadrons.
Charged hadron reconstruction benefits from precision tracking detectors, though they are limited at the highest energies. 
Neutral pions, which decay promptly to photon pairs, are reconstructed by the ECAL;  it provides excellent energy resolution and can detect photons with moderate angular resolution thanks to the relatively narrow Moli\`{e}re radius for photons.  
Other neutral hadrons (as well as charged hadrons) are detected through hadronic showers in the HCAL, which exhibit a broad angular profile from nuclear interactions.
In this study, we adopt the philosophy of particle-flow reconstruction, where information from all sub-detectors are correlated to identify and reconstruct each final-state particle, allowing one to, for example, disentangle charged and neutral hadrons in the HCAL.  
Particle-flow was first developed and used by the ALEPH experiment at LEP~\cite{Buskulic:1994wz} and is currently in use at the LHC~\cite{Sirunyan:2017ulk,Aaboud:2017aca}.

Due to isospin considerations, jets on average consist of $60\%$ charged hadrons, $30\%$ photons (from neutral pion decay), and $10\%$ neutral hadrons,\footnote{These fractions are approximate and depend more specifically on color and flavor structures, hadronization dynamics, and detector momentum thresholds.} although these fractions are subject to large jet-by-jet fluctuations~\cite{Sirunyan:2017ulk,Aaboud:2017aca}.
Because charged hadrons are the dominant jet component and because neutral hadrons suffer from the poor angular resolution of the HCAL, this has led to the study of track-only observables in jet substructure \cite{Aad:2014gea,Aad:2015cua,ATLAS-CONF-2016-055,Aad:2016oit,ATLAS:2016vmy,ATLAS:2016wzt,Krohn:2012fg,Waalewijn:2012sv,Chang:2013iba,Chang:2013rca,Schaetzel:2013vka,Larkoski:2015yqa,Spannowsky:2015eba,Bressler:2015uma}.
At a high-luminosity collider, a key advantage of using only tracks is that precise vertexing can be used to achieve minimal contamination from pileup collisions.
On the other hand, jet-to-jet fluctuations of charged content cause the distributions of track-only observables to be intrinsically wider than those based on all particles; this in turn leads to reduced discrimination power in the context of jet substructure.

The aim of this paper is to quantify the degree to which neutral information is needed for discrimination tasks using jet substructure (\Sec{sec:results}), and then determine the calorimeter performance required to improve upon tracker-only information (\Sec{sec:ImprovedDetector}).
Both studies rely on the same parton shower generator tools, described in \Sec{sec:mc}, and the same suite of jet substructure observables, listed in \Sec{sec:observables}.
We use a multivariate analysis to construct optimal discriminators for various jet types, as detailed in \Sec{sec:mvas}, and we use ROC (Receiver-Operator Characteristic) curves to quantify the discrimination power under different neutral-particle and calorimeter-performance scenarios.
For consistency, we adopt the convention that signal-background pairs are always arranged in the hierarchy $W \to Z \to t \to q \to g$, such that e.g.\ $W$ samples are always treated as signal and $g$ samples are always treated as a background.

\subsection{Parton shower samples}
\label{sec:mc}

Events samples are generated both at a proton-proton center-of-mass energy of $\sqrt{s} = 100~\tev$ to model a future collider and at $\sqrt{s} = 13~\tev$ for comparisons to LHC performance.
For both energies, parton-level $W^+ W^-$, $ZZ$, $t\bar{t}$, $q\bar{q}$, and $gg$ events are first produced at leading-order using \textsc{MadGraph5\_aMC\at NLO}~\cite{Alwall:2014hca} (version 2.3.1) with the NNPDF23LO1 parton distribution functions (PDFs)~\cite{Ball:2012cx}.
In order to focus on a relatively narrow kinematic range, the transverse momenta of the partons and undecayed gauge bosons are generated in window with $\delta \pt / \pt = 0.01$, centered at $1~\tev$ for $\sqrt{s} = 13~\tev$ and at $\pt = 5~\tev$ for $\sqrt{s} = 100~\tev$.
These parton-level events are then decayed and showered in \textsc{Pythia8}~\cite{Sjostrand:2014zea} (version 8.212) with the Monash 2013 tune~\cite{Skands:2014pea}, including the contribution from the underlying event.
For each $\pt$~bin and final state, 200,000 events are generated. 

We implement a variety of jet recombination algorithms and substructure tools via the \textsc{FastJet~3.1.3} and \textsc{FastJet~contrib~1.027} packages~\cite{fastjet:1,fastjet:2}.
As a baseline, all jets are clustered using the anti-$k_{\rm{T}}$ algorithm~\cite{Cacciari:2008gp}, with a distance parameter of $R = 0.8$.
Even though the parton-level $\pt$ distribution is narrow, the jet $\pt$ spectrum is significantly broadened by kinematic recoil from the parton shower and energy migration in and out of the jet cone.
We apply a cut on the reconstructed jet $\pt$ to remove extreme events from the analysis, vetoing those outside a window of $0.8~\tev < \pt < 1.6~\tev$ for the $\pt = 1~\tev$ bin and $4.5~\tev < \pt < 6~\tev$ for the $\pt = 5~\tev$ bin.

\subsection{Jet substructure observables}
\label{sec:observables}

The jet substructure community has developed a wide variety of observables to identify the origin of a jet based on the structure of its radiation pattern.
We focus on $W/Z/t/q/g$ discrimination.  
The Higgs boson, decaying to a bottom quark pair, is excluded as its mass and substructure are quite similar to $W$ and $Z$ bosons, and Higgs-tagging relies heavily also on the identification of bottom quarks and secondary vertex reconstruction, which we do not pursue here. 
The goal of this study is not to implement an exhaustive jet substructure catalog, but rather focus on the \emph{change} in discrimination power from adding neutral or calorimeter information.
That said, by combining several observables together into a multivariate discriminant (see \Sec{sec:mvas}), one can achieve excellent tagging identification for each pairwise discrimination task, and we do not expect the qualitative lessons from this study to change substantially by adding more jet observables.

\begin{table}[t]
\centering
\begin{tabular}{|c|c|}	\hline
	\textbf{Mass observables} & \textbf{Shape observables} \\\hline\hline \hfill & \hfill\\
	$m_{\rm plain}$							&	$\tau_{N = 1,2,3}^{\beta=1,2}$			\\
	        $m_{\rm prun}$         			&		  $\tau_{21}^{\beta=1,2}$	\\
			$m_{\rm trim}$					&      $\tau_{32}^{\beta=1,2}$	\\
	 	$m_{\rm mMDT}$  & 	$C_{N =1,2}^{\beta=1,2}$		\\
	 	$m_{\rm SD}^{\beta=1,2}$		& $D_2^{\beta=1,2}$				\\
  								&	$p_{\rm T}^D$				\\
								&	$\sum z \log z$					\\
			&	Multiplicity				\\ \hfill & \hfill \\

\hline
\end{tabular}
\caption{A summary of the observables used in the analysis, roughly divided into mass-like and shape-like categories.}
\label{tab:obslist}
\end{table}

In \Tab{tab:obslist}, we list all the observables used in this study, making a distinction between ``mass'' and ``shape'' observables. 
This label denotes a conceptual difference between mass-like observables, which are sensitive to the absolute energy scale of the jet (dimensionful), and shape-like observables, which are sensitive to relative energy scales within the jet (dimensionless).
This distinction is important for interpreting the results of this study since calorimetry is most important in cases where mass resolution drives the discrimination power.
That said, a strict distinction between these observable types is not well-defined since there are some ``mass-like'' shape observables (e.g.~$C_1^{\beta = 2}$).
Moreover, because we are working with jet samples in relatively narrow $\pt$ ranges, this blurs the distinction between dimensionful and dimensionless quantities.
If we wanted to have fully orthogonal sets of information, we would have to carefully decorrelate shape-like from mass-like information as in~\Ref{Dolen:2016kst,Shimmin:2017mfk}, which we do not pursue here.
When showing results, we define discriminants which use only mass-like observables and then also define discriminants for mass- plus shape-like observables as is typically used in LHC analyses. 

\subsubsection{Mass observables}
\begin{itemize}

\item  \textbf{Jet mass:} The resonance mass of $W/Z$ bosons and $t$ quarks sets an intrinsic jet mass scale, which is unlike a typical $q/g$ jet, such that mass can be used as a primary discriminant between $W/Z/t/q/g$ jets.
The bulk of the $W/Z$ ($t$) jet mass arises from the kinematics of the two (three) jet cores that correspond to the two (three) prongs.
By contrast, the $q$/$g$ jet mass arises mostly from soft gluon radiation, with the average mass proportional to the Casimir factor (4/3 for quarks, 3 for gluons).

\item \textbf{Groomed jet mass}: Jet grooming methods such as trimming ($m_{\rm trim}$)~\cite{Krohn:2009th}, pruning ($m_{\rm prun}$)~\cite{Ellis:2009me}, modified mass drop~\cite{Butterworth:2008iy,Dasgupta:2013ihk} ($m_{\rm mMDT}$), and soft drop~\cite{Larkoski:2014wba} ($m_{\rm SD}^{\beta}$) help remove soft and wide-angle radiation from the jet cone.
Grooming reduces the jet mass of $q/g$ jets by an $\mathcal{O}(1)$ factor, while affecting $W/Z/t$ jets less and grooming them closer to their expected intrinsic mass value.
We use each of the grooming methods above to compute the groomed jet mass.
For soft drop, we compare different values of the angular exponent $\beta$ in the soft-drop condition. We use $\beta=0$, which corresponds to the modified mass drop procedure ($m_{\rm mMDT} = m_{\rm SD}^{\beta=0}$) as well as $\beta = 1,2$, all with $z_{\rm cut} = 0.1$.
The trimmed mass is computed using $r_\text{trim} = 0.2$ and $z_\text{trim} = 0.05$.
The pruned mass is computed setting the soft radiation fraction to $z = 0.15$ and internal mass selection at $r_0 = 0.5$.
\end{itemize}

In cases where only partial jet information is used for a (groomed) jet mass observable, we rescale it by the total energy of the jet:\footnote{Very similar behavior would have been obtained from rescaling by the total jet $\pt$ instead.}
\begin{equation}
m_{\text{reported}} = m_\text{partial} \frac{E_\text{total}}{E_\text{partial}}.
\end{equation}
This corresponds to a jet reconstruction strategy where the total jet energy is known but its substructure is determined from only partial information.
In the case of track-only measurements, this procedure is sometimes referred to as track-assisted mass \cite{ATLAS:2016vmy,Schaetzel:2013vka,Larkoski:2015yqa,Spannowsky:2015eba,Bressler:2015uma}, which not only reduces the bias from charged-to-neutral fluctuations but also improves the resolution on the mass.
This is a sensible strategy even with a coarse HCAL, since determining $E_\text{total}$ does not require good angular resolution, only good energy resolution.
In the discussion below, when referring to ``track-only'' or ``track-plus-photon'' observables, we are always implicitly including this rescaling factor.

\subsubsection{Shape observables}

\begin{itemize}
\item \textbf{$\mathbf{N}$-subjettiness}: The $N$-subjettiness observable $\tau_{N}$ quantifies how consistent a jet is with having $N$ or more subjets~\cite{Thaler:2010tr}.
After identifying $N$ subjet axes, $\tau_{N}$ is computed as the $\pt$-weighted distance between each jet constituent and its nearest subjet axis. 
Two different angular exponents are considered in this study: $\beta=1, 2$.
The ratio $\tau_{\mathrm{21}}^{\beta} \equiv \tau_{\mathrm{2}}^{\beta}/\tau_{\mathrm{1}}^{\beta}$ yields considerable separation between $W/Z$ jets and $q/g$ jets.
The ratio $\tau_{\mathrm{32}}^{\beta} \equiv \tau_{\mathrm{3}}^{\beta}/\tau_{\mathrm{2}}^{\beta}$ is similarly used to identify top quarks.
To aid in $q$/$g$ discrimination, individual $\tau_{N = 1,2,3}^{\beta}$ components are also included in this study, although we do not include the case of the Les Houches Angularity with $N = 1$, $\beta = 0.5$ \cite{Gras:2017jty}.
In all cases, the subjet axes are determined using one-pass $k_{\rm {T}}$ optimization \cite{Thaler:2011gf}.

\item \textbf{Energy correlation functions}:  The observables $C_{i}^{\beta}$ \cite{Larkoski:2013eya} and $D_{i}^{\beta}$ \cite{Larkoski:2015kga} can identify $N$-prong jet substructure without requiring a subjet finding procedure, only using information about the energies and pairwise angles of particles within a jet.  Here, $C_{i}^{\beta}$ and $D_{i}^{\beta}$ are ratios involving up to $(i+1)$-point correlation functions, and $\beta$ is again an angular exponent for the pairwise particle distances.  In a typical application, $C^{\beta}_{1}$ is used for $q/g$ discrimination while $C^{\beta}_{2}$ and $D^{\beta}_{2}$ are used to identify $W/Z$ bosons.  In this study, we consider $\beta=1, 2$, and we do not include more general energy correlation functions like $N_{i=2,3}^{\beta}$ \cite{Moult:2016cvt}.  We also include the infrared and collinear unsafe case of $C^{\beta}_{1}$ with $\beta = 0$, which has the same discrimination power as $p^D_{\rm{T}}$ \cite{Pandolfi:1480598,Chatrchyan:2012sn}.  Of course, including more correlators could further aid in $W/Z/t/q/g$ separation---e.g.\ $\beta=0.5$ or $0.2$ is known to improve $q/g$ separation for $C_{i}^{\beta}$ \cite{Larkoski:2013eya} and there are powerful four-point correlator combinations like $D_{3}^{\beta}$~\cite{Larkoski:2014zma}---but we do not expect them to make a qualitative difference to this study.

\item \textbf{$\sum\mathbf{z\,\text{log}\, z}$:} The fraction of energy carried away by daughter particles within a jet, $z_i = E_i/E_{\text{parent}}$, is a measure of jet softness, and can be used to separate quark from gluon jets.  Following~\Ref{Larkoski:2014pca}, we consider a sum over the jet constituents of $\sum_i z_i \log z_i$.  In cases where only partial information is used, we define $E_{\text{parent}}$ using the reduced information such that $\sum_i z_i = 1$.

\item \textbf{Hadron multiplicity}: The number of jet constituents is powerful both for $q/g$ discrimination and for discriminating color singlet jets like $W/Z$ bosons versus colored objects like $q/g$ (see e.g.~\cite{Gallicchio:2011xq,Gallicchio:2012ez,Aad:2014gea,Larkoski:2014pca,ATLAS:2016wzt,Aad:2016oit}).

\end{itemize}

To avoid a proliferation of observables, we decided to compute all of the shape observables without applying any jet grooming techniques.  It is known that grooming can fundamentally alter the discrimination power of shape observables \cite{Moult:2016cvt,Salam:2016yht}, though we do not expect the \emph{relative} gains in including neutral or calorimeter information to be much affected by including groomed jet shapes.

\subsection{Multivariate analysis}
\label{sec:mvas}

To quantify the discrimination power of the above mass and shape observables, a boosted decision tree (BDT) classifier is trained for discriminating each pair of jet sample types.
This discriminator is close to optimal (given the information provided), allowing us to de-emphasize the power of any single observable and focus on the combined performance for a given level of neutral or calorimeter information.

The multivariate analysis is implemented via the TMVA package~\cite{Hocker:2007ht}.
The resulting BDT forest is composed of 50 trees, each with a maximum depth of 10 branches and at most, $10^6$ nodes. We apply a variety of techniques to improve the decision quality of the classifier and prevent overtraining, including gradient boosting \cite{Friedman00greedyfunction}.
More powerful multivariate and jet imaging techniques may be somewhat better in performance~\cite{deOliveira:2015xxd}, though in practice, most gains over a BDT approach are at the level of $10$--$20\%$ \cite{Baldi:2016fql,Datta:2017rhs}.

Part of the output of the BDT is a pair of distributions taking values from $-1$ to $1$, one for the signal-type jet and one for the background-type jet in the training.
These distributions are used to quantify our ability to discriminate between pairs of SM objects.
The primary way we illustrate discrimination power is through ROC curves, which plot the background efficiency for a given signal efficiency.
Here, the signal efficiency ($\epsilon_{\rm sig}$) is the cumulative density of one of the two distributions, which we label $f_s(t)$,
	\begin{align}
		\epsilon_{\rm sig} = \int_{-1}^x \textrm{d}t \, f_s(t),
	\end{align}
and the background efficiency ($\epsilon_{\rm bkg}$) is the same integral over the other distribution, $f_b(t)$.

\section{Performance impact of neutral-particle information}
\label{sec:results}

In this section, we quantify the gains in jet substructure discrimination performance from including neutral-particle information compared to using only track-based observables.
We consider three different categories of particles to calculate the mass- and shape-like observables from \Sec{sec:observables}:  (i) charged particles only, ``tracks'', (ii) charged particles plus photons, ``tracks+$\gamma$'', and (iii) all particle types, ``all particles''.
These categories roughly map onto the information available in (i) just the tracker, (ii) the tracker plus ECAL, and (iii) the whole detector.
We emphasize, however, that this is a truth-particle-level study; the impact of detector configurations is considered in \Sec{sec:ImprovedDetector}.

For our study, we considered all pairwise discrimination tasks for $W/Z/t/q/g$ jets, for a total of ten different binary BDT classifiers.
We also performed separate studies on two $\pt$~regimes: 1~TeV (at a 13 TeV machine) and 5~TeV (at a 100 TeV machine).
Given the many permutations, we focus our narrative on three tasks---$W$ vs.\ $q$, $W$ vs.\ $Z$, and $q$ vs.\ $g$---since they capture well the differing impact of neutral-particle information.
Specifically, $W$ vs.~$q$ discrimination depends on both mass and shape observables; $W$ vs.~$Z$ is affected primarily by mass observables; and $q$ vs.~$g$ exploits mostly shape observables.  
Additional comparisons including top jets are in \App{sec:appendix}.

\begin{figure}[t]
\begin{center}
\includegraphics[width=0.45\linewidth]{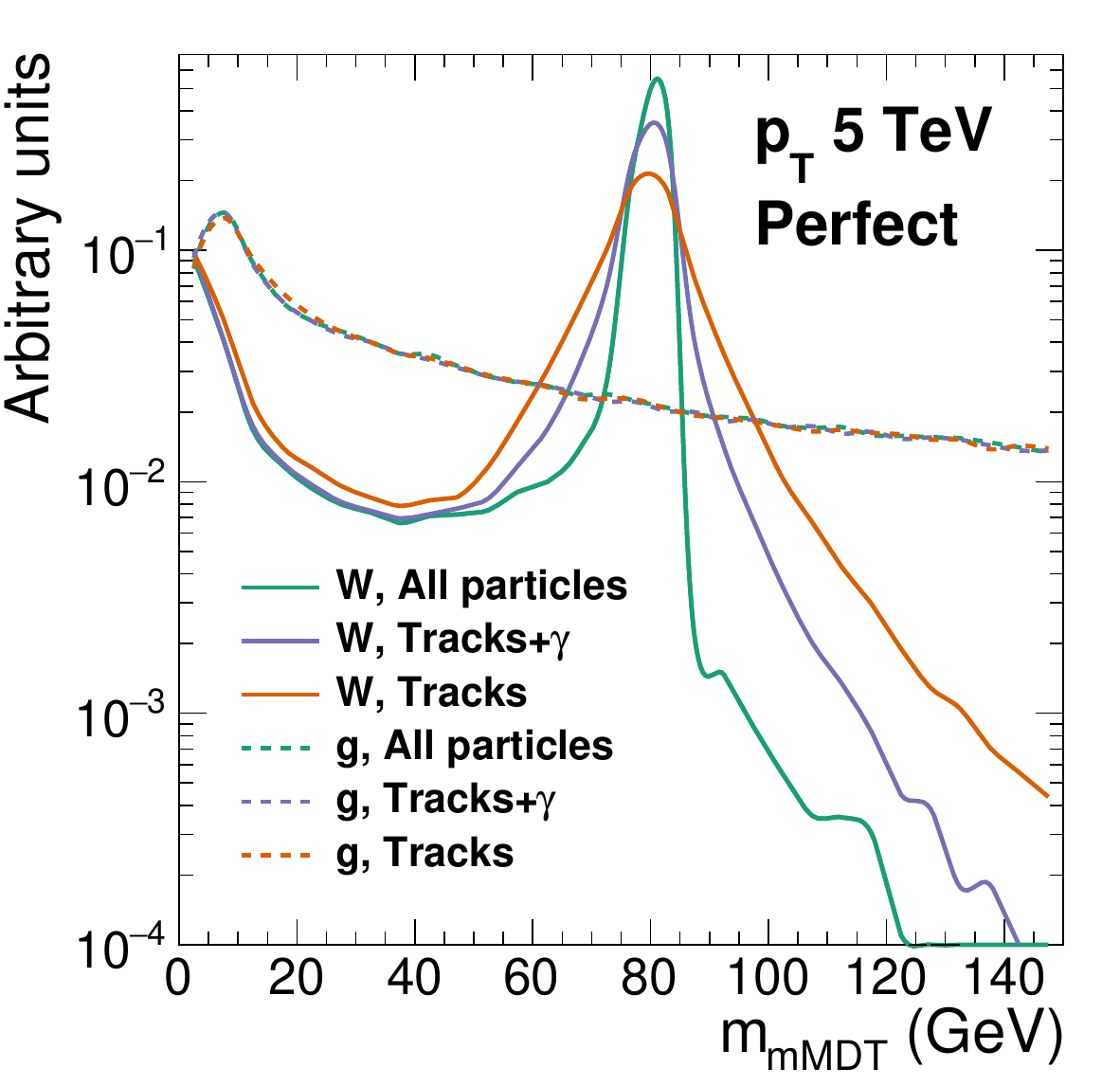}
$\qquad$
\includegraphics[width=0.45\linewidth]{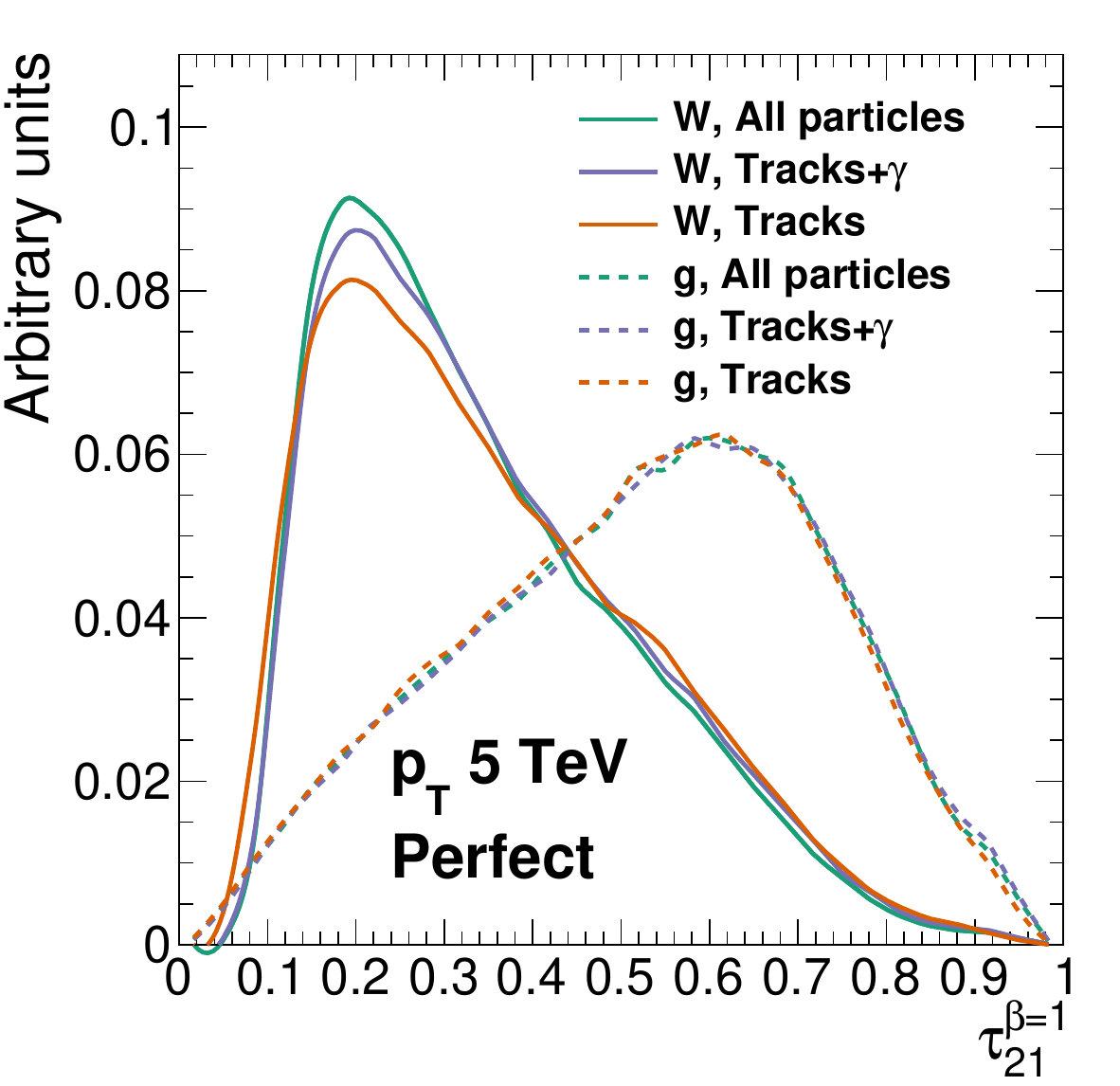}\\[-0.6em]
\end{center}
\caption{Distributions of (left) ${m}_{\rm mMDT}$ and (right) $\tau^{\beta = 1}_{21}$ for $W$ jets and gluon-initiated jets in the $5$~TeV $\pt$~bin.  Three different particle categories are considered to calculate these observables:  tracks, tracks+$\gamma$ and all particles, as defined in the text. }
\label{fig:SummaryPlots_Wvq_InfoComparison}
\end{figure}

In \Fig{fig:SummaryPlots_Wvq_InfoComparison}, we show example distributions of $m_{\rm mMDT}$ and $\tau_{21}^{\beta=1}$ obtained from the $W$ and gluon parton shower samples from \Sec{sec:mc}, assuming a perfect detector.
Immediately one notices the differing role that neutral-particle information plays for mass-like compared to shape-like observables.
For the $W$ sample with an intrinsic mass scale, the groomed mass resolution degrades quickly in going from all particle information to just tracks.
By contrast, the $N$-subjettiness shape information is much more stable as the level of neutral information is diminished.
These features will be reflected in the ROC curves below.


\subsection{\texorpdfstring{$W$}{W} jets vs.\ quark jets}

\begin{figure}[p]
\begin{center}
\includegraphics[width=0.45\linewidth]{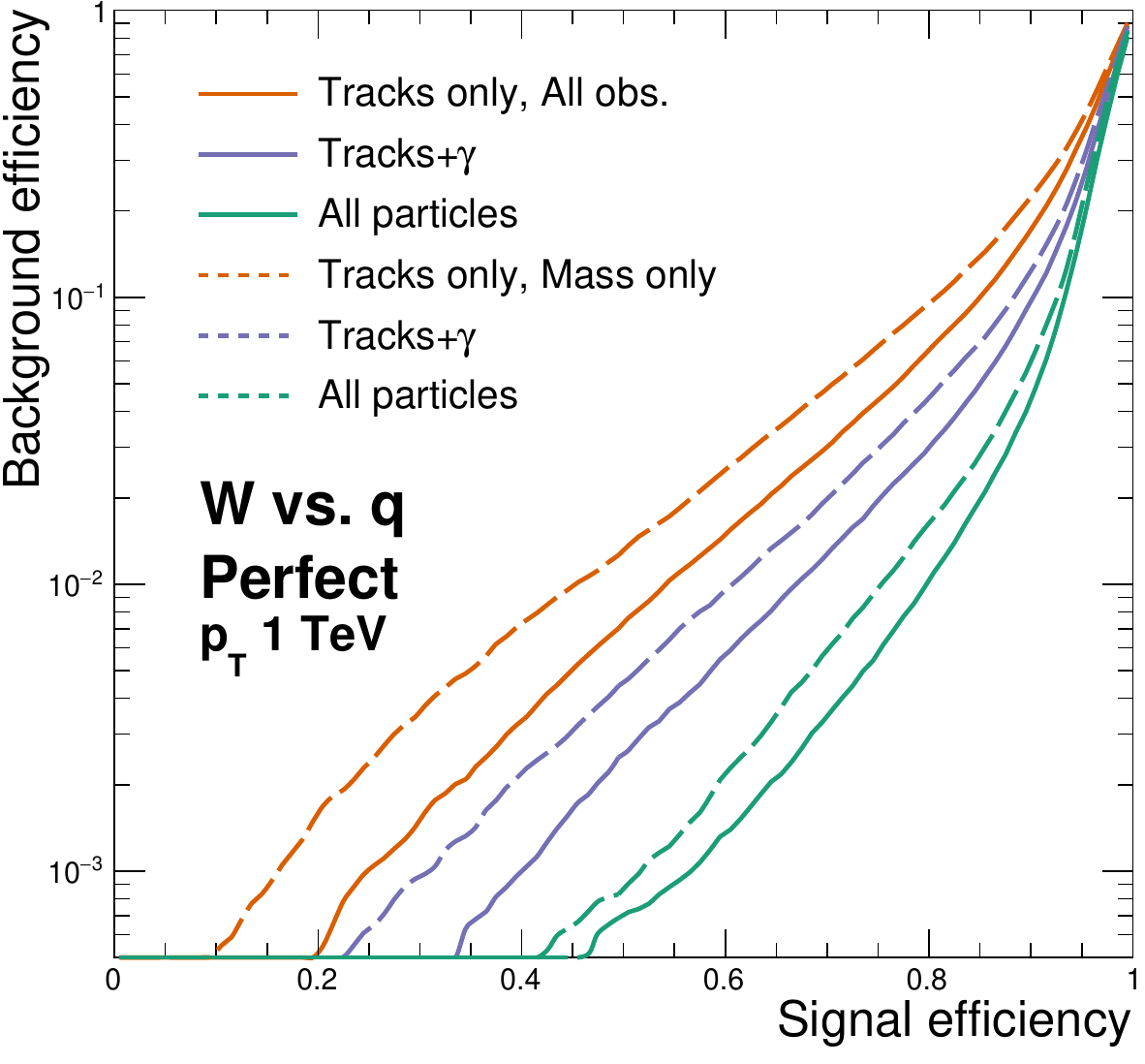}
$\qquad$
\includegraphics[width=0.45\linewidth]{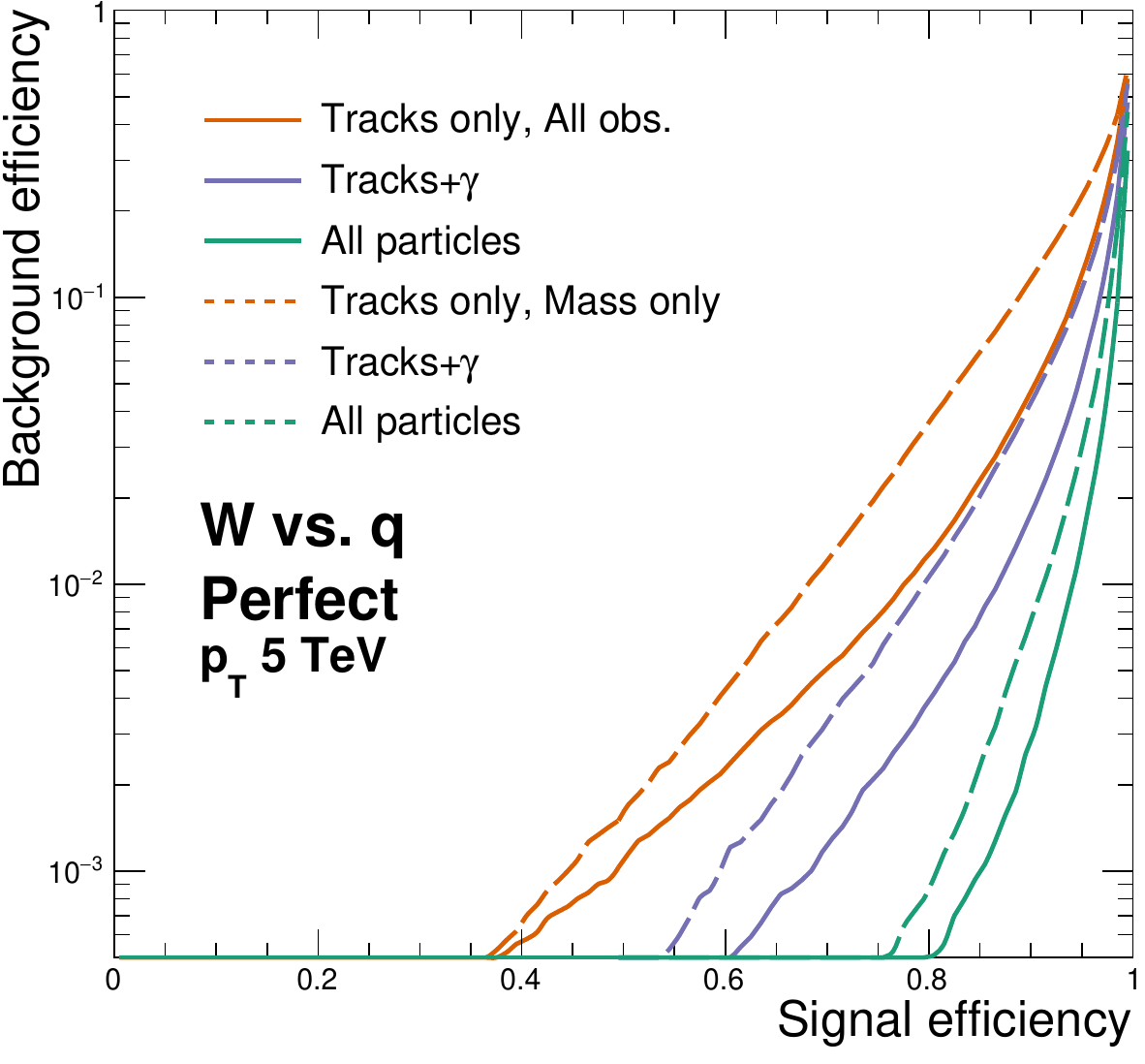}
\end{center}
\caption{ROC discrimination curves for the (left) $1$~TeV and (right) $5$~TeV~$\pt$ bins. $W$-tagging (signal) versus $q$-tagging (background) efficiencies are shown for (dashed) mass and (solid) mass-plus-shape observables, and for three different particle categories: tracks, tracks+$\gamma$ and all particles.  Here, a perfect detector is assumed.}
\label{fig:ROC_Wq_F1}
\end{figure}

\begin{figure}[p]
\begin{center}
\includegraphics[width=0.45\linewidth]{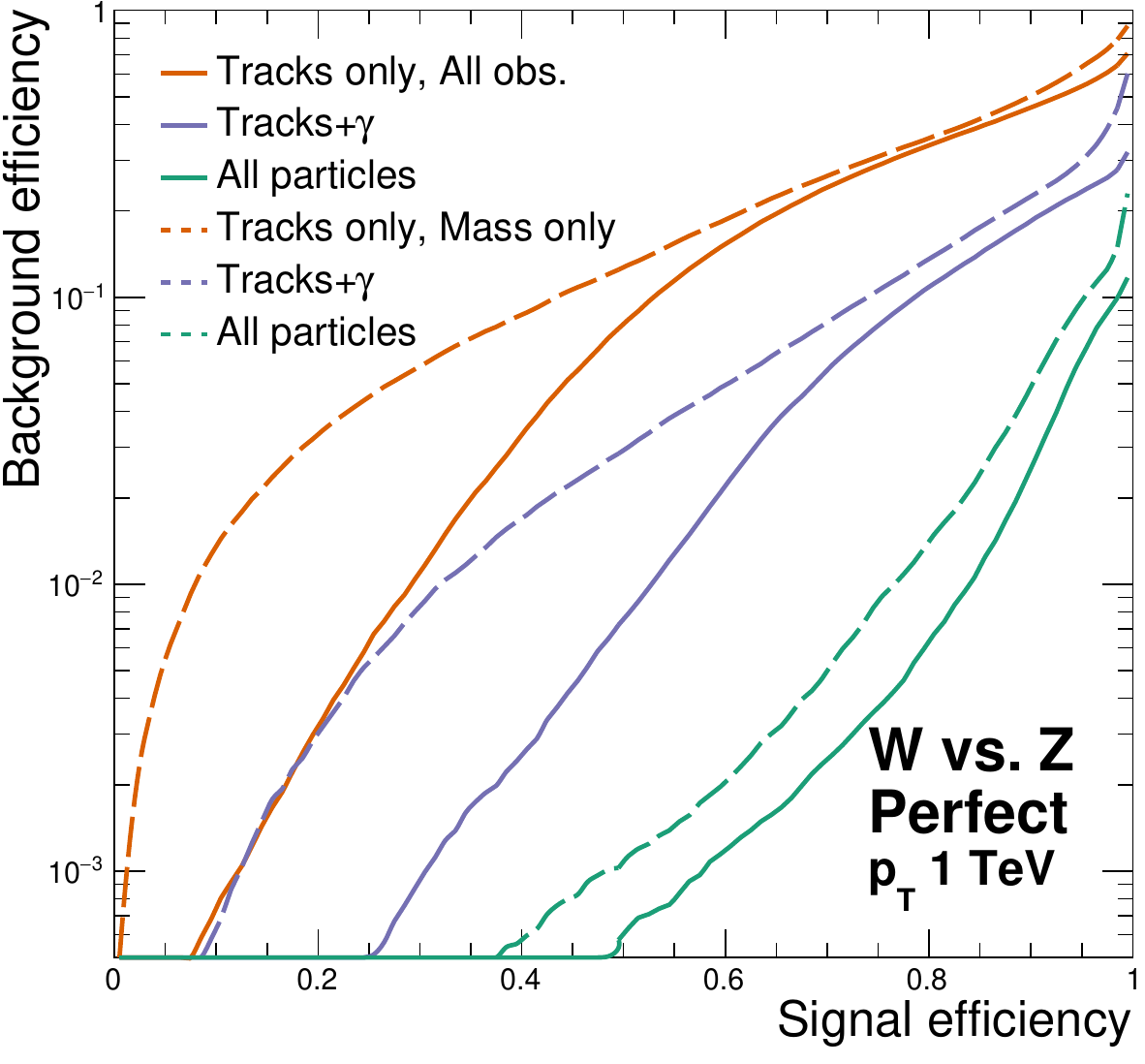}
$\qquad$
\includegraphics[width=0.45\linewidth]{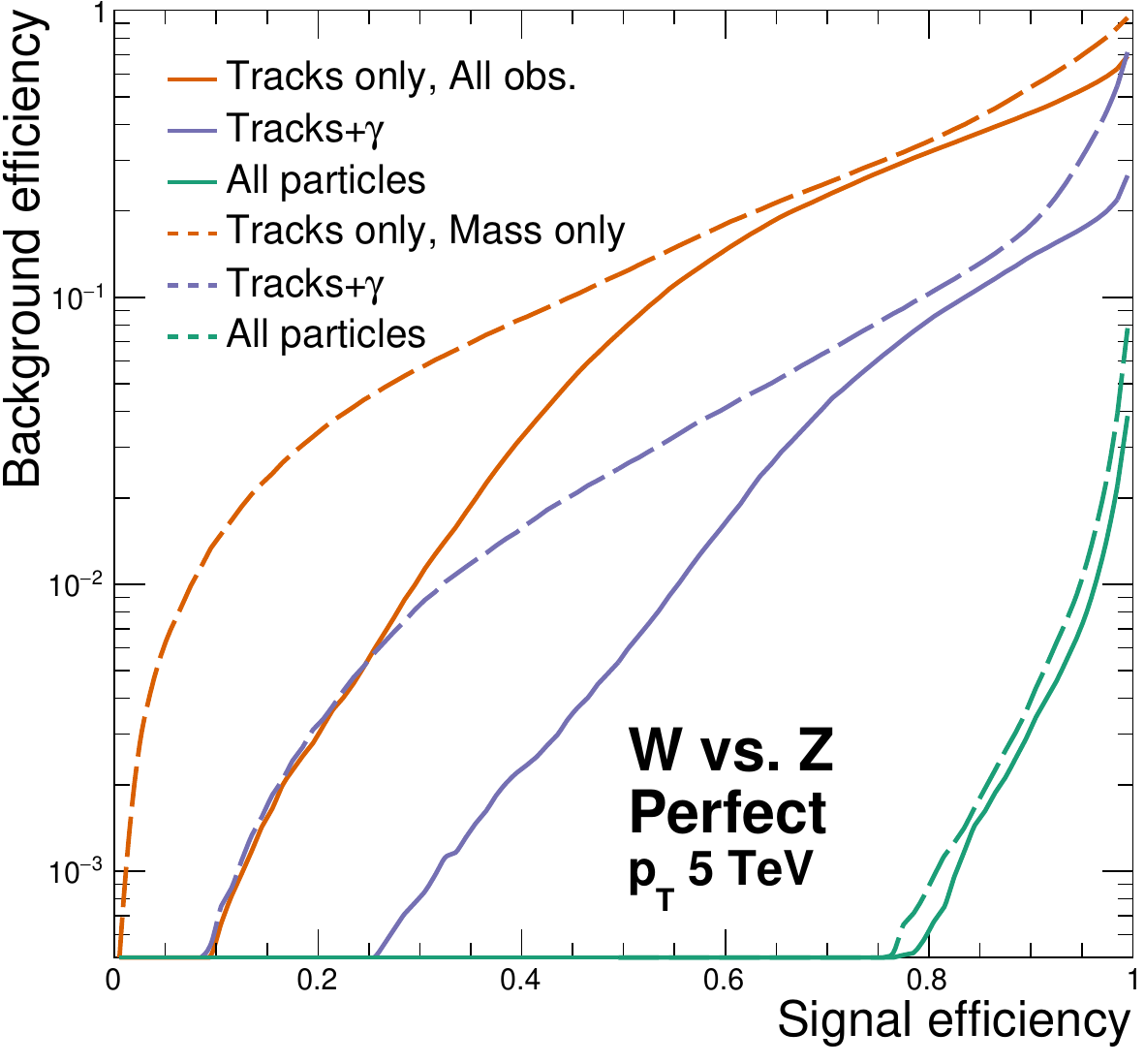}
\end{center}
\caption{Same as \Fig{fig:ROC_Wq_F1}, but for $W$ jets (signal) versus $Z$ jets (background).}
\label{fig:ROC_WZ_F1}
\end{figure}

We begin by comparing $W$ jet versus quark jet discrimination using different degrees of neutral-particle information.
In \Fig{fig:ROC_Wq_F1}, the ROC discrimination curves are reported as $W$ jet tagging efficiency versus quark jet tagging efficiency for BDT discriminants built on just mass observables and, separately, on all observables (i.e.~mass plus shape).
The reason we do not show a BDT from just shape observables is twofold:  first, almost all jet substructure analyses use mass as part of their discrimination strategies; and second, as discussed in \Sec{sec:observables}, shape observables like $C_1^{\beta = 2}$ are highly correlated with mass when considering a narrow $\pt$ bin.

As more neutral particle information is added, we see that $W/q$ discrimination power is indeed improved.
This is expected for mass-only observables, since, as already seen in the left panel of \Fig{fig:SummaryPlots_Wvq_InfoComparison}, the $W$ boson mass resolution improves as photons and neutral hadrons are included.
Since $W$ bosons and quark jets have different prong-like structures, shape information provides additional discrimination power beyond just mass.
However, the \emph{relative} improvement offered by shape information does not improve as neutral information is added, since charged particles already provide substantial information on the overall jet topology.

Finally, in comparing the left (1 TeV) and right (5 TeV) panels of \Fig{fig:ROC_Wq_F1}, one sees that the overall discrimination power increases at higher $\pt$. 
This is expected, at least when assuming a perfect detector, since more energetic quark jets radiate more copiously, whereas $W$ bosons are color singlets and therefore have a fixed radiation pattern with $\pt$.


\subsection{\texorpdfstring{$W$}{W} jets vs.\ \texorpdfstring{$Z$}{Z} jets}

We now turn to $W$ jet versus $Z$ jet discrimination.
Because $W$ and $Z$ jets are topologically quite similar, mass resolution plays a crucial role in this task. 
In \Fig{fig:ROC_WZ_F1}, the ROC discrimination curves are reported as $W$ jet tagging efficiency versus $Z$ jet tagging efficiency, using the same configurations as \Fig{fig:ROC_Wq_F1}.
When all particle information is included, shape information is quite subdominant to the mass information, as expected since the primary difference between $W$ and $Z$ is the mass peak at 80 GeV versus 91 GeV.
When neutral-particle information is removed, the discrimination power degrades dramatically, though the performance can be partially recovered by supplementing mass with shape observables.
As we will see below, the large separation power seen in the all-particle case cannot be realized in practice given calorimeter limitations.
Still, this example shows the dramatic gains possible from neutral-particle information when mass resolution drives performance.

Note that $W/Z$ separation improves at higher $\pt$ when using all particles.  The reason is that mass resolution can be degraded by underlying event and initial state radiation, which have a proportionally smaller impact as the jet $\pt$ increases.


\subsection{Quark jets vs.\ gluon jets}

\begin{figure}[t]
\begin{center}
\includegraphics[width=0.45\linewidth]{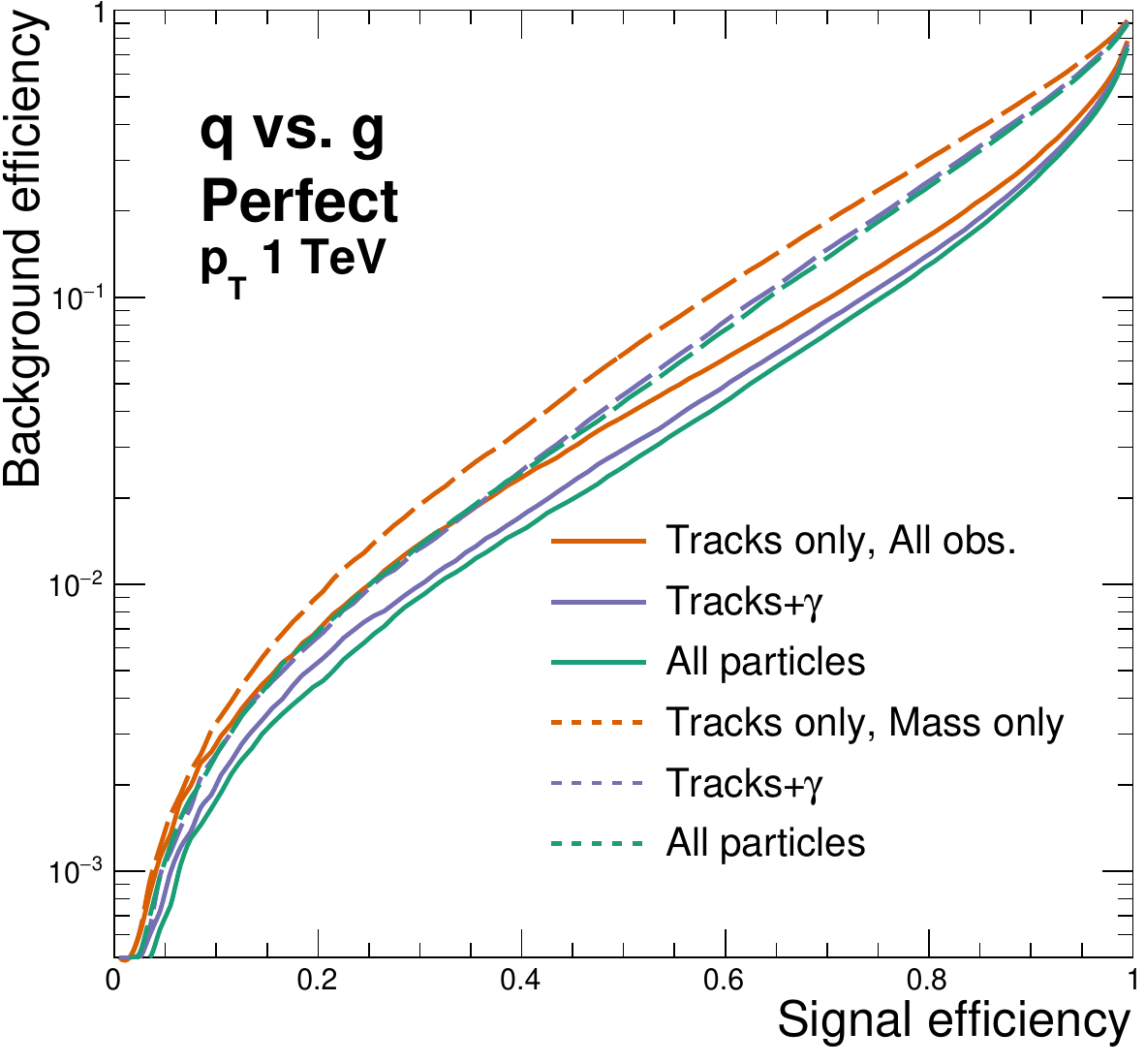}
$\qquad$
\includegraphics[width=0.45\linewidth]{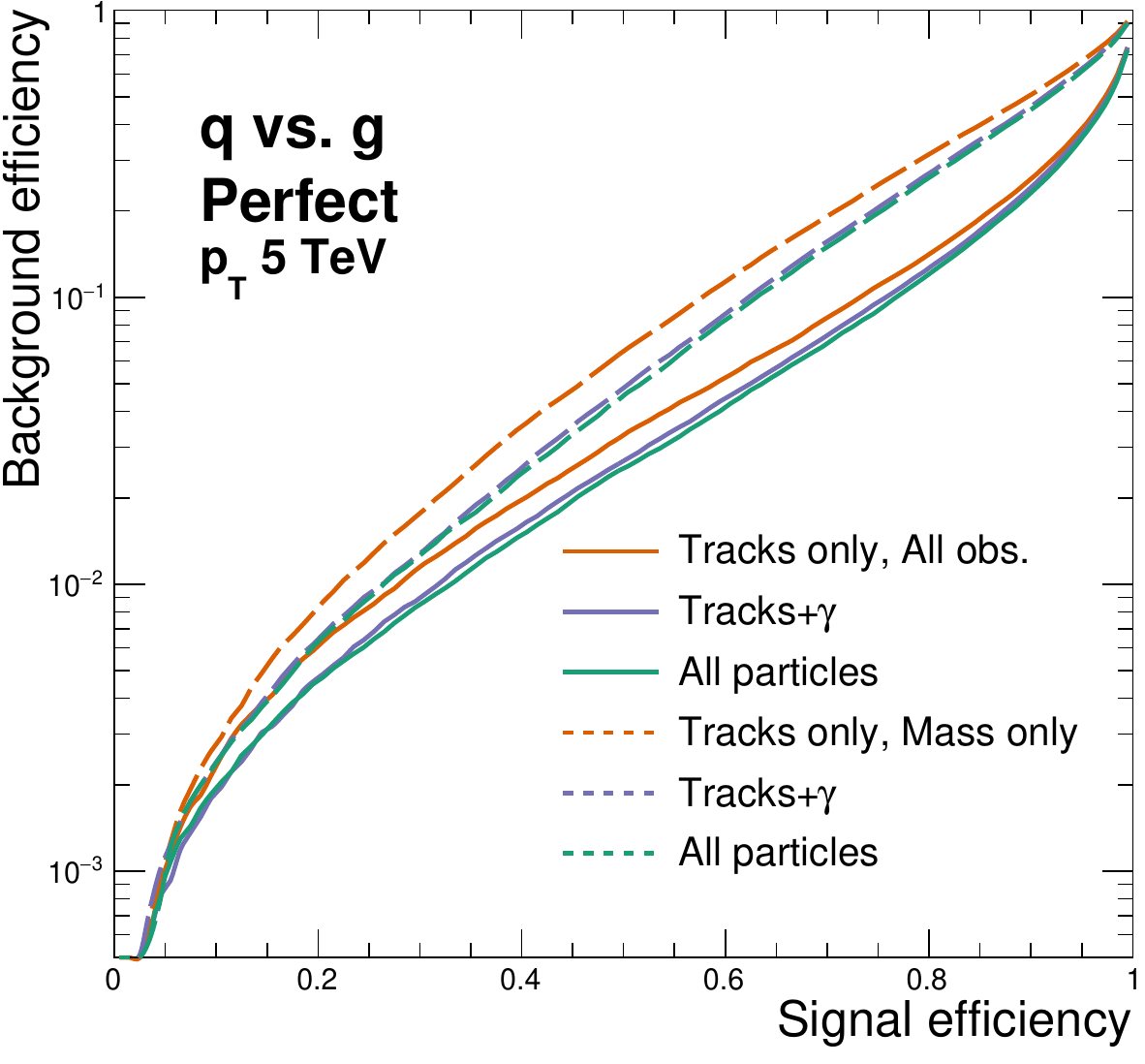}
\end{center}
\caption{Same as \Fig{fig:ROC_Wq_F1}, but for quark jets (signal) versus gluon jets (background).}
\label{fig:ROC_qg_F1}
\end{figure}

Our final case study is quark versus gluon tagging.
In \Fig{fig:ROC_qg_F1}, the ROC discrimination curves are reported as quark jet tagging efficiency versus gluon jet tagging efficiency, using the same configurations as \Fig{fig:ROC_Wq_F1}.
Because there is no intrinsic mass scale for quark and gluon jets, discrimination between the two is dominated by shape information.
While mass observables appear to perform well on their own, this is primarily due to the use of a narrow $\pt$ bin, allowing the masses to be used as a proxy for the amount of radiation in the jet.
The differences in the discrimination power between the different particle categories and observable sets are smaller than in the prior examples, and the discrimination power is fairly independent of $\pt$.
With no parametric separation in phase space between quark and gluon jets~\cite{Larkoski:2014gra}, such similarities are expected. 

Quark/gluon tagging is an example where neutral-particle information has the smallest impact since shape information about the jets can be adequately captured using only partial jet information.
From a perspective of cost over performance optimization, there is a noticeable gain in going from tracks to tracks plus photons, though perhaps not much to be gained beyond that in using all particles.
In cases like this where absolute discrimination power is fairly independent of the input particle collections, calorimeters are expected to play a subdominant role. 

\subsection{Summary of truth-level study}
\label{sec:ratiospartial}

The three case studies above are just a sampling of the ten pairwise tasks we tested.
It is instructive to summarize the other seven tasks, as well as summarize the gain in discrimination power when adding neutral-particle information.
To do so, we compute the background rejection (i.e.~one over the background efficiency) for each discriminant at an operating point with fixed signal efficiency, $S(\%)$.
We then take the ratios of the background rejection factors between scenarios $A$ and $B$, representing different particle categories:
\begin{equation}
\mathcal{R}_{S} = \left(\frac{\epsilon_\text{bkg}^A( \epsilon_\text{sig}^A = S(\%))}{\epsilon_\text{bkg}^B( \epsilon_\text{sig}^B = S(\%))} \right)^{-1}.
\label{eq:ratio}
\end{equation}
The ratio illustrates the performance gained (in background rejection) by adding neutral-particle information.

\begin{figure}[t]
\begin{center}
\includegraphics[width=0.45\linewidth]{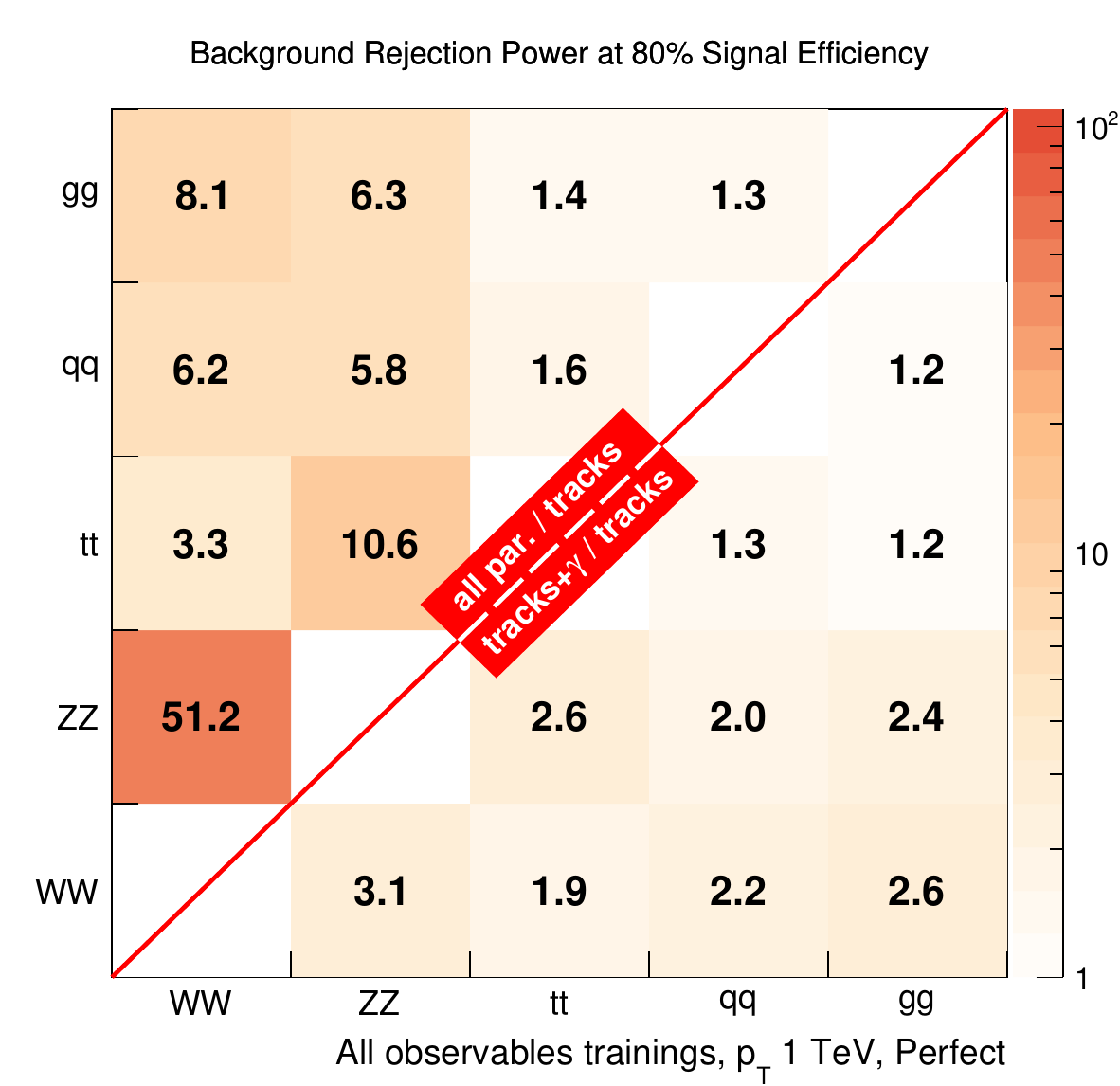}
$\qquad$
\includegraphics[width=0.45\linewidth]{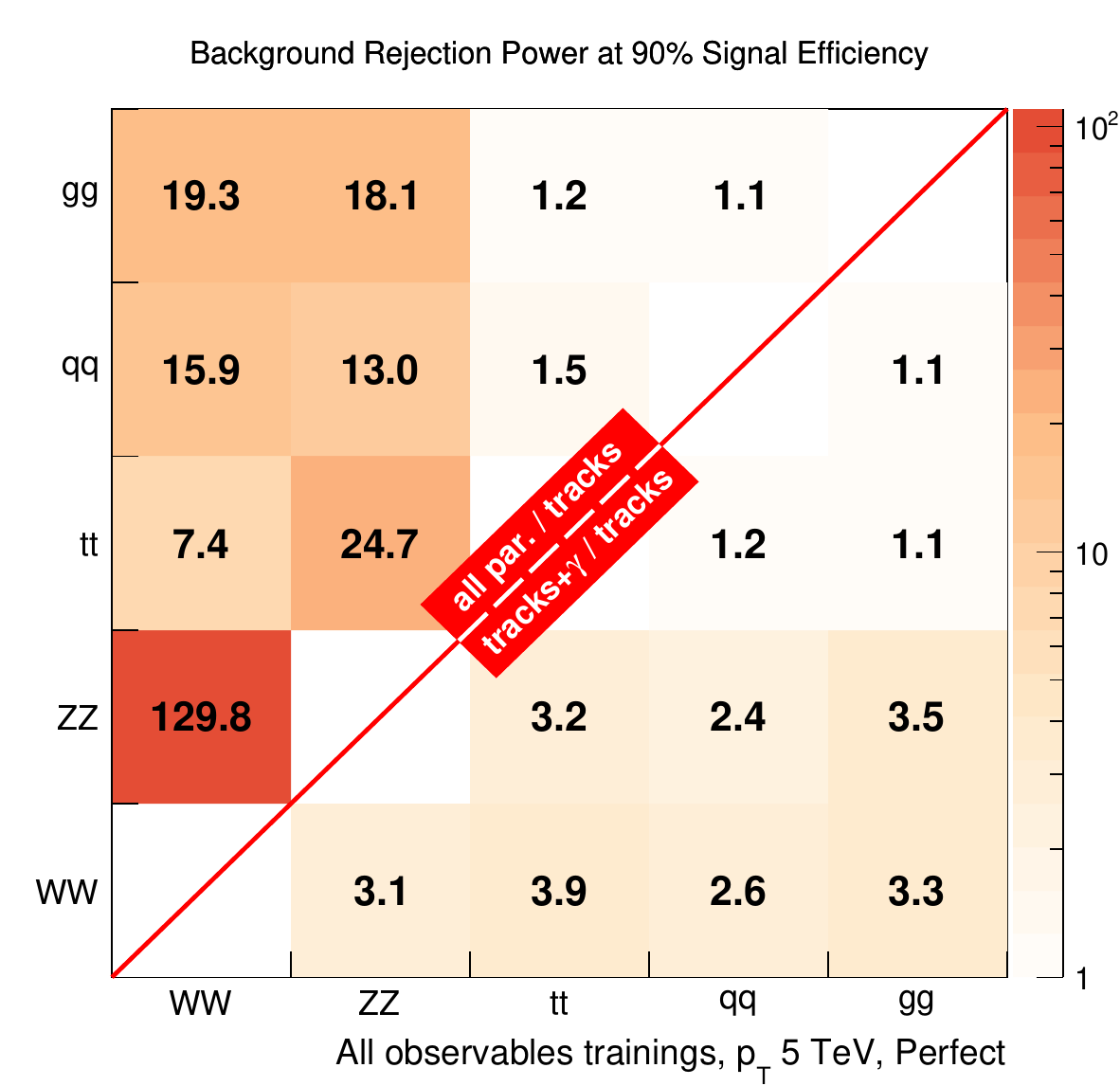}
\end{center}
\caption{Background rejection ratio (left) $\mathcal{R}_{80}$ for the $\pt =$~1 TeV bin and (right) $\mathcal{R}_{90}$ for the $\pt =$~5 TeV bins, for all pairwise discrimination tasks among $W$/$Z$/$t$/$q$/$g$ jets, using mass-plus-shape observables and a perfect detector.  Values in the lower right triangle refer to the ratio (tracks$+\gamma$)/(tracks), while those in the upper left triangle refer to (all~particle)/(tracks).   All such plots in this paper share the same color scale.}
\label{fig:MAPSPerfect}
\end{figure}

In \Fig{fig:MAPSPerfect}, we display these ratios for the $\pt=1~(5)$~TeV bin for $S(\%)=80\%~(90\%)$, arranged in a $5 \times 5$ matrix for each pairwise comparison of $W$/$Z$/$t$/$q$/$g$ jets.
As the scenario $A$ baseline, we take the track-only particle category.
We then compare to scenario $B$ with either tracks plus photons (lower right triangle) or all particles (upper left triangle).
As expected from the case studies above, the improvement in adding neutral-particle information is marginal for quark vs.\ gluon jets, but is more significant for $W$/$Z$ vs.\ $q$/$g$ jets.
The gains are largest when mass resolution is most important such as in $W$ vs.~$Z$ discrimination.  
In all cases, using tracks+$\gamma$ yields a noticeable improvement compared to just tracks, but the gains are always better using all particle information.

When inspecting the full $5 \times 5$ discrimination matrix, we find an interesting feature in top vs. $q$/$g$ jet discrimination.
Contrary to naive expectations, the gains when including neutral-particle information are not as strong as in the $W$/$Z$ vs. $q$/$g$ cases, despite the fact that the top quark has an intrinsic mass scale.
It is important to note that the top jet case is more complication due to the colored, 3-prong nature of the top quark, and we suspect that there are two effects that render absolute mass information less relevant.
The first is that the absolute mass scale for the top jet ($m_\text{top}$) is much larger than the $W$ and $Z$ jet cases, leading to greater baseline signal/background separation.  
The second is that the top jet, being colored, tends to have a worse intrinsic mass resolution compared to $W$/$Z$ color singlets.
This ultimately results in mass resolution being less important for top vs. $q$/$g$ jet discrimination, and therefore the degradation in mass resolution from the lack of neutral-particle information is less pronounced.


\section{Performance impact of improved calorimetry}
\label{sec:ImprovedDetector}

The results in the previous section demonstrate the importance of neutral-particle information for hadronic jet identification at a general-purpose detector.
Of course, the above analysis assumed a perfect detector, neglecting the significant degradation due to the granularity and smearing that are present in real detector systems.
To place our findings in a more realistic context, we now study the effects of improved calorimeter granularity and resolution on the discrimination of $W/Z/t/q/g$ jets.

\subsection{Detector configurations}
\label{sec:detectors}


\begin{table}[t]
   \centering
   \begin{tabular}{|r|c|l|l|l|}\hline
   \textbf{Scenario:}& \textbf{Perfect} & \textbf{CMS-like} & \textbf{Future 1} & \textbf{Future 2}  \\\hline
   \hline
   $\sigma_{\text{HCAL}}^{\eta}/E=\sigma_{\text{HCAL}}^{\phi}/E$ & \multirow{5}{*}{0} & 0.022  & 0.01 & 0.002 \\\cline{1-1}\cline{3-5}
   $\sigma_{\text{ECAL}}^{\eta}/E=\sigma_{\text{ECAL}}^{\phi}/E$ & & 0.0175 & 0.005 & 0.001 \\\cline{1-1}\cline{3-5}
   $\sigma_{\text{charged particles}}^{E}/E$ & & $0.00025\, \pt \oplus 0.015$ & \multicolumn{2}{c|}{\multirow{3}{*}{$\Bigg\}\times \,\,\frac{1}{2}$}} \\ \cline{1-1}
   $\sigma_{\text{photons}}^{E}/E$ & & $0.021/\sqrt{E} \oplus 0.094/E \oplus 0.005$ & \multicolumn{2}{c|}{} \\\cline{1-1}
   $\sigma_{\text{neutral hadrons}}^{E}/E$ & & $0.45/\sqrt{E} \oplus 0.05$ & \multicolumn{2}{c|}{}\\\hline
   $E_{\text{track}}^{\text{max}}$ & $\infty$ & \multicolumn{3}{c|}{220 GeV} \\\hline
   \end{tabular}
   \caption{Smearing resolutions and calorimeter granularities for the detector components in each hypothetical detector scenario. Unless otherwise specified, all quantities are dimensionless. Resolutions which depend on $\pt$ and $E$ are scaled with coefficients measured in GeV$^a$, for appropriate $a$. }
   \label{tab:scenarios}
\end{table}

We develop a custom detector simulation which reproduces the main resolution effects relevant for jet substructure reconstruction
through particle level smearing and granularization described in detail below.  
%
This simulation is representative of current and future detector concepts employing particle-flow-based reconstruction, such as the CMS~\cite{Sirunyan:2017ulk} or ATLAS~\cite{Aaboud:2017aca} detectors at the LHC.

In the simulation, we first categorize the generated particles into charged particles, photons (including $\pi_0\to\gamma\gamma$), and neutral hadrons.
Tracking inefficiencies occur at high particle momenta and within high momentum jets, where the tracking detector granularity is not sufficient to reconstruct highly-collimated particles.
Since both inefficiencies are correlated within high momentum jets, they are simulated together by treating charged particles with momenta above a threshold $E_{\text{track}}^{\text{max}}$ as neutral hadrons.
CMS and ATLAS also have tracking inefficiencies at lower momenta and there is an ambiguity in the mapping of calorimeter deposits to charged and neutral particles which is enhanced in high-momentum jets.
Those effects are not explicitly simulated, though, jet substructure reconstruction performance was found to be well-reproduced approximating all tracking-related effects with a simple $E_{\text{track}}^{\text{max}}$ threshold.
The generated neutral hadrons are then discretized to simulate the spatial resolution $\sigma_{\text{CAL}}^{\eta}=\sigma_{\text{CAL}}^{\phi}$ of the ECAL and HCAL.
Finally, all particles are smeared according to parametrized resolutions $\sigma_{\text{particle}}^{E}$ for each particle type.
To match onto the study of \Sec{sec:results}, we sometimes include results for a perfect detector as well.

As a baseline, we consider a scenario that represents the future performance of the CMS detector~\cite{CMS}.
The threshold $E_{\text{track}}^{\text{max}}$ for this scenario is chosen such that it matches the jet mass resolution at high momenta obtained from simulations of the current CMS detector~\cite{CMS-PAS-JME-14-002}, and then increased by a factor 2 as suggested by simulation studies for the HL-LHC Phase-II upgrade of the CMS tracker~\cite{CMS-TDR-17-001}.
This improvement comes from a higher granularity pixel detector which will better distinguish hits from nearby high $\pt$~tracks.
Using $W$ and $q$/$g$ jets with transverse momenta in the range of 300 GeV and 3.5 TeV, we compared the resolution in jet mass and substructure observables obtained from this scenario to CMS public results and found them to be compatible~\cite{CMS-PAS-JME-16-003, Khachatryan:2014vla, Sirunyan:2016cao}.
The $W$ and  $q$/$g$ jet selection efficiencies also agree with the CMS performance, which are similar to ATLAS as well.

Beyond the CMS-like configuration, we consider two more detector scenarios, which include improved calorimeter performance as proposed for future detectors~\cite{Chekanov:2016ppq}.
These two scenarios are summarized in Table~\ref{tab:scenarios} as ``Future 1'' (F1) and ``Future 2'' (F2).
Both scenarios feature a factor of two improvement in the calorimeter energy resolution compared to a CMS-like configuration.
The calorimeter granularity is also improved by at least a factor of two (for F1) and ten (for F2), with more significant improvements in the ECAL compared to the HCAL.

\begin{figure}[p]
\begin{center}
\includegraphics[width=0.42\linewidth]{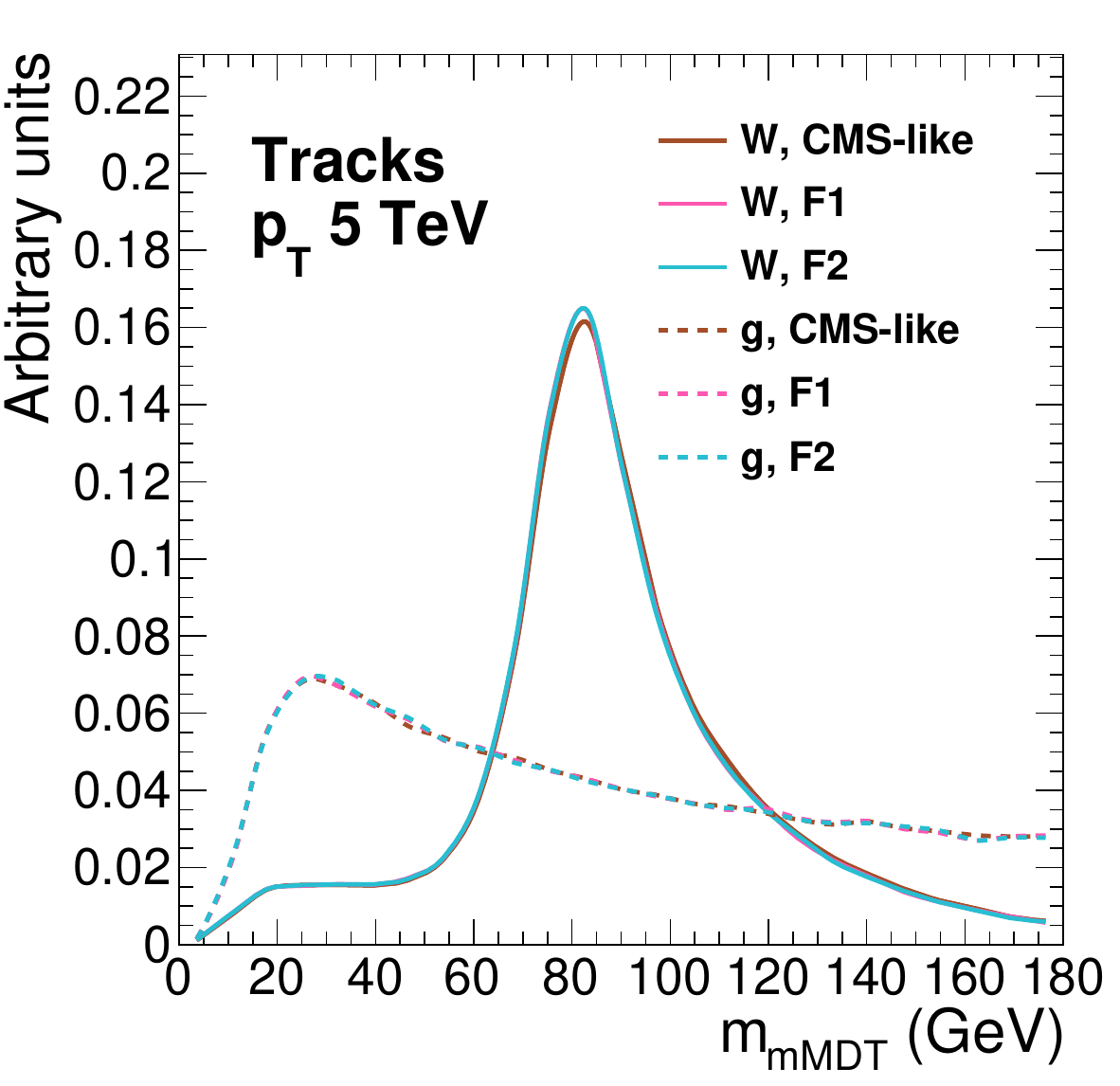}
$\qquad$
\includegraphics[width=0.42\linewidth]{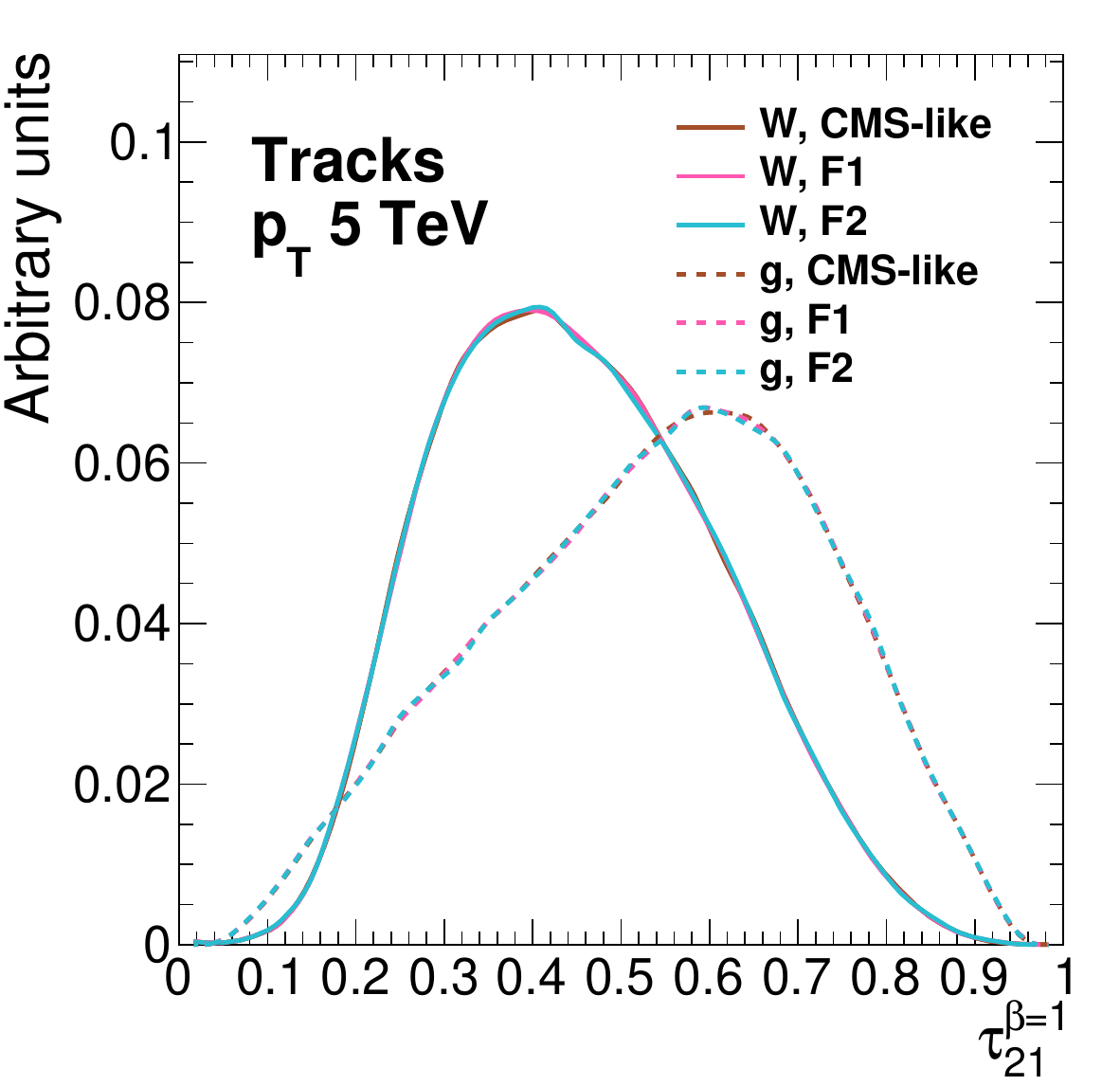}\\
\includegraphics[width=0.42\linewidth]{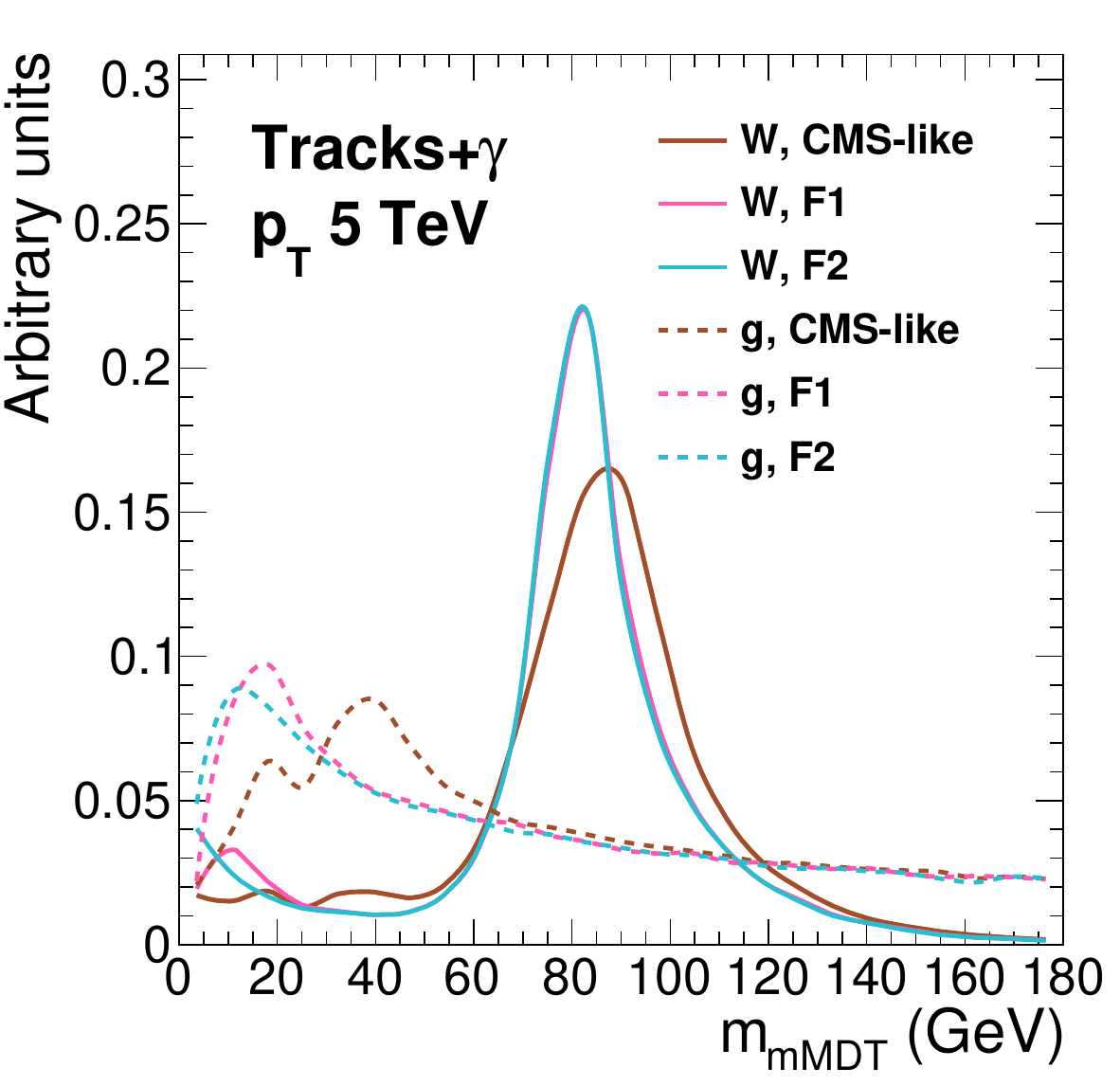}
$\qquad$
\includegraphics[width=0.42\linewidth]{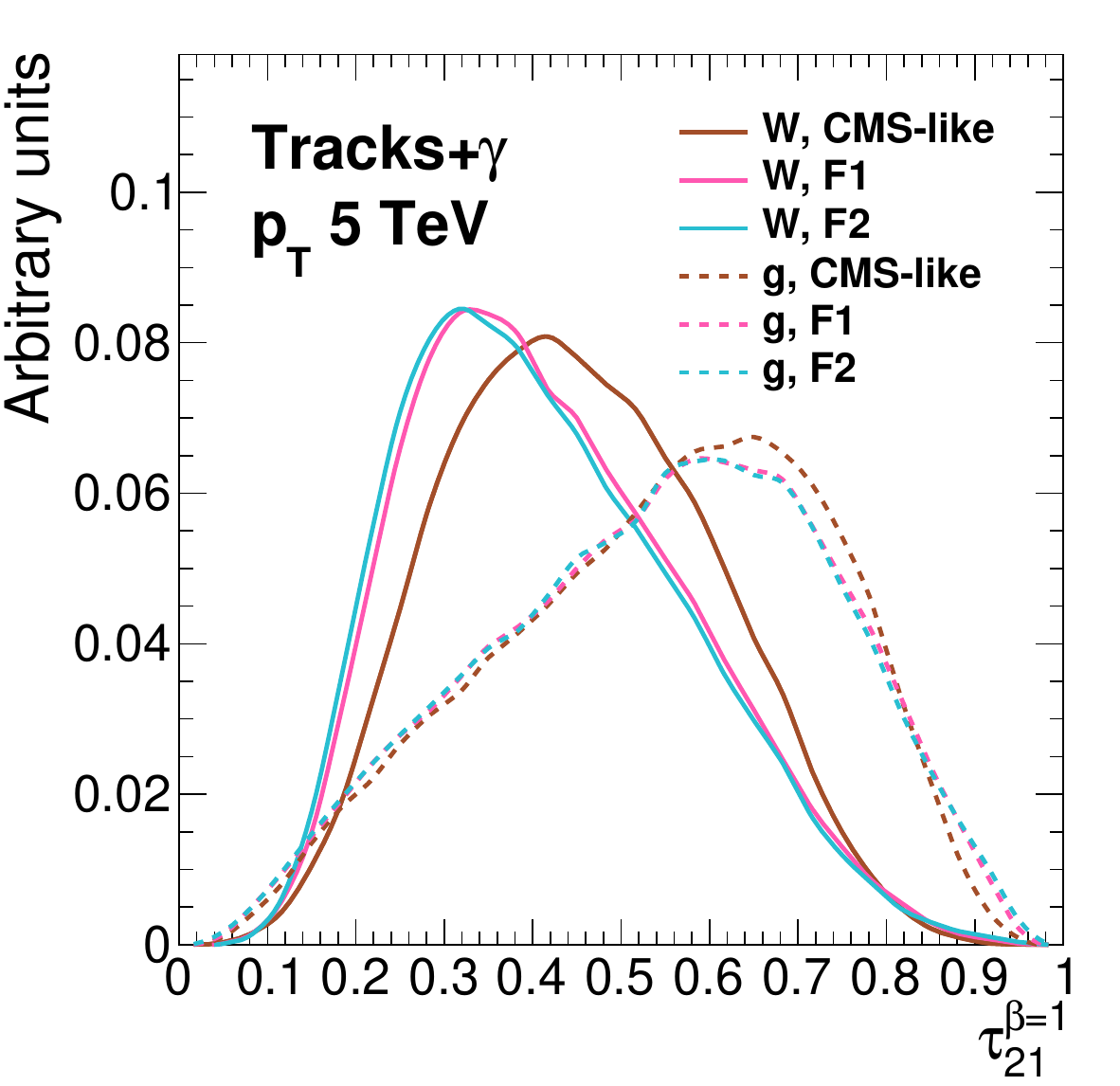}\\
\includegraphics[width=0.42\linewidth]{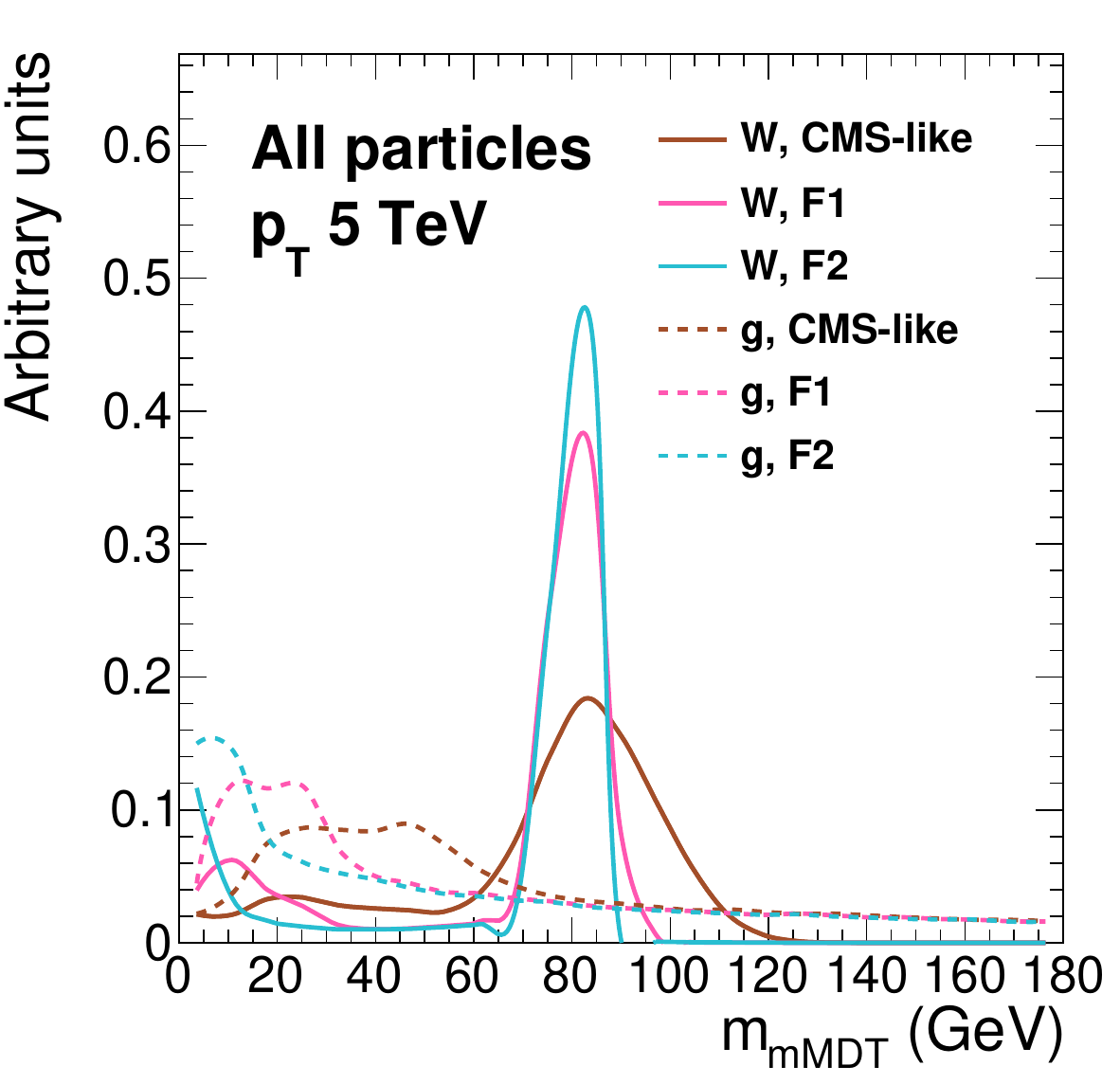}
$\qquad$
\includegraphics[width=0.42\linewidth]{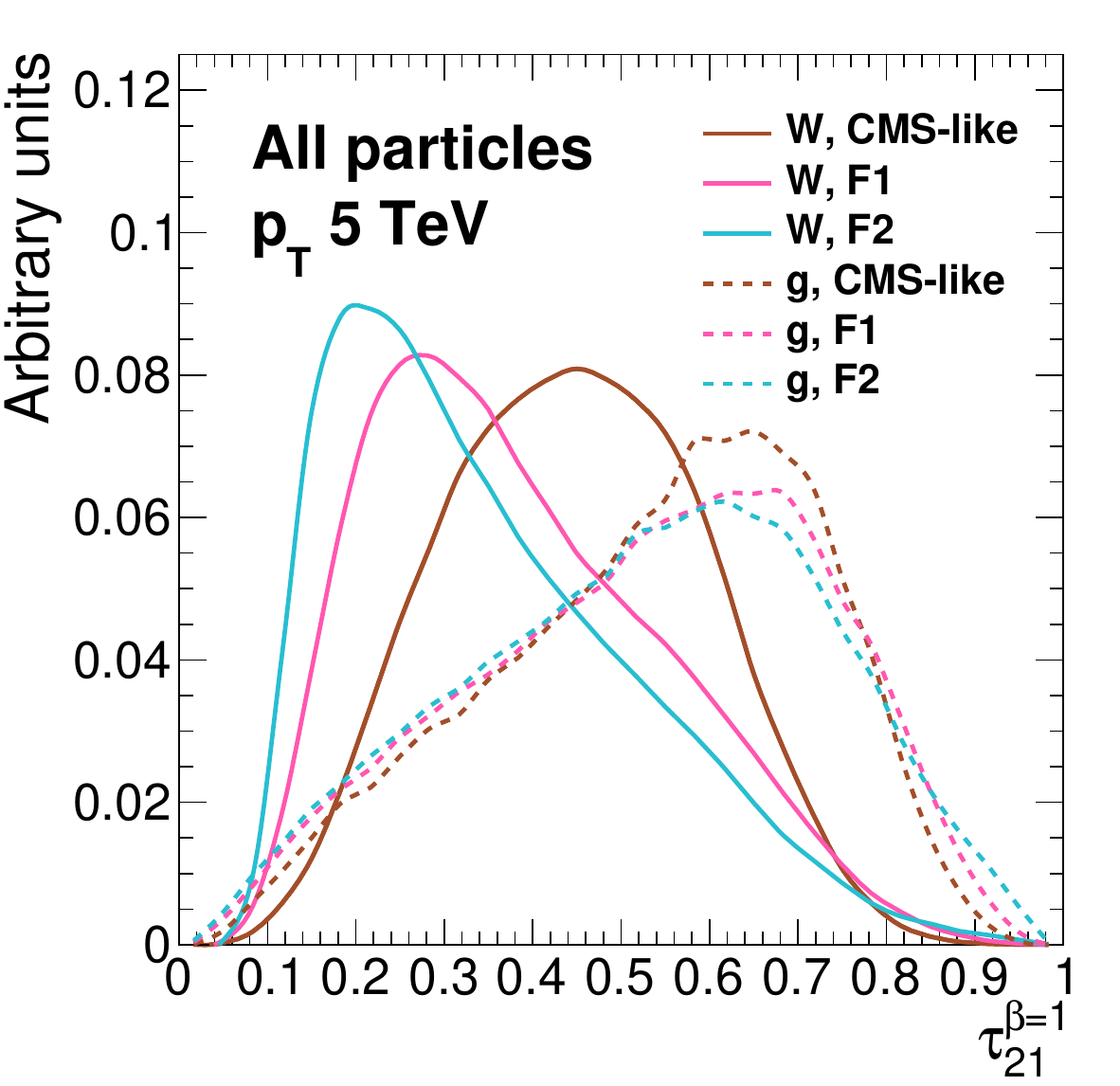}\\
\end{center}
\caption{Distributions of (left) ${m}_{\rm mMDT}$ and (right) $\tau_{21}^{\beta=1}$ for $W$ jets and gluon jets at $\pt = 5$ TeV in each detector configuration. From top to bottom, three different particle categories are considered: tracks, tracks+$\gamma$, and all particles.}
\label{fig:SummaryPlots_Wvq_DetectorComparison}
\end{figure}

\begin{figure}[tp]
\begin{center}
\includegraphics[width=0.42\linewidth]{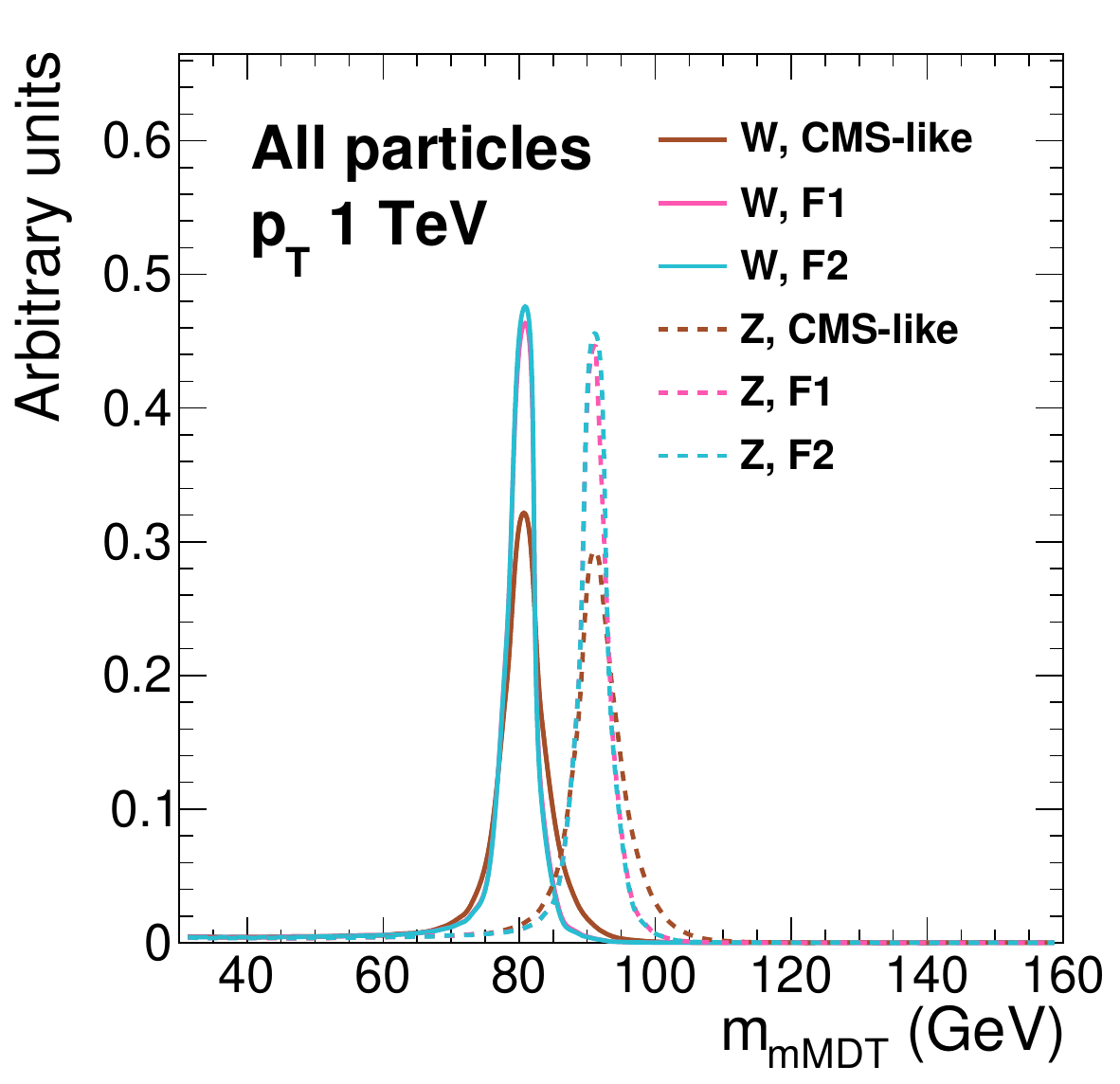}
\includegraphics[width=0.42\linewidth]{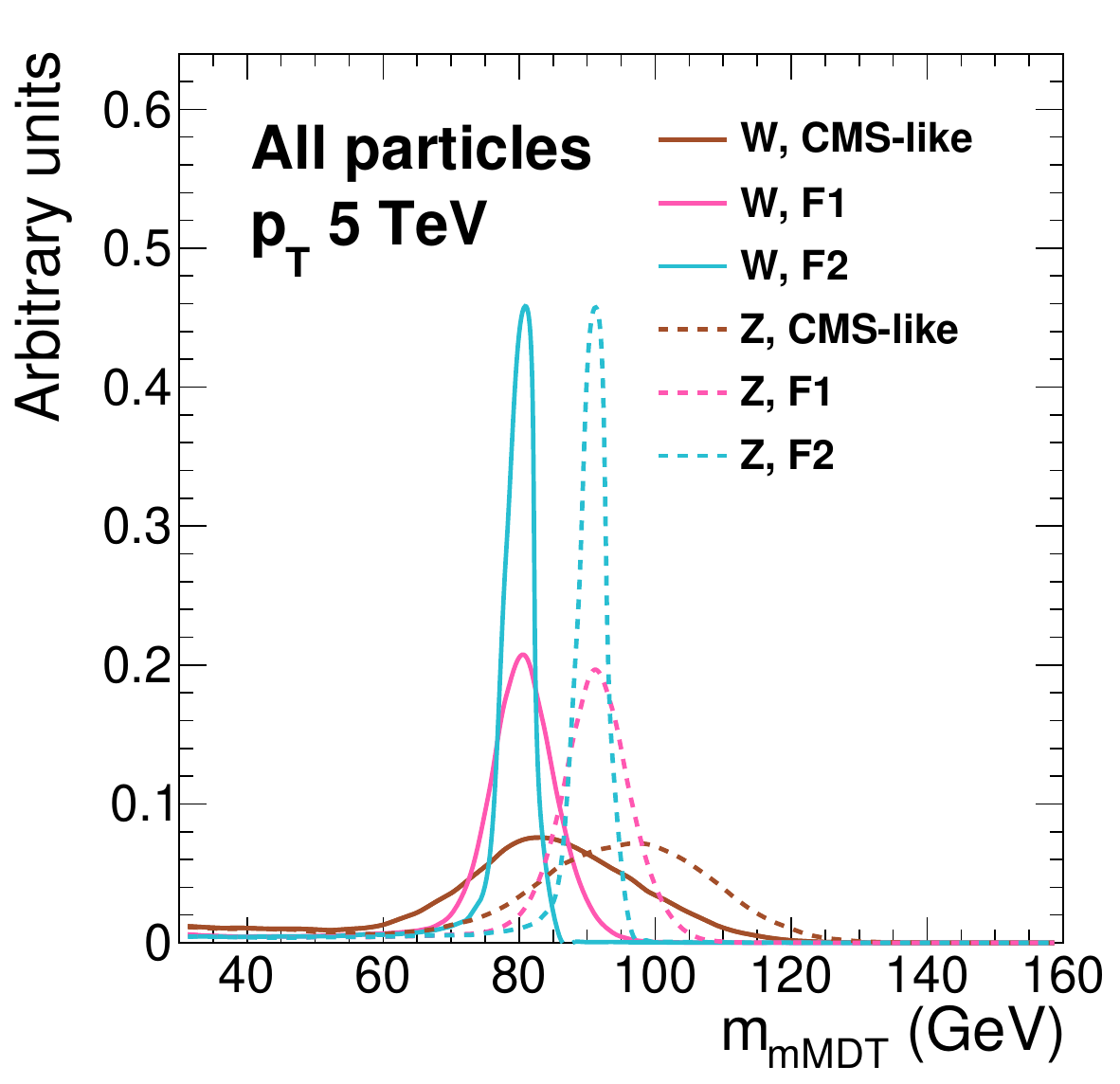}
\end{center}
\caption{Distributions of ${m}_{\rm mMDT}$ for $W$ jets and $Z$ jets at (left) $\pt = 1$ TeV  and (right) $5$ TeV in each detector configuration.  We consider the all particles category.}
\label{fig:SummaryPlots_WvZ_DetecComparison}
\end{figure}

We illustrate the impact of detector performance in \Fig{fig:SummaryPlots_Wvq_DetectorComparison}, showing (left column) ${m}_{\rm mMDT}$ and (right column) $\tau_{21}^{\beta=1}$ distributions for $W$ and gluon jets.
The various curves in each panel correspond to the three different detector configurations, while the rows correspond to different levels of neutral-particle information.
Obviously, calorimeter performance has the biggest impact when neutral-particle information is used, and both the jet mass resolution and discrimination based on $\tau_{21}^{\beta=1}$ improve with the detector performance.
In the $W$ versus $Z$ jet discrimination scenario, the effect of mass resolution is further magnified, as shown in \Fig{fig:SummaryPlots_WvZ_DetecComparison}.
This is especially true in the case of $p_T$ = 5~TeV where the improved calorimeter granularity results in a drastic improvement on the mass resolution.
A key question, to be discussed further in \Sec{sec:conclusions}, is whether the performance gains seen in the F1 and F2 configurations are sufficiently dramatic to justify their application in a future multi-purpose detector.

Both at the LHC and at future hadron colliders, pileup is an ever-present challenge that degrades jet performance.
For the studies here, though, the effect of pileup is ignored.
One justification for this choice is that we focus on very high $\pt$~jets, which are less susceptible to pileup present at a significantly lower energy scales.
More to the point, there are a number of pileup mitigation techniques that have been shown to greatly reduce the effects of pileup on jet substructure observables~\cite{Krohn:2013lba,puppi,softkiller,shapesub,Komiske:2017ubm}.
We, therefore, assume that the effects of pileup can be factorized from the impact of calorimeter granularity and resolution, at least for the high $\pt$ jets studies here.
To verify the above assumption about pileup, we performed a validation study where neutral energy equivalent to roughly 200 pileup collisions was injected into $W$ jet events.  
(Charged pileup can be safely ignored since the vast majority of it can be removed using vertexing information.) 
We found that $W$ jet tagging performance and mass resolution were negligibly affected by pileup, both for jet $\pt$~of 1 TeV and 5 TeV.
Of course, pileup is a much larger effect for more moderately boosted objects in the range of a few hundred GeV.
For future work, it would be interesting to study the effect of calorimeter resolution and granularity on pileup mitigation techniques at more moderate energies.

\subsection{Results using all particles}
\label{sec:results:all}

\begin{figure}
\begin{center}
\includegraphics[width=0.42\linewidth]{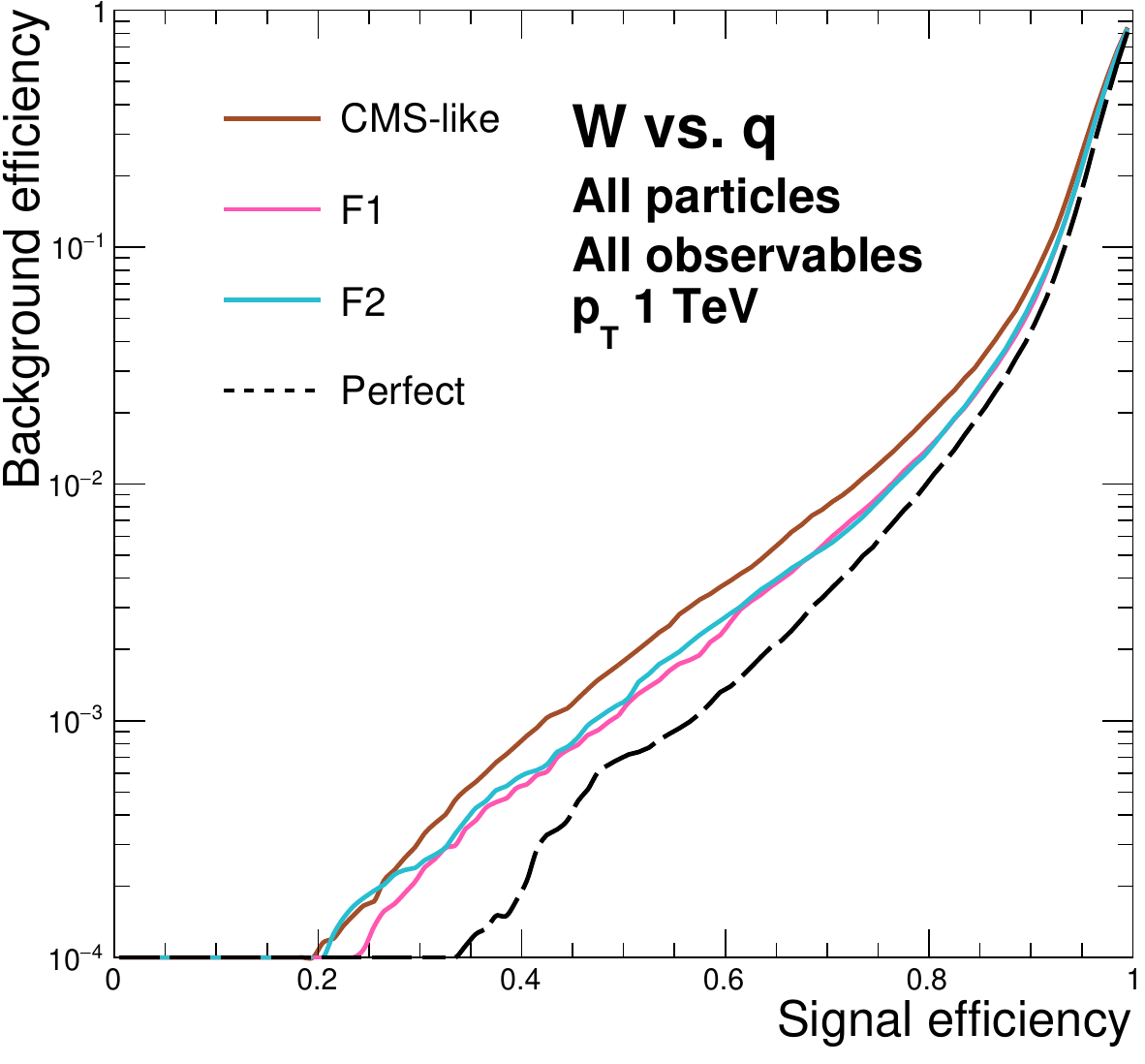}
$\qquad$
\includegraphics[width=0.42\linewidth]{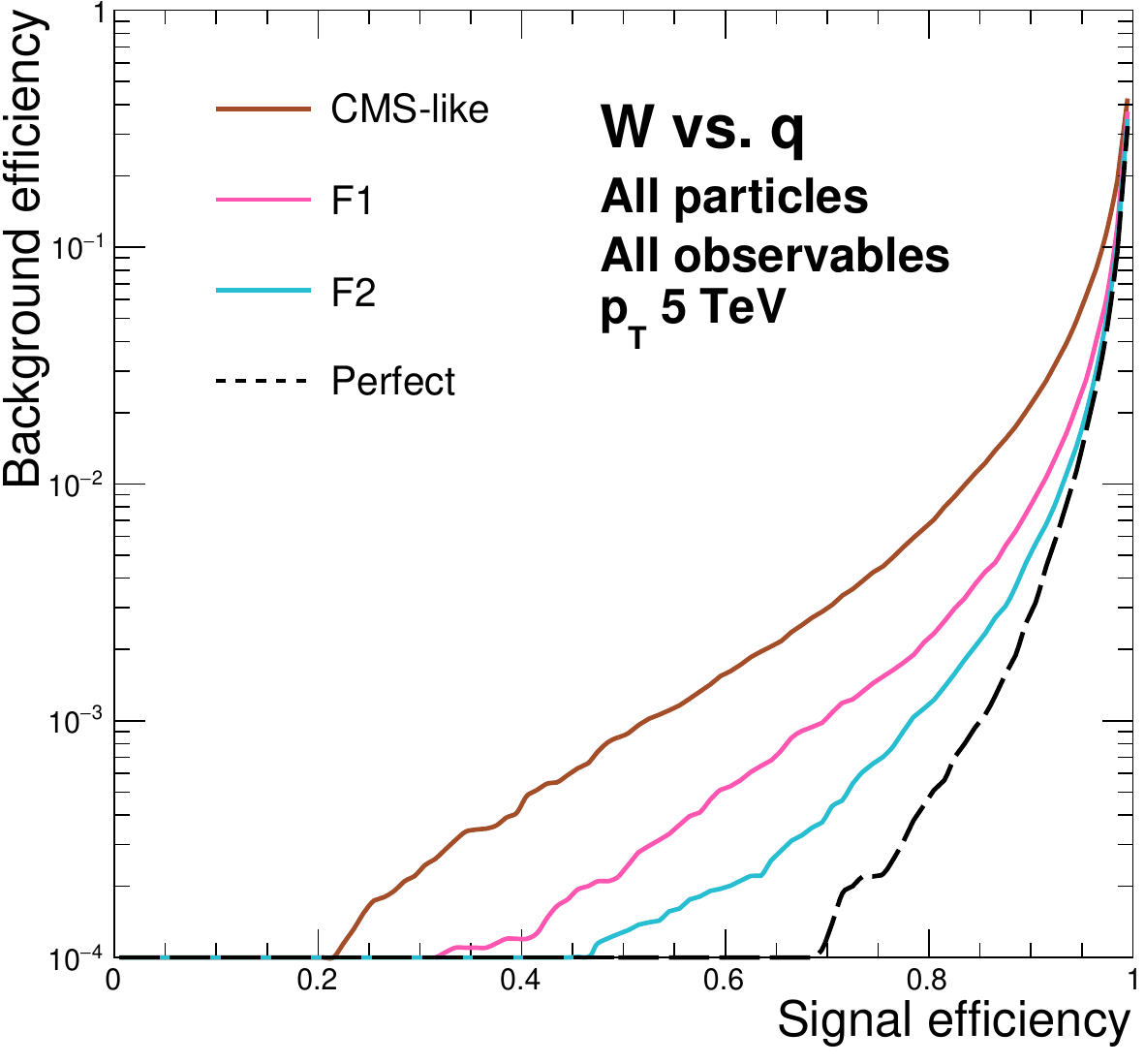}\\
\includegraphics[width=0.42\linewidth]{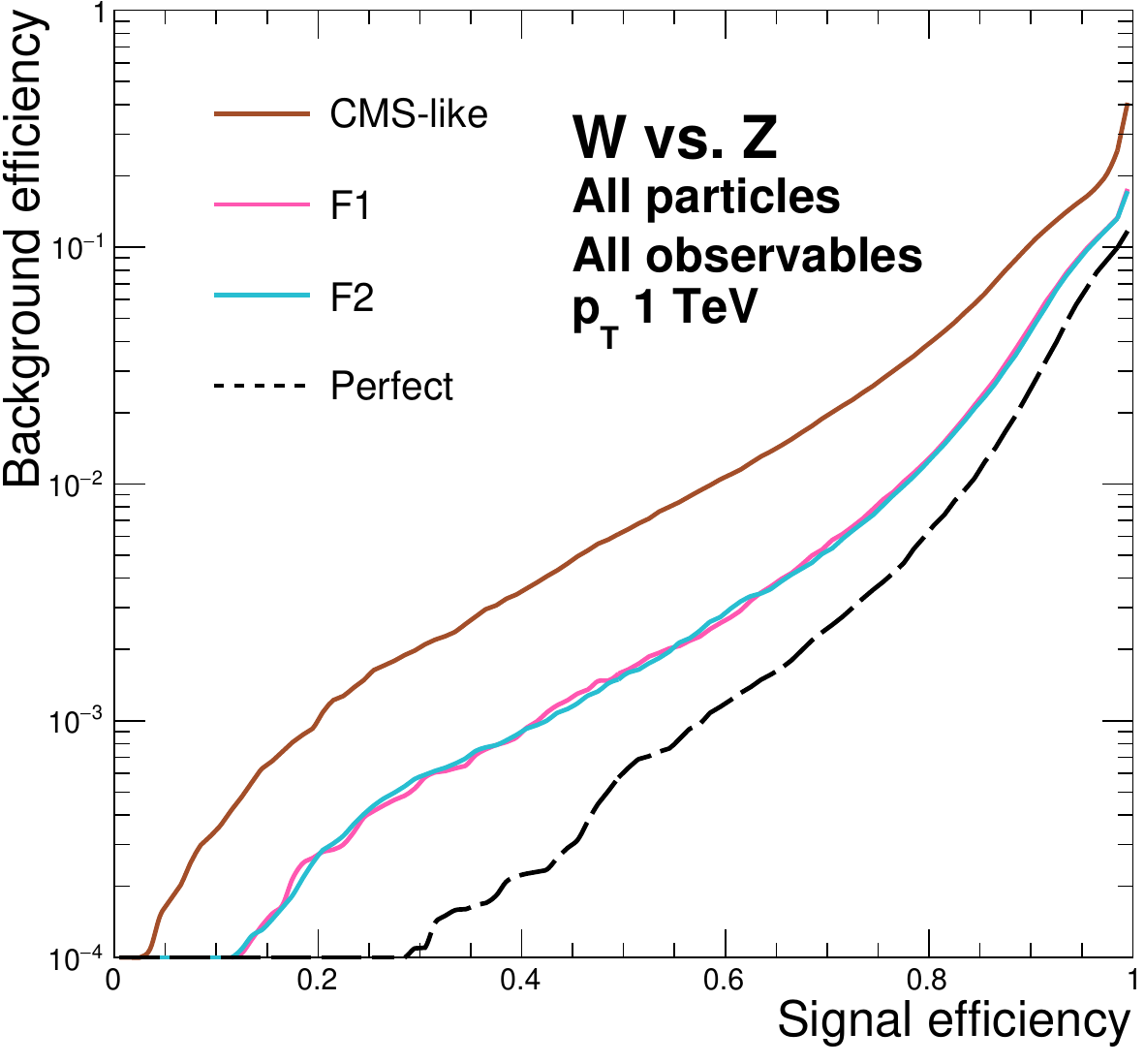}
$\qquad$
\includegraphics[width=0.42\linewidth]{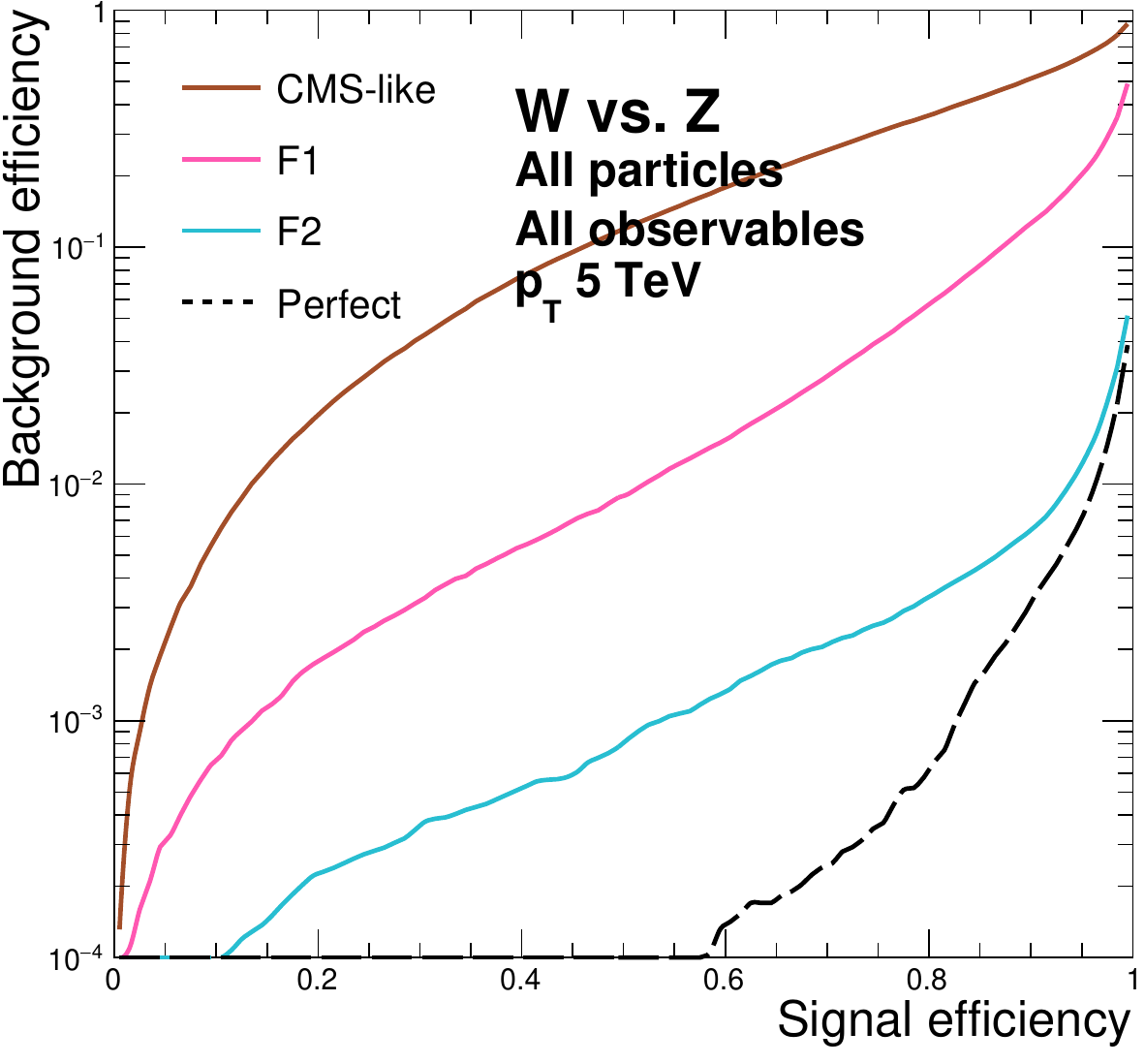}\\
\includegraphics[width=0.42\linewidth]{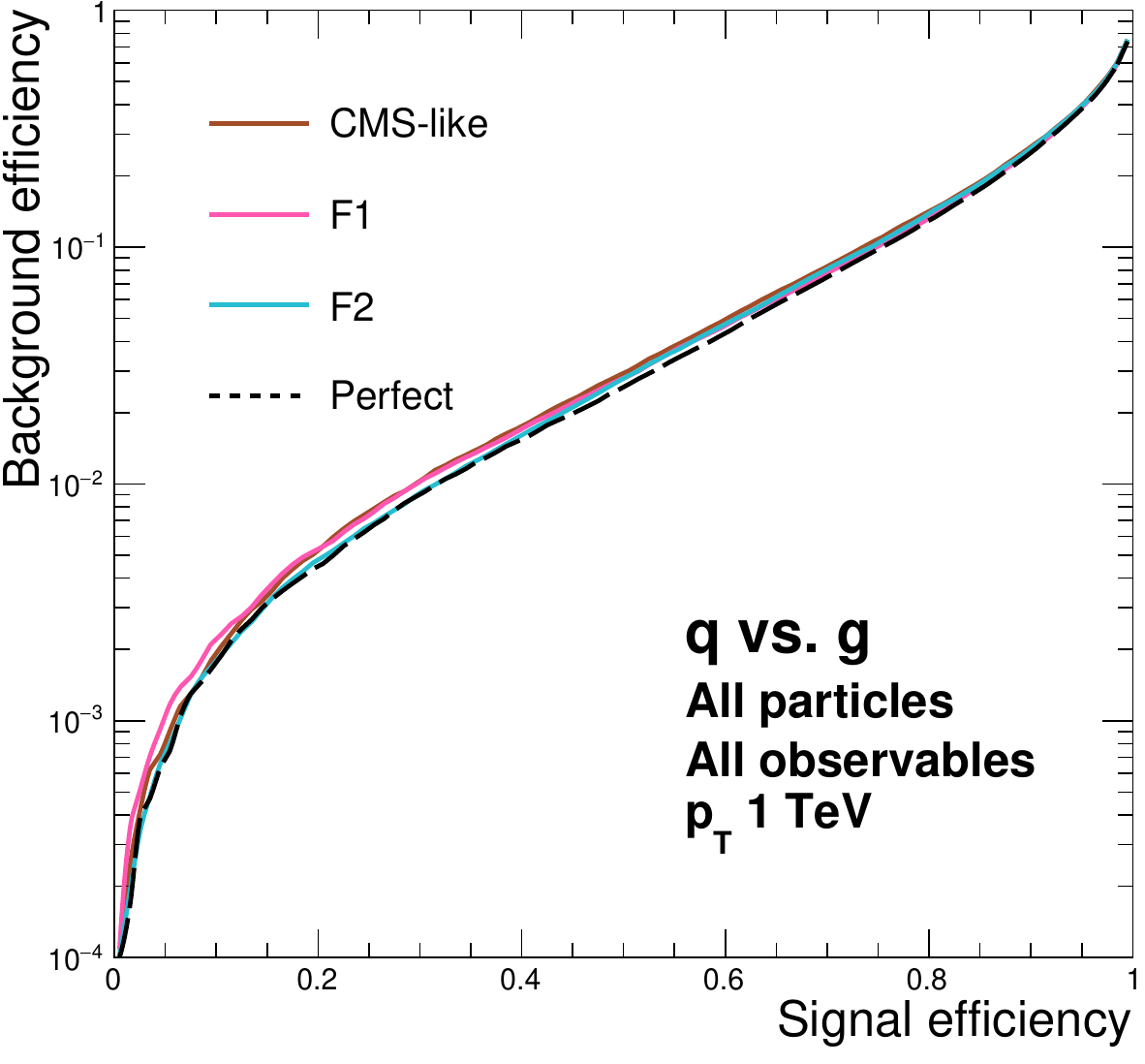}
$\qquad$
\includegraphics[width=0.42\linewidth]{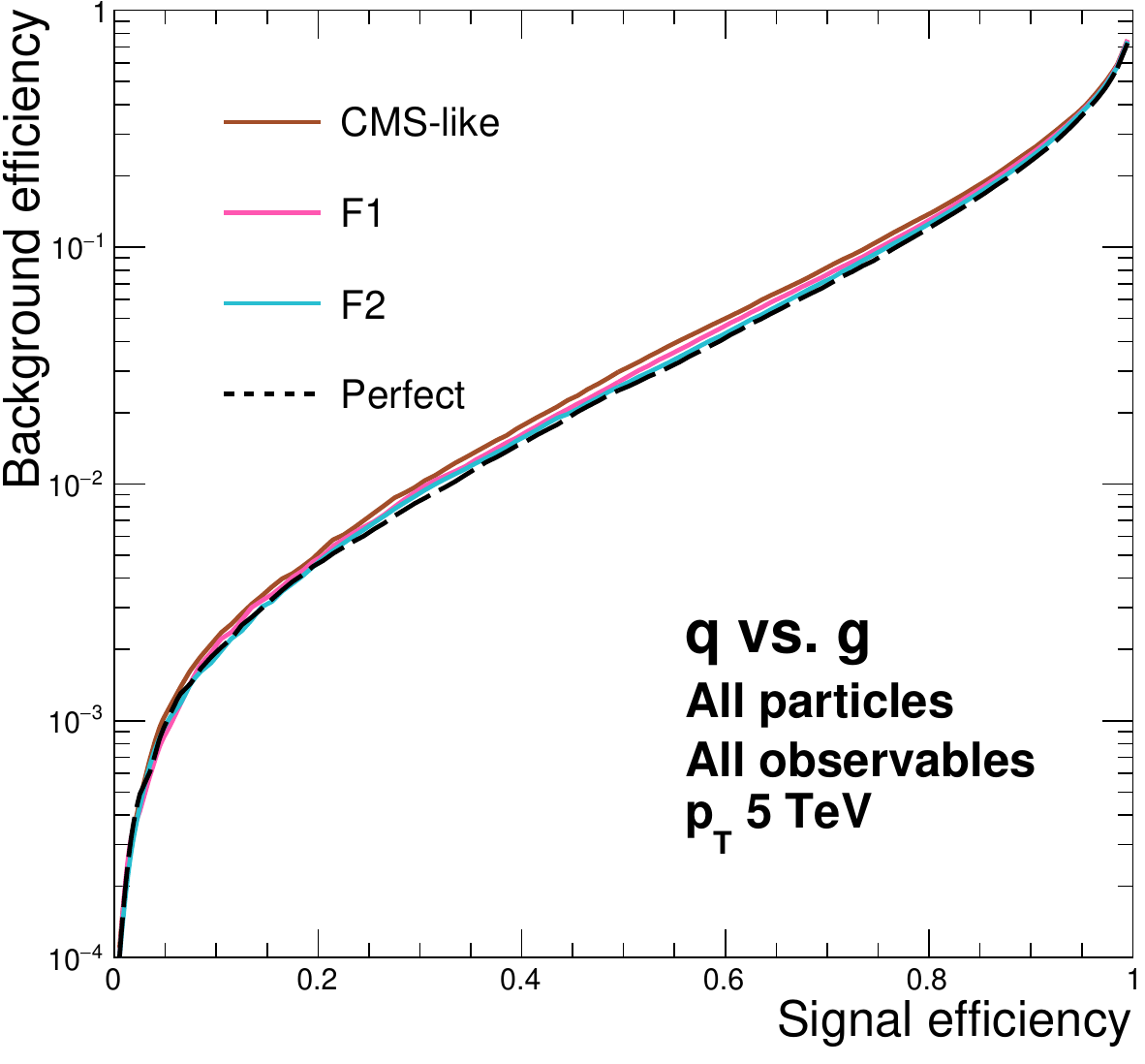}
\end{center}
\caption{All-particle discrimination performance using realistic detectors.  Shown are ROC curves for (top row) $W$ vs.~q, (middle row) $W$ vs.~$Z$, and (bottom row) $q$ vs.~$g$ discrimination in the (left column) $\pt=1$ TeV and (right column) $\pt=5$ TeV bins.  The ``perfect'' detector reference matches the all-particle curves in \Figss{fig:ROC_Wq_F1}{fig:ROC_WZ_F1}{fig:ROC_qg_F1}.}
\label{fig:ROC_DetectorComparison_detectors}
\end{figure}

At a general-purpose detector with an effective particle-flow reconstruction algorithm, it typically makes sense to use all particle information for jet substructure reconstruction.
In \Fig{fig:ROC_DetectorComparison_detectors}, we consider the same three discrimination tasks from \Sec{sec:results}---$W$ vs.~$q$, $W$ vs.~$Z$, and $q$ vs.~$g$---now comparing the performance across the CMS-like, F1, and F2 detector configurations using all available information.
Results are shown both for the (left column) 1~TeV and (right column) 5~TeV $\pt$ bins.
We also include perfect reconstruction for reference.

In all cases, having better detector performance improves (or maintains) discrimination power. 
The most (un)striking feature is that $q$ vs.~$g$ jet separation only slightly improves with the detector performance, as could be anticipated since these are shape-based discrimination tasks which already leverage the great tracking performance.
When considering $W$ vs.~$q$ or $W$ vs.~$Z$ discrimination, however, a sizable gain comes from using an improved detector.
In the $W$ vs.~$q$ case, where shape information is also powerful, the improvement in background rejection going from a CMS~$\to$~F1~$\to$~F2~detector at $\pt = 5$ TeV is around a factor of $5$ at a signal efficiency of $50\%$.
This improvement can most easily be seen by looking at the Fig.~\ref{fig:SummaryPlots_Wvq_DetectorComparison} (bottom left) which shows the mass resolution improvement of the $W$ jet signal.
Even larger gains are seen for $W$ vs.~$Z$ discrimination, where going from a CMS~$\to$~F2~detector results in two orders of magnitude improvement in background rejection power, again primarily because of the improved mass resolution which can be best seen in Fig.~\ref{fig:SummaryPlots_WvZ_DetecComparison}.

It is interesting to compare the different behaviors in the 1~TeV and 5~TeV samples.
At $\pt = 1$ TeV, going from an F1~$\to$~F2~detector shows no significant improvement for any of the discrimination tasks, whereas, at $\pt = 5$ TeV, there are still gains in going from F1~$\to$~F2.
This indicates there is an asymptotic detector granularity for a given $\pt$~such that any further improvements do not have a corresponding physics gain.
The reason this happens is that calorimeter energy resolution (which is assumed to be the same in the F1 and F2 scenarios) also plays a role in discrimination power.
This suggests the possibility of jointly optimizing granularity and resolution to achieve balanced jet substructure performance.

\subsection{Results using partial information}
\label{sec:DetecCompResults}

\begin{figure}[p]
\begin{center}
\includegraphics[width=0.42\linewidth]{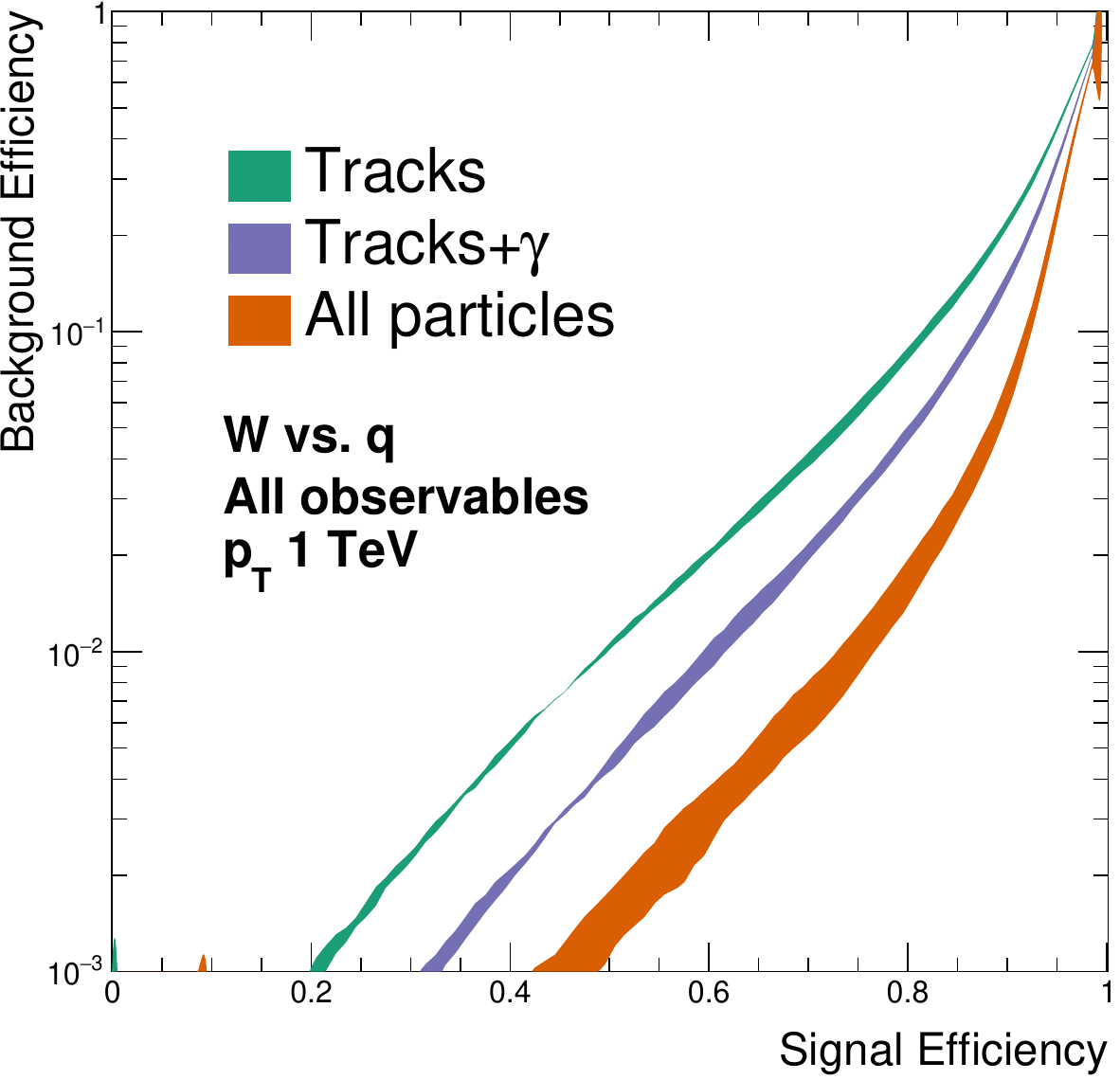}
$\qquad$
\includegraphics[width=0.42\linewidth]{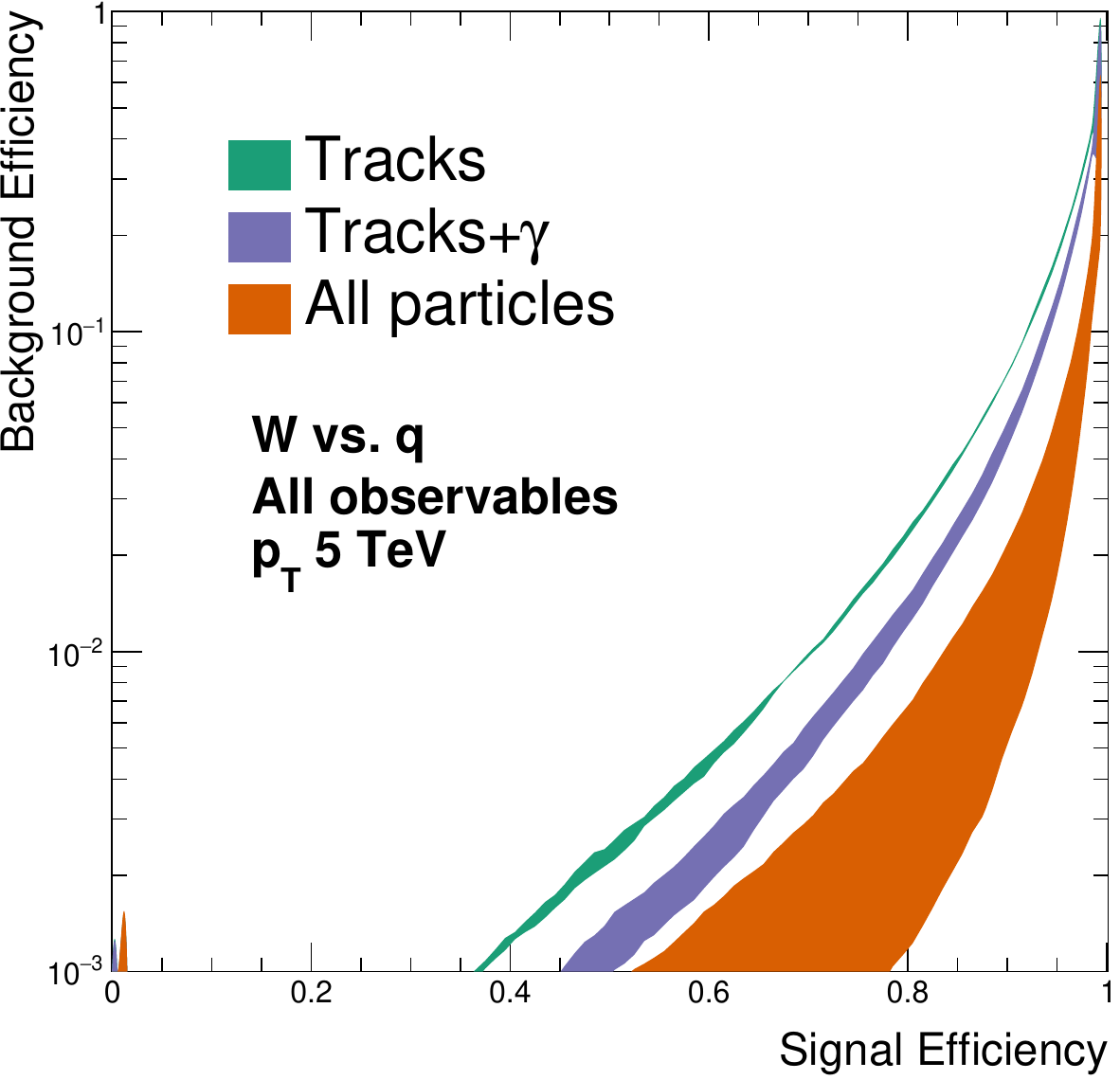} \\
\includegraphics[width=0.42\linewidth]{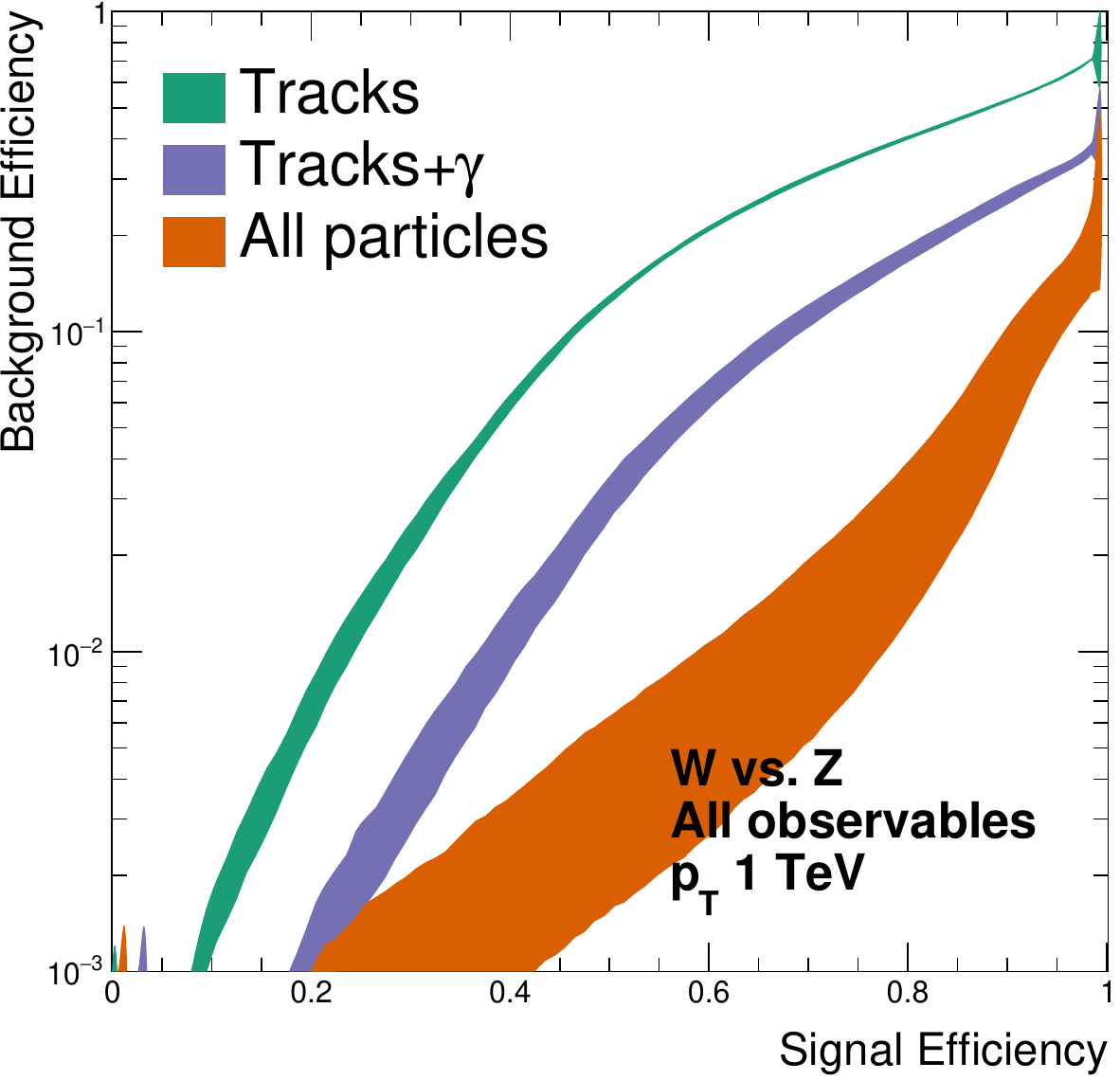}
$\qquad$
\includegraphics[width=0.42\linewidth]{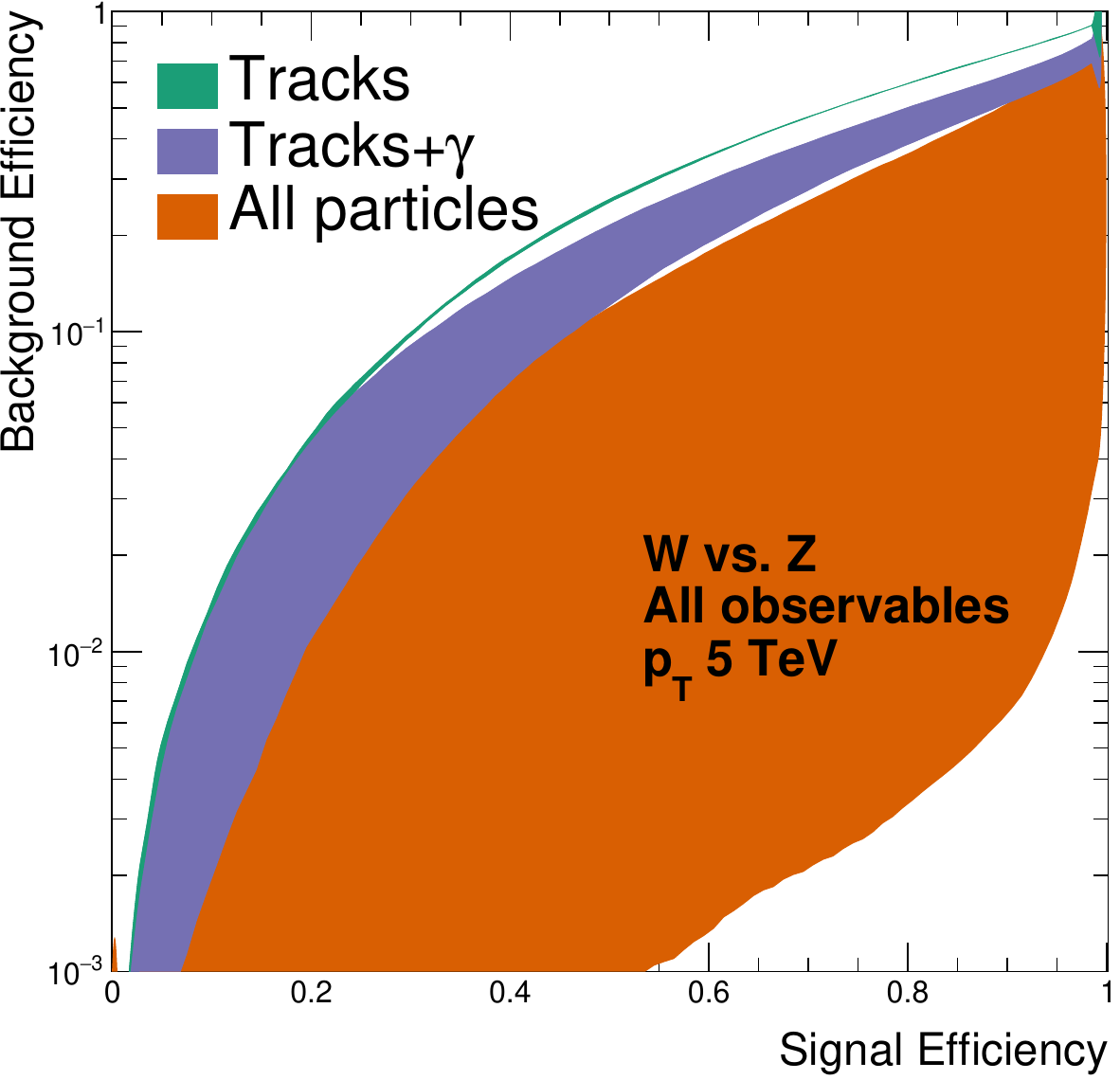} \\
\includegraphics[width=0.42\linewidth]{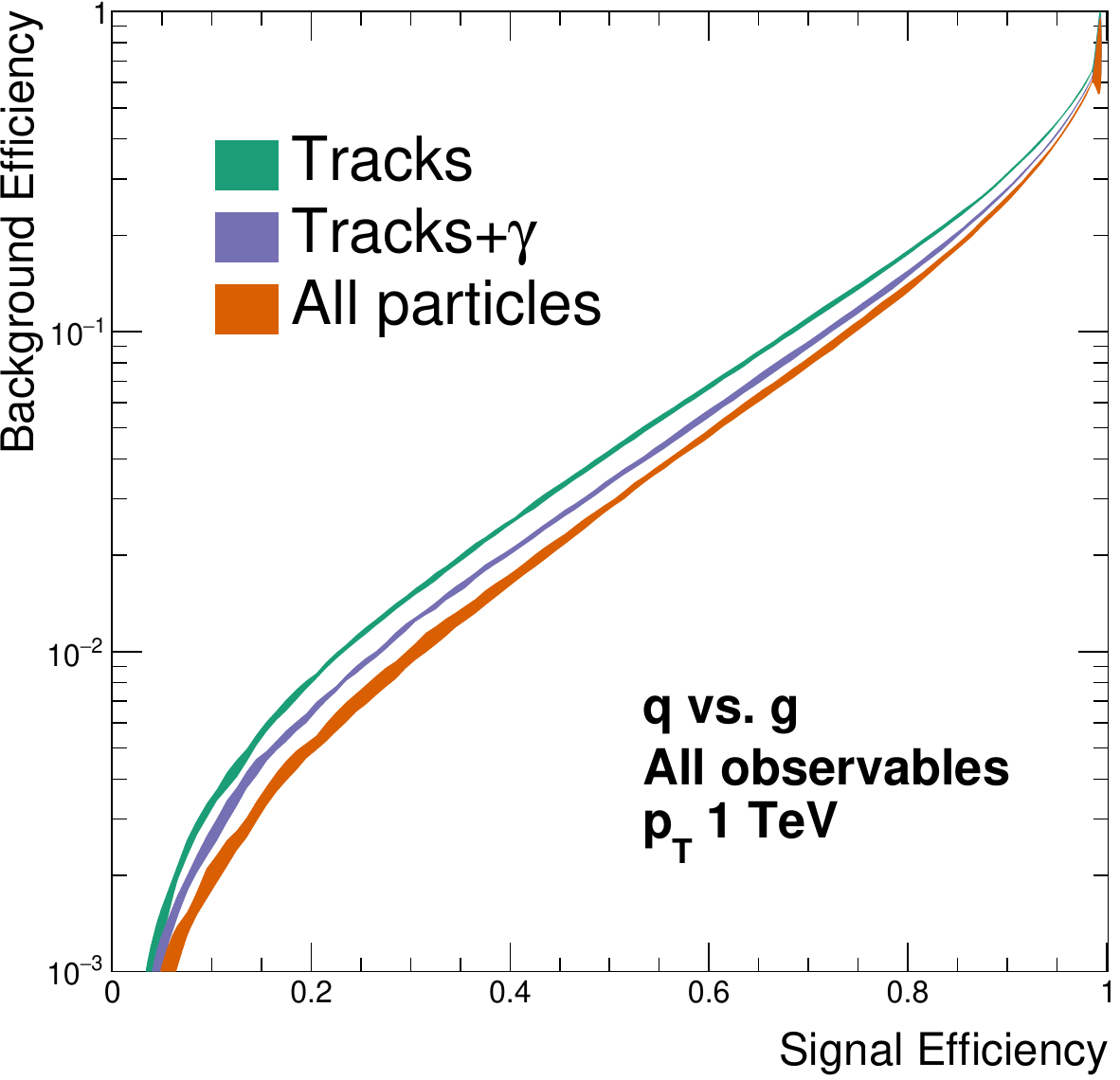}
$\qquad$
\includegraphics[width=0.42\linewidth]{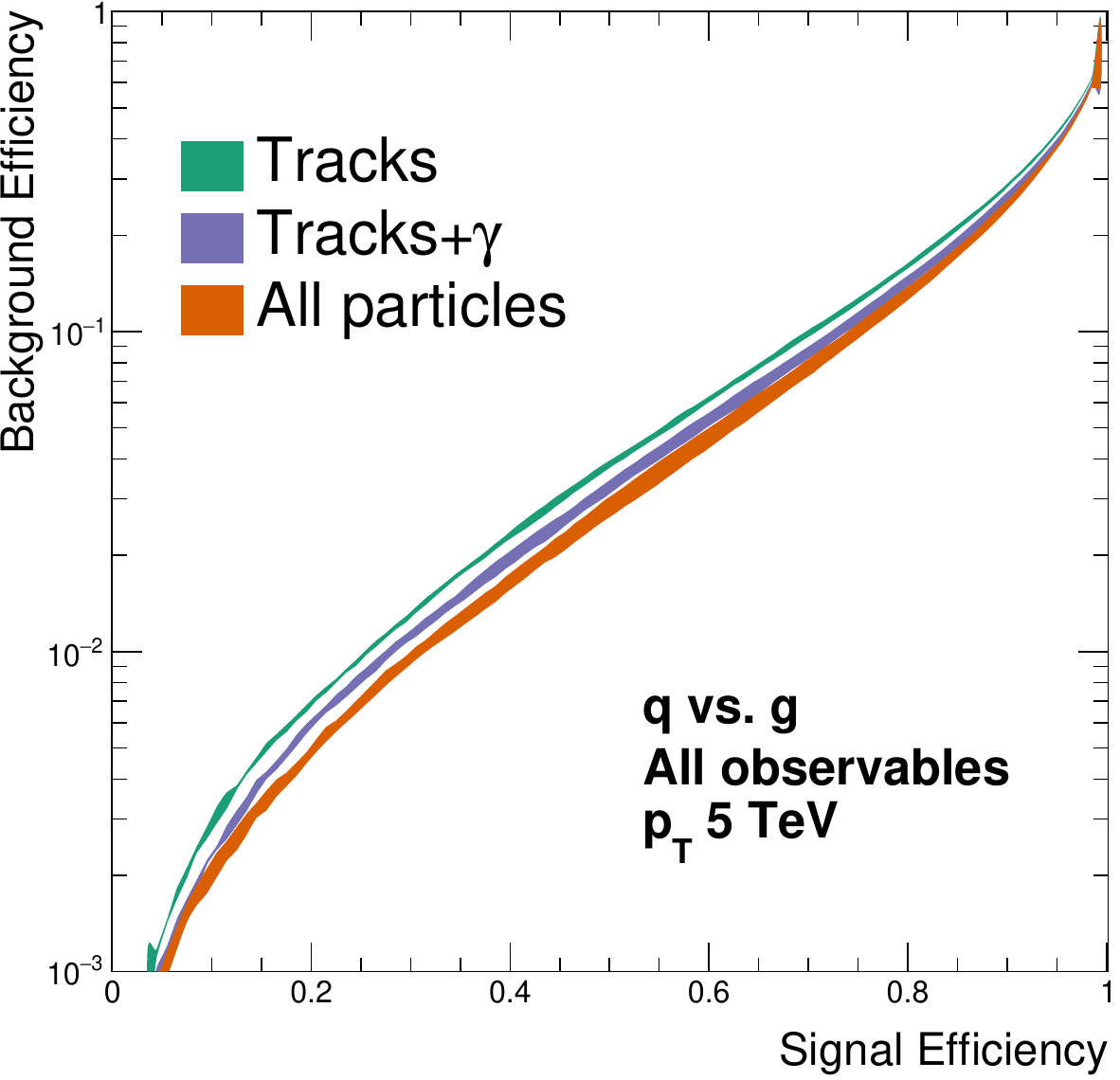}
\end{center}
\caption{Discrimination performance as the input particles are varied, taking the ROC envelope of the CMS-like, F1, and F2 detector configurations.  As in \Fig{fig:ROC_DetectorComparison_detectors}, the layout is (top row) $W$ vs.~$q$, (middle row) $W$ vs.~$Z$, and (bottom row) $q$ vs.~$g$, with (left column) $\pt=1$ TeV and (right column) $\pt=5$ TeV.}
\label{fig:ROC_DetectorComparison_inputs}
\end{figure}

In some contexts, it may be beneficial to use partial jet information for jet substructure studies, for example using track-only measurements to help mitigate pileup.
In \Fig{fig:ROC_DetectorComparison_inputs}, we vary the input particle categories---tracks, tracks$+\gamma$, or all particles---and plot the envelope of the CMS-like, F1 and F2 detector results in the same layout as \Fig{fig:ROC_DetectorComparison_detectors}.

In all cases, as we include more neutral-particle information, we improve (or maintain) discrimination power.
As expected from \Sec{sec:results}, the gain is minimal (at most a factor of 2) when considering the $q$ vs.~$g$ jet discrimination, which relies primarily on jet shape information.
For $W$ vs.~$q$ and $W$ vs.~$Z$ discrimination, adding neutral-particle information greatly improves the discrimination power.
The effect is extreme for $W$ vs.~$Z$ jets where mass information is dominant and degrading the mass resolution, both through removing neutral-particle information and degrading detector performance, has a very significant impact on the discrimination. 

From the size of the envelopes, one can also see that the difference between the CMS-like and F2 performance is most pronounced when all particles are included since those depend on HCAL resolution.
When using tracks plus photons, the differences between detector configurations are more modest, suggesting that ECAL granularity is not the primary limiting factor for jet substructure performance.

\subsection{Summary of detector study}
\label{sec:detector:summary}

\begin{figure}[p]
\begin{center}
\includegraphics[width=0.45\linewidth]{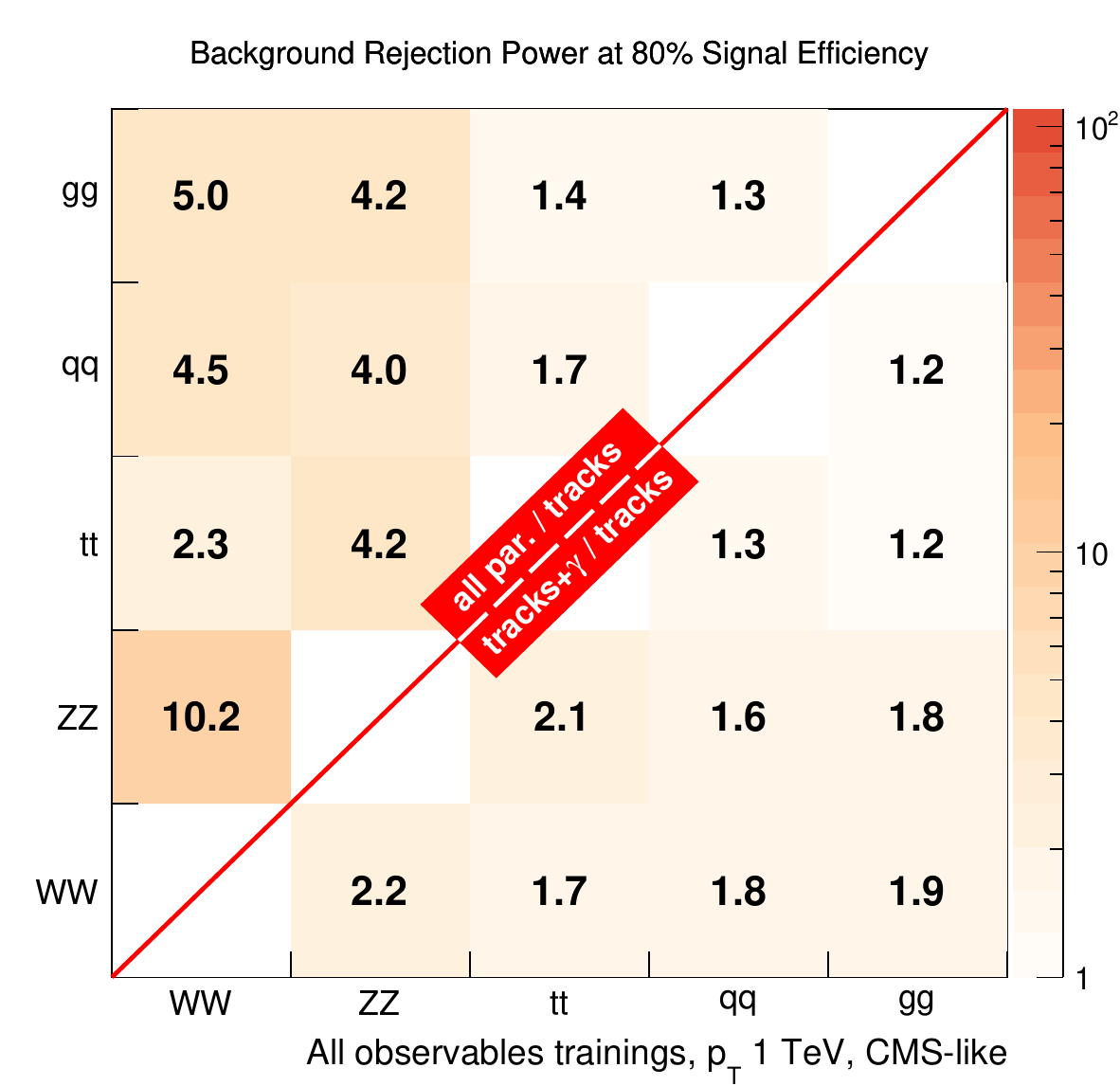}
\includegraphics[width=0.45\linewidth]{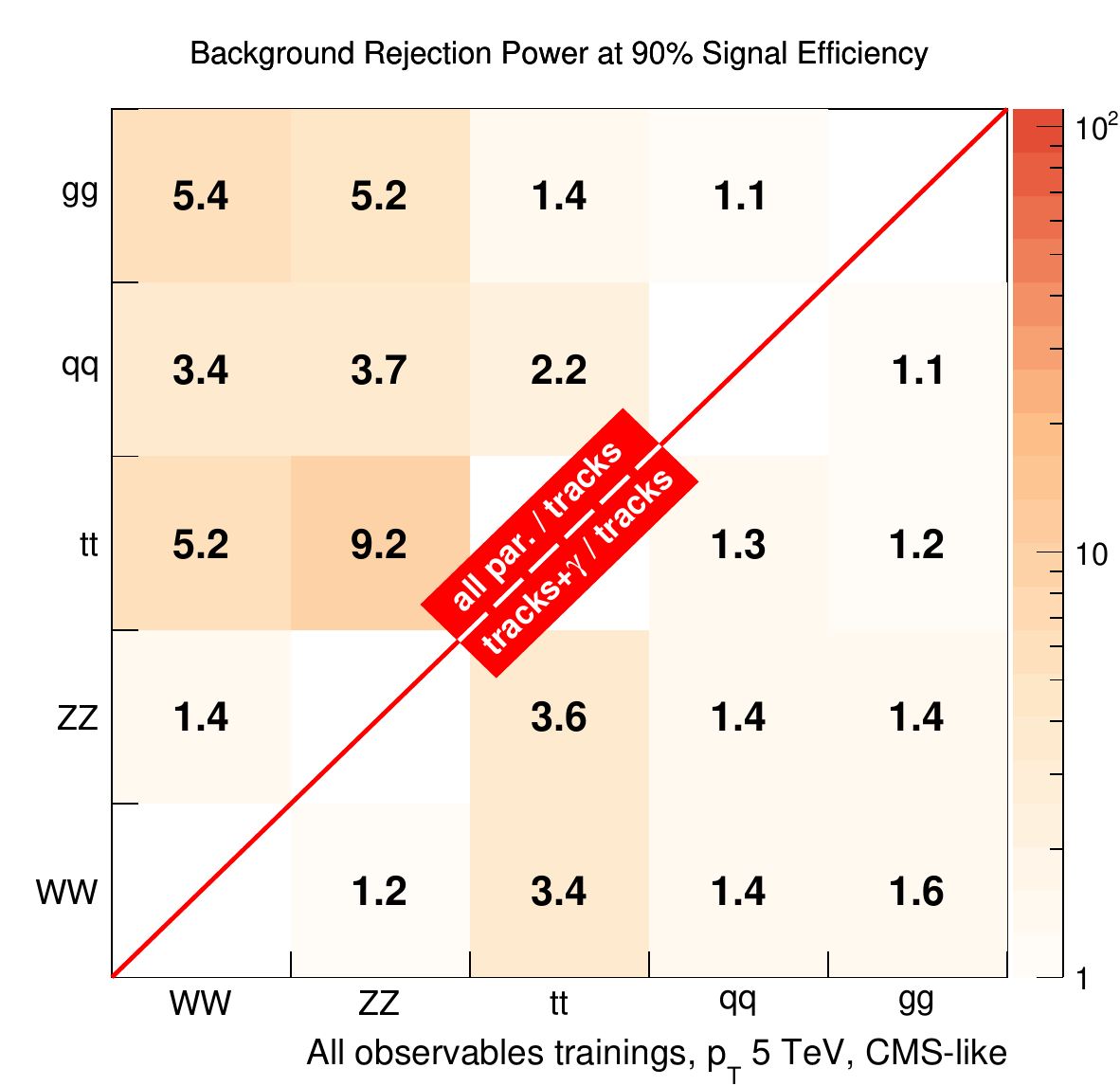}
\end{center}
\caption{Background rejection ratio (left) $\mathcal{R}_{80}$ for the $\pt =$~1 TeV bin and (right) $\mathcal{R}_{90}$ for the $\pt =$~5 TeV bins, for all pairwise discrimination tasks among $W$/$Z$/$t$/$q$/$g$ jets using all particle information.  This can be compared to \Fig{fig:MAPSPerfect}, where values in the lower right triangle refer to the ratio (tracks$+\gamma$)/(tracks), while those in the upper left triangle refer to (all~particle)/(tracks).}
\label{fig:MAPS5TeV_1}
\end{figure}

\begin{figure}[p]
\begin{center}
\includegraphics[width=0.45\linewidth]{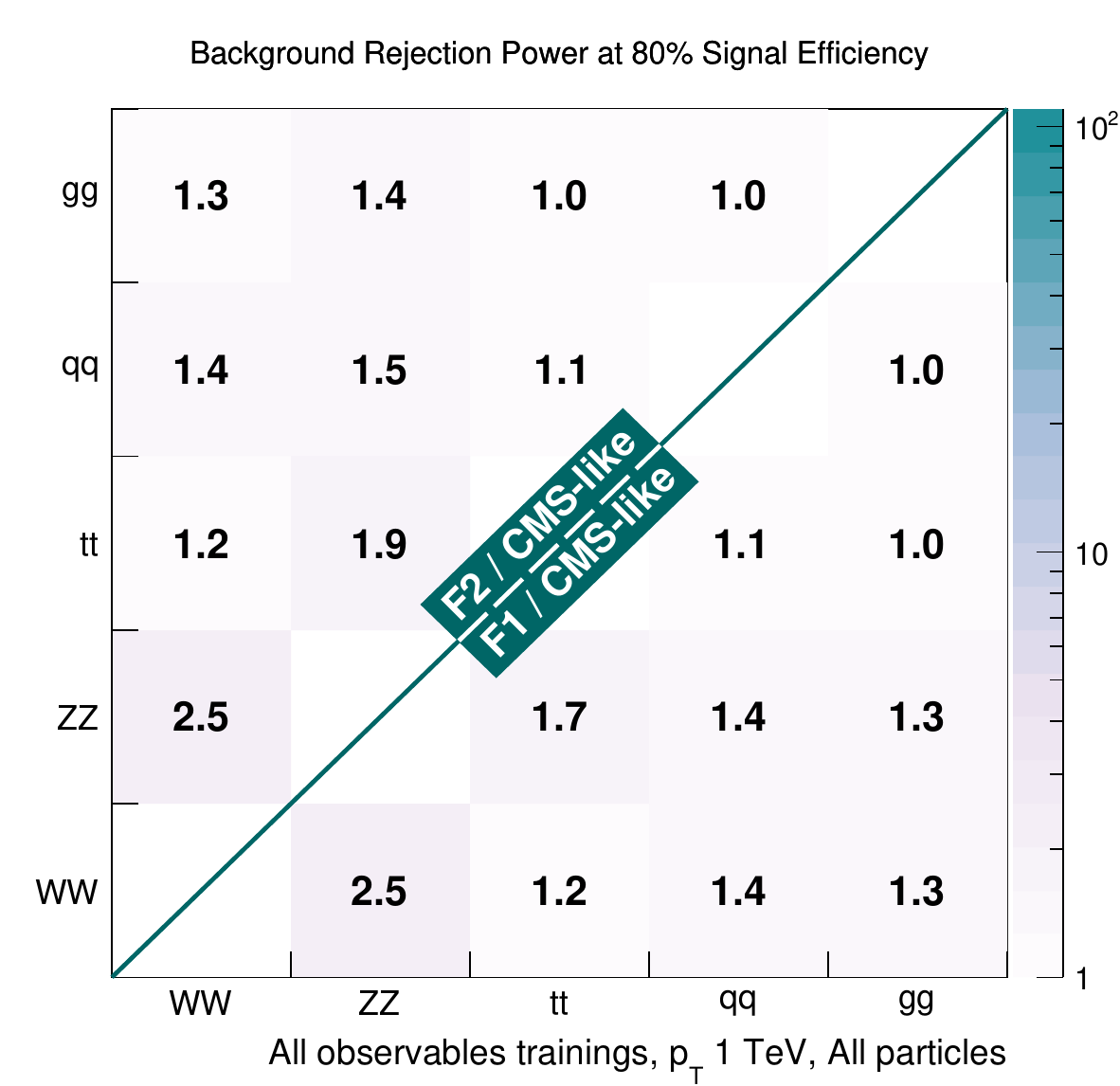}
\includegraphics[width=0.45\linewidth]{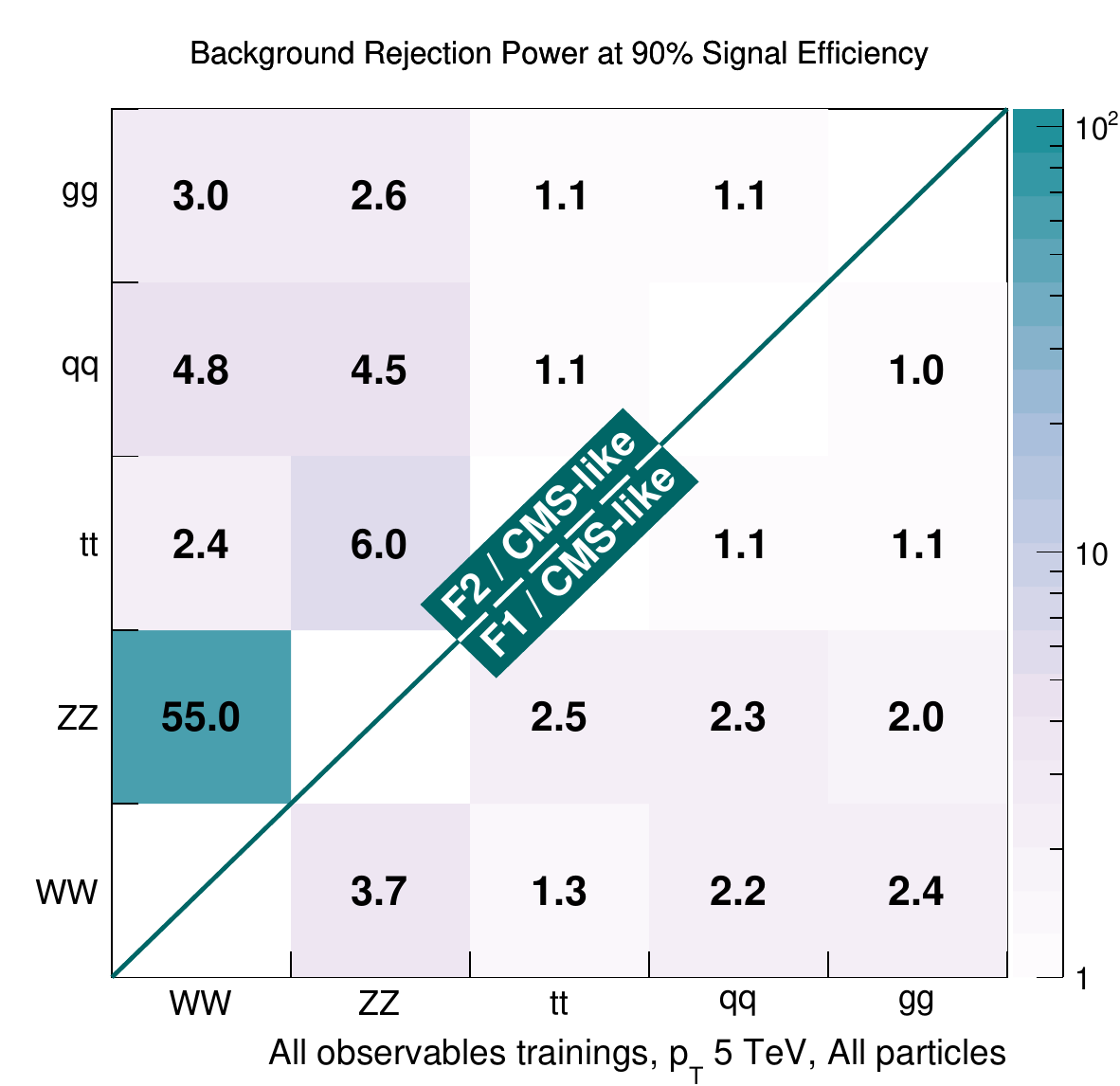}
\end{center}
\caption{Same as \Fig{fig:MAPS5TeV_1}, but using all particle information and showing $\mathcal{R}_{S}$ values for (Future 2)/(CMS-like) in the upper left triangle and (Future 1)/(CMS-like) in the lower right triangle.}
\label{fig:MAPS5TeV_2}
\end{figure}

Finally, we summarize the change in discrimination power as we vary both the particle input categories and detector configurations.
Following \Eq{eq:ratio}, we define $\mathcal{R}_S$ as the ratio of the background rejection power for two different scenarios for fixed signal efficiency $S\%$.
%

In \Fig{fig:MAPS5TeV_1}, we fix the CMS-like detector configuration and illustrate the gain in background rejection in including more neutral-particle information.
The $5 \times 5$ grids show $\mathcal{R}_{80}$ ($\mathcal{R}_{90}$) for 1~(5)~TeV on the left (right) for each pairwise comparison of $W$/$Z$/$t$/$q$/$g$ jets, with (tracks$+\gamma$)/(tracks) in the lower right triangle and (all particle)/(tracks) in the upper left triangle.
Comparing to the study in \Fig{fig:MAPSPerfect}, which assumed a perfect detector, the qualitative features are quite similar.
The main difference is that the importance of using all particle information is muted when using a realistic detector compared to a perfect one, but the gains are still substantial.
One exception is the $W$/$Z$ discrimination at 5 TeV where the CMS-like detector is not well-suited to handle such high $\pt$ jets, and we will see benefits below from improved detector granularity and resolution.

In \Fig{fig:MAPS5TeV_2}, we consider only all particle information and illustrate the gains in going from a CMS-like detector to an F1 or F2 detector.
For the extreme case of $W$ vs.~$Z$ jets at $\pt =$~5 TeV,  the Future 2 detector scenario improves upon the CMS-like scenario by almost two orders of magnitude in background rejection power. 
This is an extreme example, though any discrimination task involving $W/Z$ identification benefits substantially from the CMS~$\to$~F1 or CMS~$\to$~F2 improvement.
This example highlights the importance of understanding physical thresholds to achieve major performance improvements and should be a central consideration for future detector research efforts. 
Since $q$ vs.~$g$ discrimination is dominated by shape information, it is not surprising that the future detectors do not offer much performance improvement.

An important note is that top vs. $q$/$g$ discrimination is not affected much by the detector configuration, as explored further in \App{sec:appendix}.
This is because the mass information, while dominant, is already maximally discriminant due to the large top mass and is less susceptible to mass resolution effects, as previously discussed in \Sec{sec:ratiospartial}.
For all tasks, the impact of calorimeter performance is stronger when increasing the jet $\pt$~from 1~TeV (top) to 5~TeV (bottom) because the F1 and F2 detector configurations were selected for their applicability to the higher-energy regime.

\section{Conclusions and outlook}
\label{sec:conclusions}

Highly-boosted hadronic topologies and jet substructure techniques have proven to be an important part of the LHC physics program, and they will only become more important at future higher-energy experiments.
It is therefore vital that we understand how best to utilize current multipurpose detectors, as well as to design new ones.
In this paper, we studied the importance of neutral-particle information and calorimetry for jet substructure performance.
We used binary BDT classifiers to understand differences in the jet discrimination power achievable in different detector regimes.
While recent studies like \Refs{Larkoski:2015yqa,Spannowsky:2015eba,Bressler:2015uma} have focused on the advantages of using track-only information to study boosted scenarios, our results highlight cases where calorimetric information is necessary to maximize physics performance.
Using a toy detector simulation with particle-level smearing and discretization tuned for jet substructure performance, we quantified the potential gains from deploying high-resolution calorimetry at a hypothetical future high-energy multi-TeV-scale collider, which could ultimately improve our ability to find new physics.
We also explored the intermediate possibility of using tracks plus photons, which recovers a large fraction of the neutral-particle information with reduced demands on calorimeter performance.

By studying the discrimination power between $W$/$Z$/$t$/$q$/$g$ jets at very high $\pt$, we were able to disentangle which jet features are required for different jet discrimination tasks.
We found that for discrimination based primarily on dimensionless jet shape information, such as $q$ vs.\ $g$ jets, adding neutral-particle information or improving calorimeter granularity adds very little to performance.
When mass resolution is important, such as in $W$ vs.\ $q$ jets or maximally in $W$ vs.\ $Z$ jet discrimination, adding neutral-particle information and improving calorimeter granularity are extremely important. 
For example, in the more typical case of $W$ vs.\ $q$ jet discrimination, improving the ECAL and HCAL granularity by a factor of 5 or 10 can improve background rejection by a factor of 10 for 5~TeV jets; a similar factor can be lost if one does not include neutral-particle information at all. 
The effect is even more striking for $W$ vs.\ $Z$ jet discrimination, which could be a very important tool to characterize potential new physics signals.
There, background rejection can be improved by 2 orders of magnitude using all-particle information with the highest granularity calorimetry.

One surprising outcome of this study is that mass resolution does not play as crucial of a role for boosted top identification, even though jet mass is a powerful discriminant.
This finding is consistent with the conclusions of~\Ref{Larkoski:2015yqa}, but nevertheless highlights the interplay between mass and shape information as well as the impact of signal topology and color structure.
Though not part of this study, it would be fruitful to consider reconstructing boosted objects beyond the SM, such as recently explored in \Refs{Aguilar-Saavedra:2017zuc,Sirunyan:2017dnz,Aguilar-Saavedra:2017rzt}, which can have different masses, color representations, and decay topologies compared to SM resonances.

In exploring the impact of detector design on jet substructure performance, we have set aside two very important factors: reconstruction feasibility and cost.
While there have been studies which explore potential fundamental calorimeter limitations~\cite{Bressler:2015uma},
we leave it to other studies such as~\cite{Chekanov:2016ppq,Chekanov:talk1,Chekanov:talk2} to determine if high-granularity calorimetry and new reconstruction techniques can deliver even better per-particle separation and performance.
This study, emphasizing the effects on physics performance, is meant to be complementary to those detailed reconstruction studies.  Assuming that such reconstruction performance can indeed be achieved, we explicitly quantify the physics gains possible from improved calorimetry.
Ultimately, one needs to simultaneously optimize cost, feasibility, and physics potential, and the particle physics community needs to decide whether, for example, robust jet discrimination is an important enough goal to drive detector design.
Future work may include studies of more detector configurations and higher jet energies.
It is also worth considering that, especially in a multivariate context, it may be possible to achieve the same discrimination power using observables with less demanding resolution requirements.
In this way, the best balance of cost and performance might be achieved through a combination of track-based, track-plus-photon, and all-particle observables. 

\section*{Acknowledgements}

We thank Sergei Chekanov, Ashutosh Kotwal, Andrew Larkoski, Benjamin Nachman, and Tilman Plehn for useful discussions.  The work of M.F. is supported by the U.S. Department of Energy (DOE) under grant contract number DE-SC-00011640. The work of J.T. is supported by the DOE under grant contract number DE-SC-00012567.  The work of N.T. and C.V. is supported by Fermi Research Alliance, LLC under Contract No. DE-AC02-07CH11359 with the U.S. Department of Energy, Office of Science, Office of High Energy Physics. A.H. gratefully acknowledges funding in the Emmy-Noether program (HI 1952/1-1) of the German Research Foundation DFG. M.N. and E.C. acknowledge the support of the DOE via the contract DE-SC0010010. E.C. was partially supported by a Karen T. Romer Undergraduate Teaching and Research Award through Brown University.

\appendix
\section{Top jets vs.\ gluon jets}
\label{sec:appendix}

\begin{figure}[tp]
\begin{center}
\includegraphics[width=0.42\linewidth]{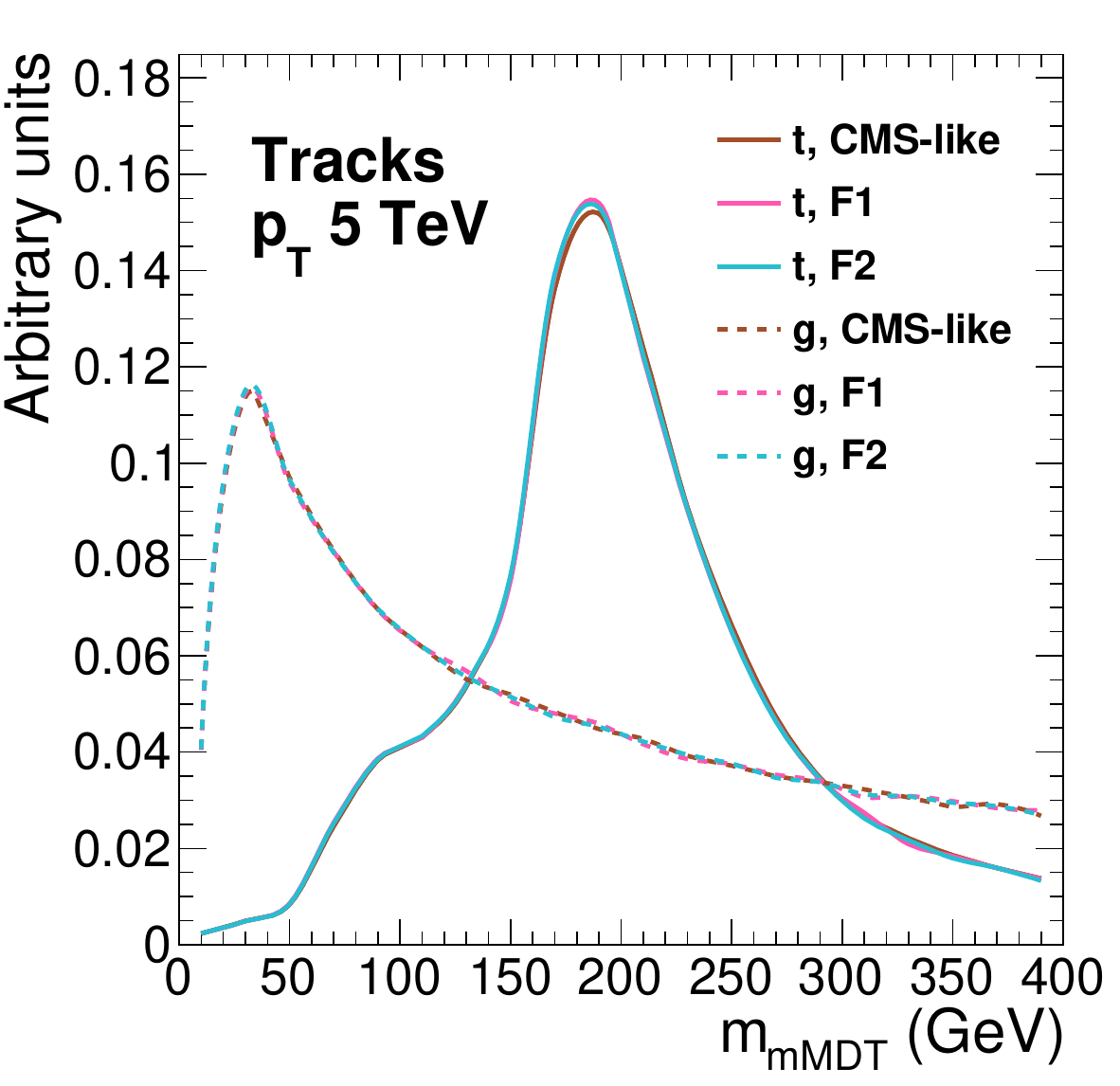}
$\qquad$
\includegraphics[width=0.42\linewidth]{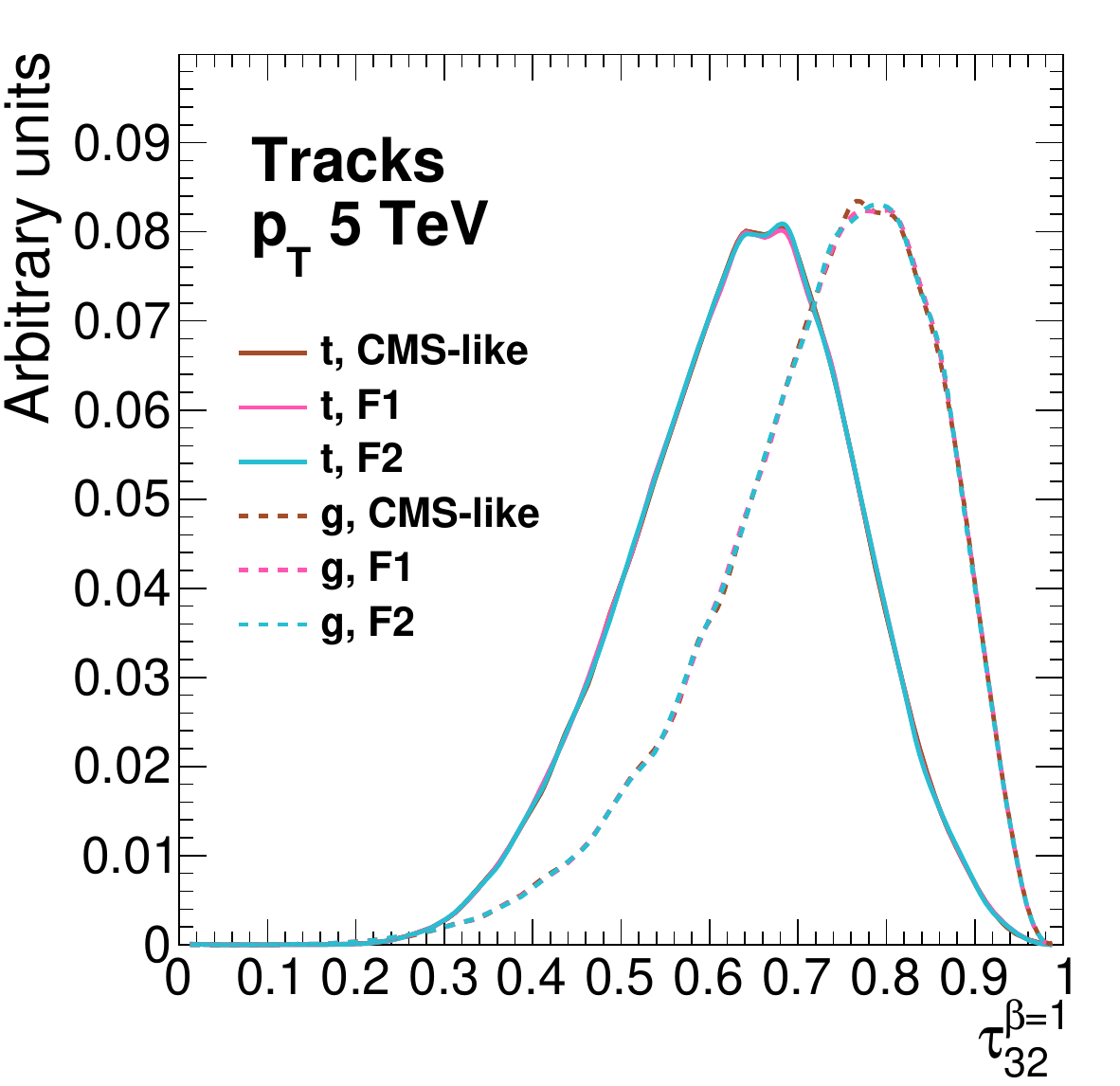} \\
\includegraphics[width=0.42\linewidth]{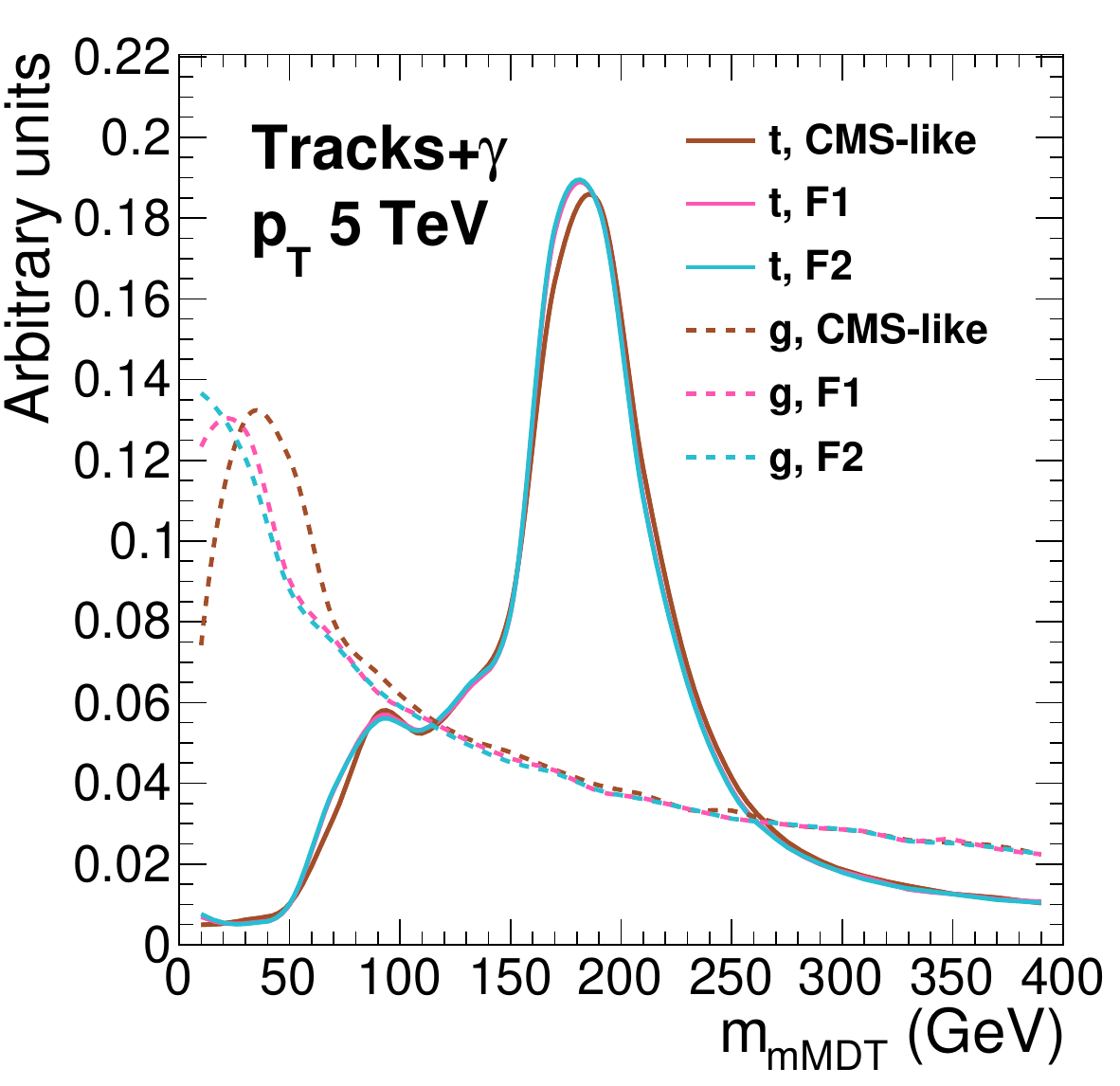}
$\qquad$
\includegraphics[width=0.42\linewidth]{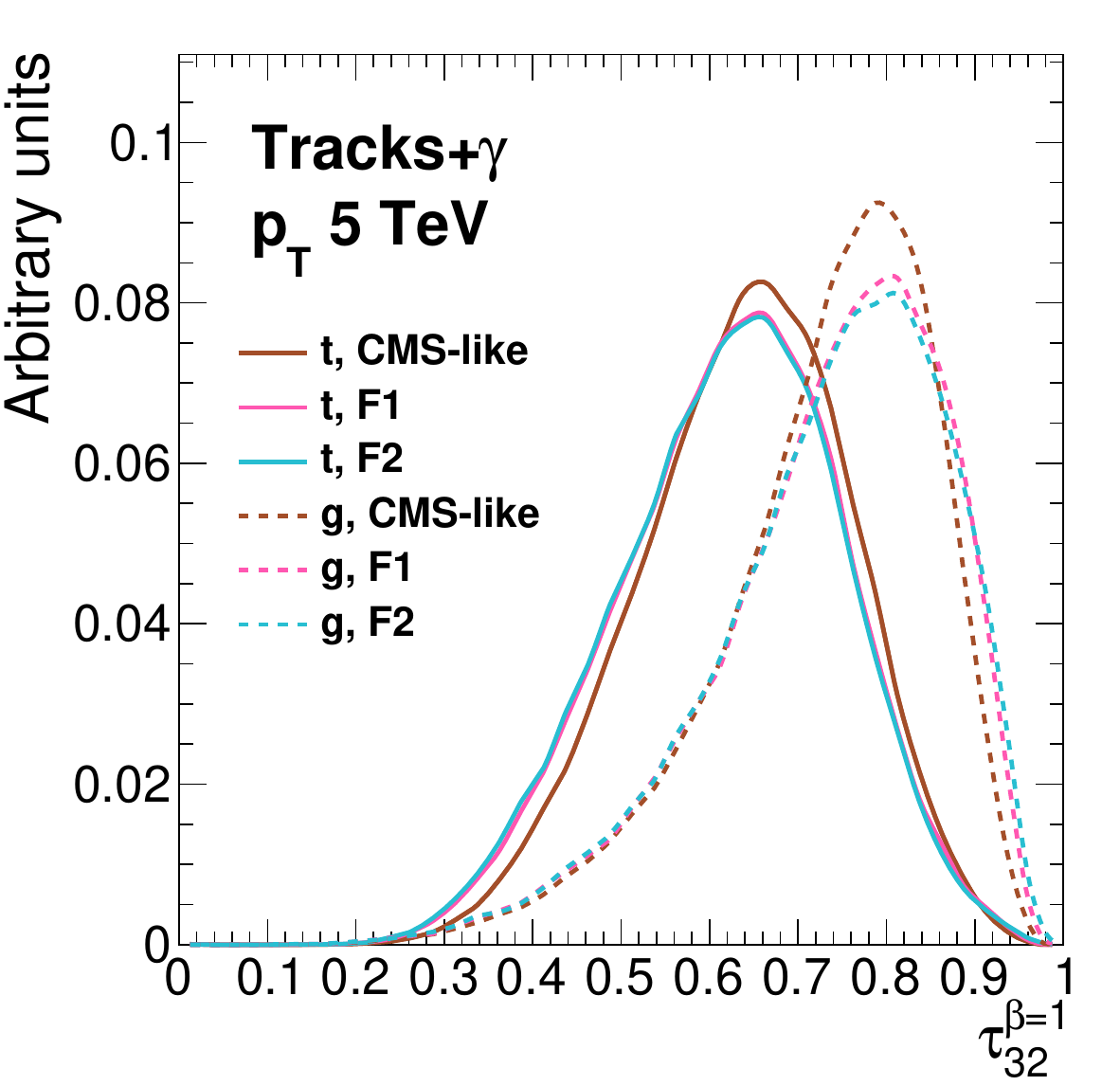} \\
\includegraphics[width=0.42\linewidth]{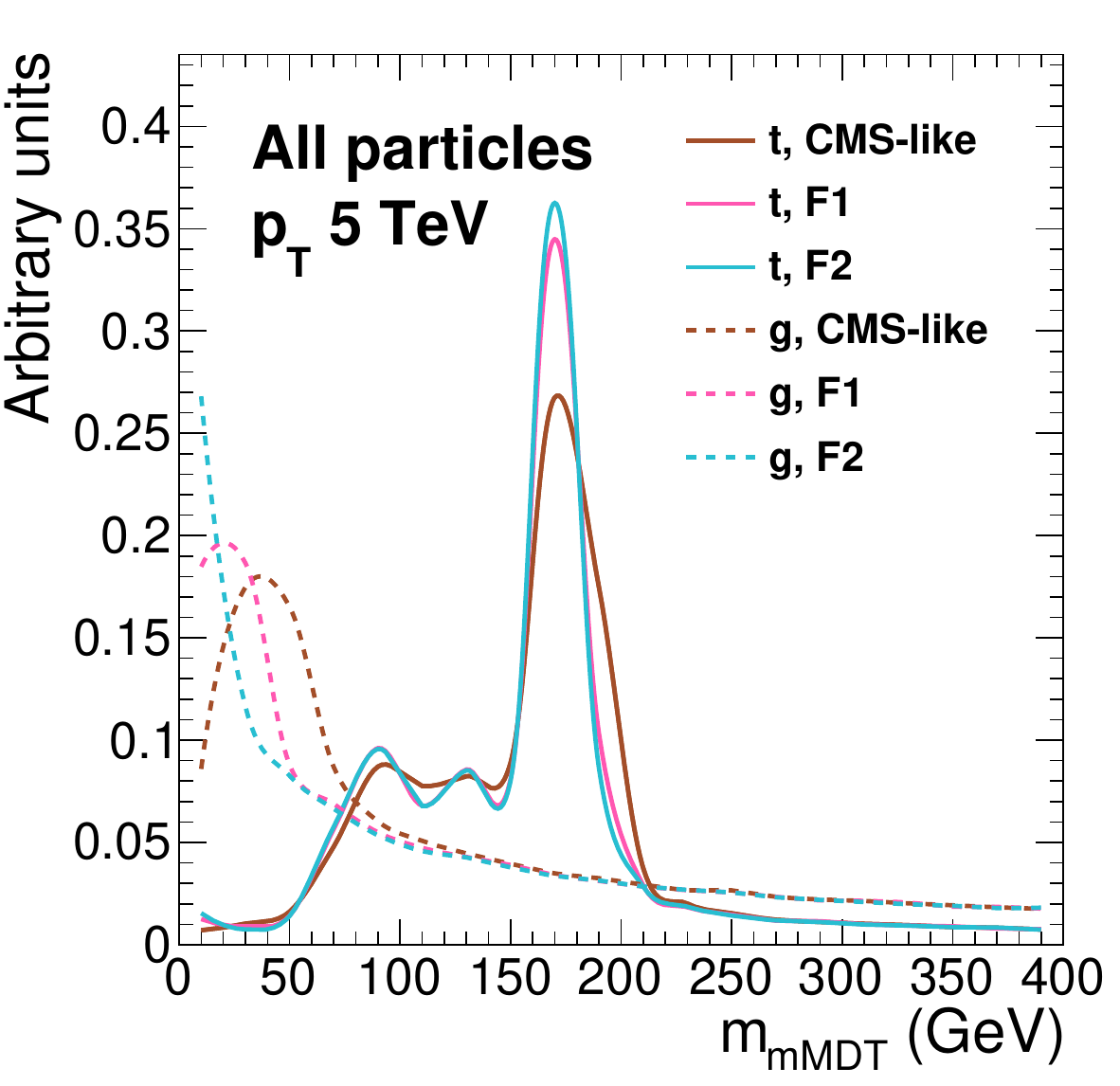}
$\qquad$
\includegraphics[width=0.42\linewidth]{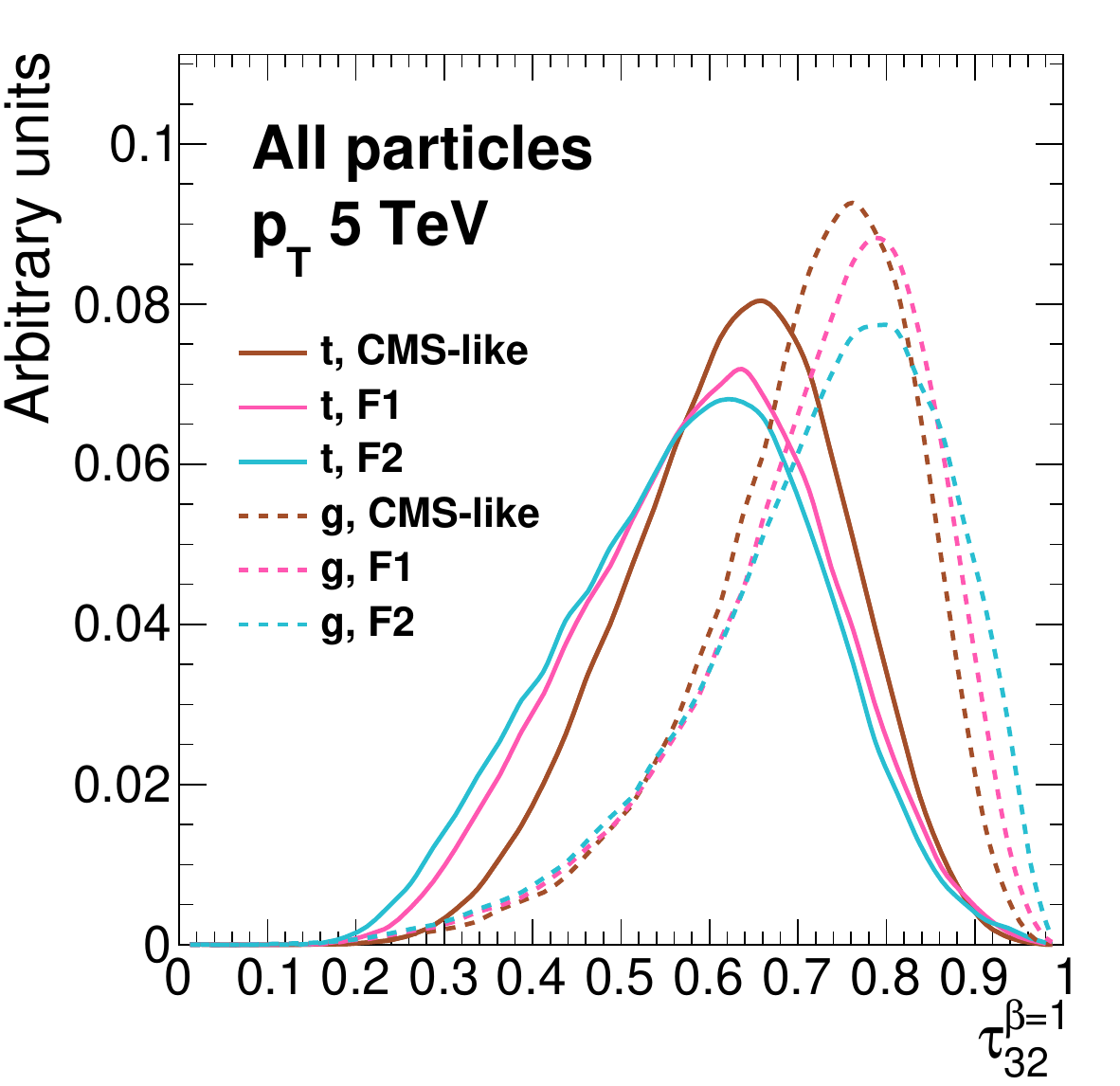}
\end{center}
\caption{Same layout as \Fig{fig:SummaryPlots_Wvq_DetectorComparison} but now comparing top jets to gluon jets with (left column) $m_{\rm mMDT}$ and (right column) $\tau_{32}^{\beta=1}$.}
\label{fig:SummaryPlots_gvt_DetecComparison}
\end{figure}

\begin{figure}[tp]
\begin{center}
\includegraphics[width=0.45\linewidth]{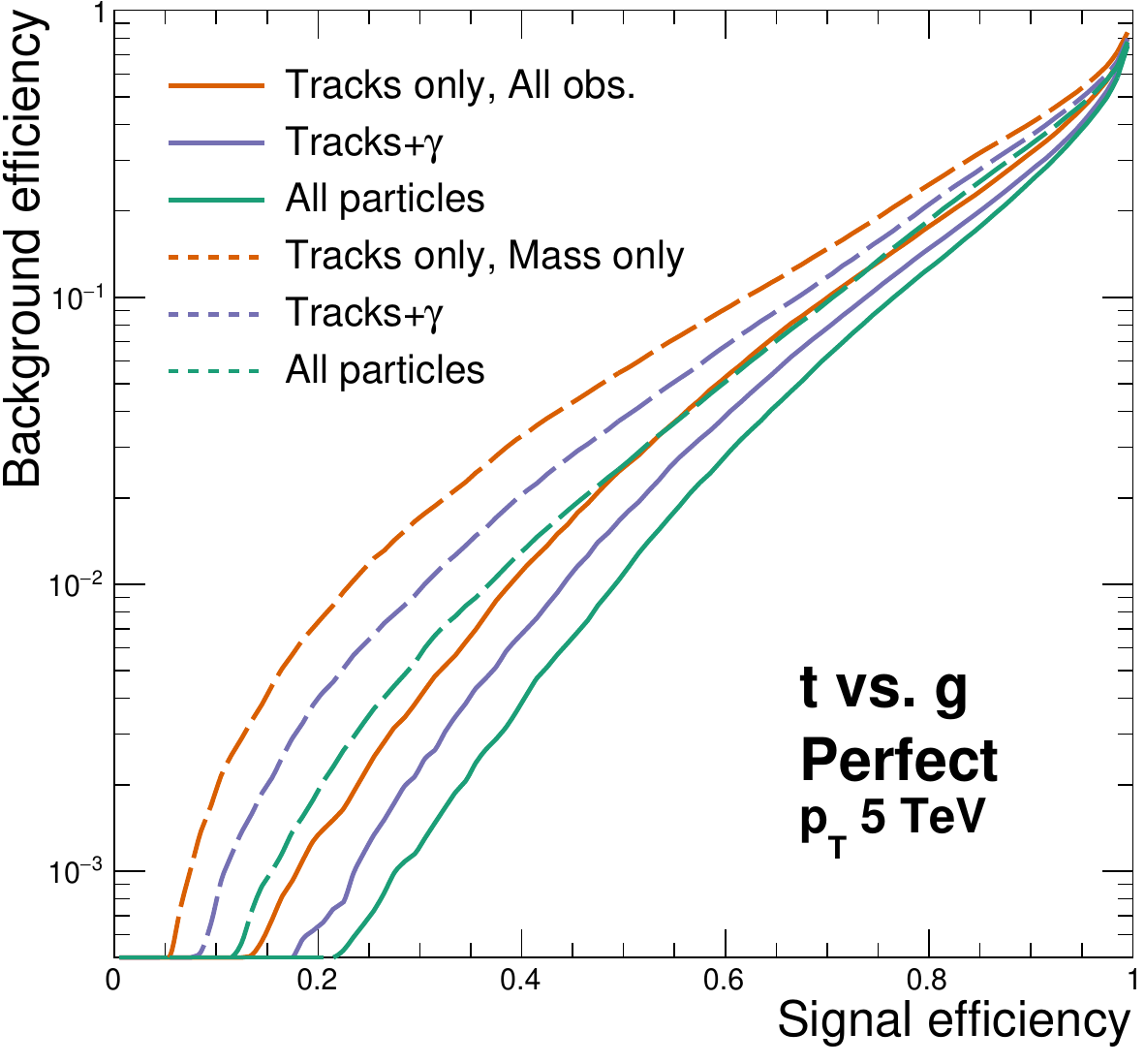}
$\qquad$
\includegraphics[width=0.45\linewidth]{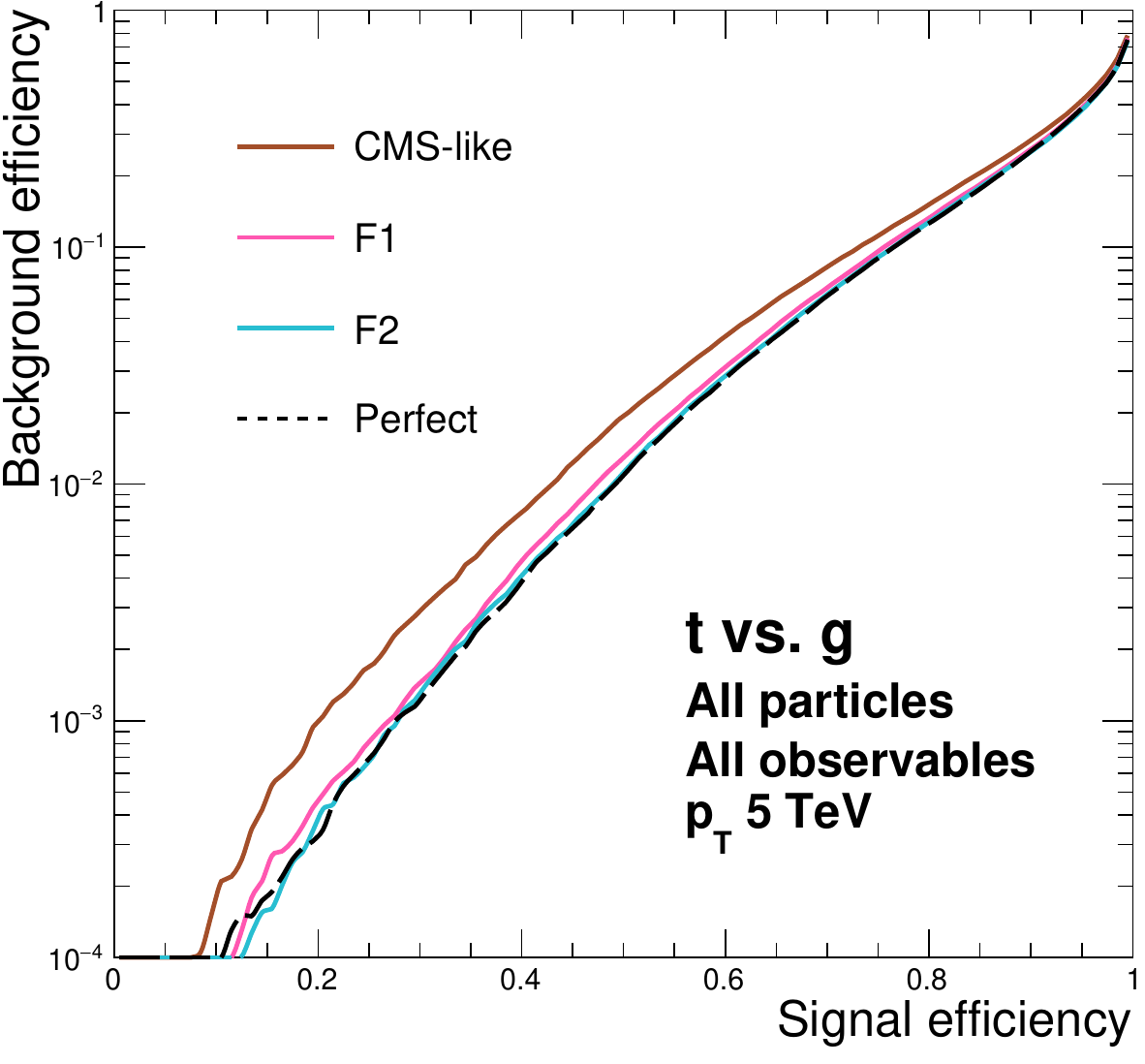}
\end{center}
\caption{ROC discrimination curves for top vs.~gluon discrimination in the $\pt = 5$ TeV bin.  The left panel for a perfect detector has the same format as \Fig{fig:ROC_Wq_F1}, and the right panel for realistic detectors has the same format as \Fig{fig:ROC_DetectorComparison_detectors}.}
\label{fig:SummaryPlots_gvt_ROCs}
\end{figure}

A surprising outcome of this study is that mass resolution does not play as dominant a role for top jet vs.~$g$ jet separation as it does for, say, $W$ jet vs.~$g$ jet separation.
We can see this in \Fig{fig:SummaryPlots_gvt_DetecComparison}, which shows the $m_{\rm mMDT}$ and $\tau_{32}^{\beta=1}$ distributions for different detector configurations and particle categories.
Compared to the $W$ vs.~$g$ case in \Fig{fig:SummaryPlots_Wvq_DetectorComparison}, the signal/background separation in the top vs.~$g$ case is more stable as the detector configuration changes.
Similarly, in the ROC curves in \Fig{fig:SummaryPlots_gvt_ROCs}, there is relatively little spread in the performance as the amount of neutral and calorimeter information is varied. 

It is important to emphasize that the mass resolution for top jets does degrade as neutral-particle information is lost.
The point made in \Secs{sec:ratiospartial}{sec:detector:summary} is that the \emph{separation} of the observable distributions is not heavily impacted by the degree of neutral-particle information, yielding values of $\mathcal{R}_S$ in \Fig{fig:MAPSPerfect} that are always less than $1.5$ for (tracks$+\gamma$)/(tracks) and (all~particles)/(tracks).
In this sense, the discrimination power offered by mass observables is maximal for top quarks, even as the mass resolution degrades.

\bibliographystyle{JHEP}
\bibliography{trackObs}

\providecommand{\href}[2]{#2}\begingroup\raggedright\begin{thebibliography}{10}

\bibitem{CMS}
{\scshape CMS} collaboration, S.~Chatrchyan et~al., \emph{{The CMS Experiment
  at the CERN LHC}},
  \href{http://dx.doi.org/10.1088/1748-0221/3/08/S08004}{\emph{JINST}
  {\bfseries 3} (2008) S08004}.

\bibitem{Aad:2008zzm}
{\scshape ATLAS} collaboration, G.~Aad et~al., \emph{{The ATLAS Experiment at
  the CERN Large Hadron Collider}},
  \href{http://dx.doi.org/10.1088/1748-0221/3/08/S08003}{\emph{JINST}
  {\bfseries 3} (2008) S08003}.

\bibitem{Seymour:1991cb}
M.~H. Seymour, \emph{{Tagging a heavy Higgs boson}},  in \emph{{ECFA Large
  Hadron Collider Workshop, Aachen, Germany, 4-9 Oct 1990: Proceedings.2.}},
  pp.~557--569, 1991.

\bibitem{Seymour:1993mx}
M.~H. Seymour, \emph{{Searches for new particles using cone and cluster jet
  algorithms: A Comparative study}},
  \href{http://dx.doi.org/10.1007/BF01559532}{\emph{Z. Phys.} {\bfseries C62}
  (1994) 127--138}.

\bibitem{Butterworth:2002tt}
J.~M. Butterworth, B.~E. Cox and J.~R. Forshaw, \emph{{$W W$ scattering at the
  CERN LHC}}, \href{http://dx.doi.org/10.1103/PhysRevD.65.096014}{\emph{Phys.
  Rev.} {\bfseries D65} (2002) 096014},
  [\href{https://arxiv.org/abs/hep-ph/0201098}{{\ttfamily hep-ph/0201098}}].

\bibitem{Butterworth:2007ke}
J.~M. Butterworth, J.~R. Ellis and A.~R. Raklev, \emph{{Reconstructing
  sparticle mass spectra using hadronic decays}},
  \href{http://dx.doi.org/10.1088/1126-6708/2007/05/033}{\emph{JHEP} {\bfseries
  05} (2007) 033}, [\href{https://arxiv.org/abs/hep-ph/0702150}{{\ttfamily
  hep-ph/0702150}}].

\bibitem{Butterworth:2008iy}
J.~M. Butterworth, A.~R. Davison, M.~Rubin and G.~P. Salam, \emph{{Jet
  substructure as a new Higgs search channel at the LHC}},
  \href{http://dx.doi.org/10.1103/PhysRevLett.100.242001}{\emph{Phys. Rev.
  Lett.} {\bfseries 100} (2008) 242001},
  [\href{https://arxiv.org/abs/0802.2470}{{\ttfamily 0802.2470}}].

\bibitem{Abdesselam:2010pt}
A.~Abdesselam et~al., \emph{{Boosted objects: A Probe of beyond the Standard
  Model physics}},
  \href{http://dx.doi.org/10.1140/epjc/s10052-011-1661-y}{\emph{Eur. Phys. J.}
  {\bfseries C71} (2011) 1661},
  [\href{https://arxiv.org/abs/1012.5412}{{\ttfamily 1012.5412}}].

\bibitem{Altheimer:2012mn}
A.~Altheimer et~al., \emph{{Jet Substructure at the Tevatron and LHC: New
  results, new tools, new benchmarks}},
  \href{http://dx.doi.org/10.1088/0954-3899/39/6/063001}{\emph{J. Phys.}
  {\bfseries G39} (2012) 063001},
  [\href{https://arxiv.org/abs/1201.0008}{{\ttfamily 1201.0008}}].

\bibitem{Altheimer:2013yza}
A.~Altheimer et~al., \emph{{Boosted objects and jet substructure at the LHC.
  Report of BOOST2012, held at IFIC Valencia, 23rd-27th of July 2012}},
  \href{http://dx.doi.org/10.1140/epjc/s10052-014-2792-8}{\emph{Eur. Phys. J.}
  {\bfseries C74} (2014) 2792},
  [\href{https://arxiv.org/abs/1311.2708}{{\ttfamily 1311.2708}}].

\bibitem{Adams:2015hiv}
D.~Adams et~al., \emph{{Towards an Understanding of the Correlations in Jet
  Substructure}},
  \href{http://dx.doi.org/10.1140/epjc/s10052-015-3587-2}{\emph{Eur. Phys. J.}
  {\bfseries C75} (2015) 409},
  [\href{https://arxiv.org/abs/1504.00679}{{\ttfamily 1504.00679}}].

\bibitem{Larkoski:2017jix}
A.~J. Larkoski, I.~Moult and B.~Nachman, \emph{{Jet Substructure at the Large
  Hadron Collider: A Review of Recent Advances in Theory and Machine
  Learning}},  \href{https://arxiv.org/abs/1709.04464}{{\ttfamily 1709.04464}}.

\bibitem{Aad:2014gea}
{\scshape ATLAS} collaboration, G.~Aad et~al., \emph{{Light-quark and gluon jet
  discrimination in $pp$ collisions at $\sqrt{s}=7\mathrm {\ TeV}$ with the
  ATLAS detector}},
  \href{http://dx.doi.org/10.1140/epjc/s10052-014-3023-z}{\emph{Eur. Phys. J.}
  {\bfseries C74} (2014) 3023},
  [\href{https://arxiv.org/abs/1405.6583}{{\ttfamily 1405.6583}}].

\bibitem{Aad:2015cua}
{\scshape ATLAS} collaboration, G.~Aad et~al., \emph{{Measurement of jet charge
  in dijet events from $\sqrt{s}$=8  TeV pp collisions with the ATLAS
  detector}}, \href{http://dx.doi.org/10.1103/PhysRevD.93.052003}{\emph{Phys.
  Rev.} {\bfseries D93} (2016) 052003},
  [\href{https://arxiv.org/abs/1509.05190}{{\ttfamily 1509.05190}}].

\bibitem{ATLAS-CONF-2016-055}
{\scshape ATLAS} collaboration, \emph{{Search for resonances with boson-tagged
  jets in 15.5 fb$^{-1}$ of $pp$ collisions at $\sqrt{s} = 13$ TeV collected
  with the ATLAS detector}},  Tech. Rep. ATLAS-CONF-2016-055, CERN, Geneva,
  Aug, 2016.

\bibitem{Aad:2016oit}
{\scshape ATLAS} collaboration, G.~Aad et~al., \emph{{Measurement of the
  charged-particle multiplicity inside jets from $\sqrt{s}=8$ TeV $pp$
  collisions with the ATLAS detector}},
  \href{http://dx.doi.org/10.1140/epjc/s10052-016-4126-5}{\emph{Eur. Phys. J.}
  {\bfseries C76} (2016) 322},
  [\href{https://arxiv.org/abs/1602.00988}{{\ttfamily 1602.00988}}].

\bibitem{ATLAS:2016vmy}
{\scshape ATLAS} collaboration, T.~A. collaboration, \emph{{Jet mass
  reconstruction with the ATLAS Detector in early Run 2 data}},  tech. rep.,
  2016.

\bibitem{ATLAS:2016wzt}
{\scshape ATLAS} collaboration, T.~A. collaboration, \emph{{Discrimination of
  Light Quark and Gluon Jets in $pp$ collisions at $\sqrt{s} = 8$ TeV with the
  ATLAS Detector}}, .

\bibitem{Krohn:2012fg}
D.~Krohn, M.~D. Schwartz, T.~Lin and W.~J. Waalewijn, \emph{{Jet Charge at the
  LHC}}, \href{http://dx.doi.org/10.1103/PhysRevLett.110.212001}{\emph{Phys.
  Rev. Lett.} {\bfseries 110} (2013) 212001},
  [\href{https://arxiv.org/abs/1209.2421}{{\ttfamily 1209.2421}}].

\bibitem{Waalewijn:2012sv}
W.~J. Waalewijn, \emph{{Calculating the Charge of a Jet}},
  \href{http://dx.doi.org/10.1103/PhysRevD.86.094030}{\emph{Phys. Rev.}
  {\bfseries D86} (2012) 094030},
  [\href{https://arxiv.org/abs/1209.3019}{{\ttfamily 1209.3019}}].

\bibitem{Chang:2013iba}
H.-M. Chang, M.~Procura, J.~Thaler and W.~J. Waalewijn, \emph{{Calculating
  Track Thrust with Track Functions}},
  \href{http://dx.doi.org/10.1103/PhysRevD.88.034030}{\emph{Phys. Rev.}
  {\bfseries D88} (2013) 034030},
  [\href{https://arxiv.org/abs/1306.6630}{{\ttfamily 1306.6630}}].

\bibitem{Chang:2013rca}
H.-M. Chang, M.~Procura, J.~Thaler and W.~J. Waalewijn, \emph{{Calculating
  Track-Based Observables for the LHC}},
  \href{http://dx.doi.org/10.1103/PhysRevLett.111.102002}{\emph{Phys. Rev.
  Lett.} {\bfseries 111} (2013) 102002},
  [\href{https://arxiv.org/abs/1303.6637}{{\ttfamily 1303.6637}}].

\bibitem{Schaetzel:2013vka}
S.~Schaetzel and M.~Spannowsky, \emph{{Tagging highly boosted top quarks}},
  \href{http://dx.doi.org/10.1103/PhysRevD.89.014007}{\emph{Phys. Rev.}
  {\bfseries D89} (2014) 014007},
  [\href{https://arxiv.org/abs/1308.0540}{{\ttfamily 1308.0540}}].

\bibitem{Larkoski:2015yqa}
A.~J. Larkoski, F.~Maltoni and M.~Selvaggi, \emph{{Tracking down hyper-boosted
  top quarks}}, \href{http://dx.doi.org/10.1007/JHEP06(2015)032}{\emph{JHEP}
  {\bfseries 06} (2015) 032},
  [\href{https://arxiv.org/abs/1503.03347}{{\ttfamily 1503.03347}}].

\bibitem{Spannowsky:2015eba}
M.~Spannowsky and M.~Stoll, \emph{{Tracking New Physics at the LHC and
  beyond}}, \href{http://dx.doi.org/10.1103/PhysRevD.92.054033}{\emph{Phys.
  Rev.} {\bfseries D92} (2015) 054033},
  [\href{https://arxiv.org/abs/1505.01921}{{\ttfamily 1505.01921}}].

\bibitem{Bressler:2015uma}
S.~Bressler, T.~Flacke, Y.~Kats, S.~J. Lee and G.~Perez, \emph{{Hadronic
  Calorimeter Shower Size: Challenges and Opportunities for Jet Substructure in
  the Superboosted Regime}},
  \href{http://dx.doi.org/10.1016/j.physletb.2016.02.068}{\emph{Phys. Lett.}
  {\bfseries B756} (2016) 137--141},
  [\href{https://arxiv.org/abs/1506.02656}{{\ttfamily 1506.02656}}].

\bibitem{Buskulic:1994wz}
{\scshape ALEPH} collaboration, D.~Buskulic et~al., \emph{{Performance of the
  ALEPH detector at LEP}},
  \href{http://dx.doi.org/10.1016/0168-9002(95)00138-7}{\emph{Nucl. Instrum.
  Meth.} {\bfseries A360} (1995) 481--506}.

\bibitem{Sirunyan:2017ulk}
{\scshape CMS} collaboration, A.~M. Sirunyan et~al., ``{Particle-flow
  reconstruction and global event description with the CMS detector}.'' 2017.

\bibitem{Aaboud:2017aca}
{\scshape ATLAS} collaboration, M.~Aaboud et~al., \emph{{Jet reconstruction and
  performance using particle flow with the ATLAS Detector}},
  \href{http://dx.doi.org/10.1140/epjc/s10052-017-5031-2}{\emph{Eur. Phys. J.}
  {\bfseries C77} (2017) 466},
  [\href{https://arxiv.org/abs/1703.10485}{{\ttfamily 1703.10485}}].

\bibitem{Alwall:2014hca}
J.~Alwall, R.~Frederix, S.~Frixione, V.~Hirschi, F.~Maltoni, O.~Mattelaer
  et~al., \emph{{The automated computation of tree-level and next-to-leading
  order differential cross sections, and their matching to parton shower
  simulations}}, \href{http://dx.doi.org/10.1007/JHEP07(2014)079}{\emph{JHEP}
  {\bfseries 07} (2014) 079},
  [\href{https://arxiv.org/abs/1405.0301}{{\ttfamily 1405.0301}}].

\bibitem{Ball:2012cx}
R.~D. Ball et~al., \emph{{Parton distributions with LHC data}},
  \href{http://dx.doi.org/10.1016/j.nuclphysb.2012.10.003}{\emph{Nucl. Phys.}
  {\bfseries B867} (2013) 244--289},
  [\href{https://arxiv.org/abs/1207.1303}{{\ttfamily 1207.1303}}].

\bibitem{Sjostrand:2014zea}
T.~Sj{\"o}strand, S.~Ask, J.~R. Christiansen, R.~Corke, N.~Desai, P.~Ilten
  et~al., \emph{{An Introduction to PYTHIA 8.2}},
  \href{http://dx.doi.org/10.1016/j.cpc.2015.01.024}{\emph{Comput. Phys.
  Commun.} {\bfseries 191} (2015) 159--177},
  [\href{https://arxiv.org/abs/1410.3012}{{\ttfamily 1410.3012}}].

\bibitem{Skands:2014pea}
P.~Skands, S.~Carrazza and J.~Rojo, \emph{{Tuning PYTHIA 8.1: the Monash 2013
  Tune}}, \href{http://dx.doi.org/10.1140/epjc/s10052-014-3024-y}{\emph{Eur.
  Phys. J.} {\bfseries C74} (2014) 3024},
  [\href{https://arxiv.org/abs/1404.5630}{{\ttfamily 1404.5630}}].

\bibitem{fastjet:1}
M.~Cacciari, G.~P. Salam and G.~Soyez, \emph{Fastjet user manual},
  \href{http://dx.doi.org/10.1140/epjc/s10052-012-1896-2}{\emph{Eur. Phys. J.}
  {\bfseries C72} (2012) 1896},
  [\href{https://arxiv.org/abs/1111.6097}{{\ttfamily 1111.6097}}].

\bibitem{fastjet:2}
M.~Cacciari and G.~P. Salam, \emph{Dispelling the $n^{3}$ myth for the $k_t$
  jet-finder},
  \href{http://dx.doi.org/10.1016/j.physletb.2006.08.037}{\emph{Phys. Lett.}
  {\bfseries B641} (2006) 57--61},
  [\href{https://arxiv.org/abs/hep-ph/0512210}{{\ttfamily hep-ph/0512210}}].

\bibitem{Cacciari:2008gp}
M.~Cacciari, G.~P. Salam and G.~Soyez, \emph{{The Anti-k(t) jet clustering
  algorithm}},
  \href{http://dx.doi.org/10.1088/1126-6708/2008/04/063}{\emph{JHEP} {\bfseries
  04} (2008) 063}, [\href{https://arxiv.org/abs/0802.1189}{{\ttfamily
  0802.1189}}].

\bibitem{Dolen:2016kst}
J.~Dolen, P.~Harris, S.~Marzani, S.~Rappoccio and N.~Tran, \emph{{Thinking
  outside the ROCs: Designing Decorrelated Taggers (DDT) for jet
  substructure}}, \href{http://dx.doi.org/10.1007/JHEP05(2016)156}{\emph{JHEP}
  {\bfseries 05} (2016) 156},
  [\href{https://arxiv.org/abs/1603.00027}{{\ttfamily 1603.00027}}].

\bibitem{Shimmin:2017mfk}
C.~Shimmin, P.~Sadowski, P.~Baldi, E.~Weik, D.~Whiteson, E.~Goul et~al.,
  \emph{{Decorrelated Jet Substructure Tagging using Adversarial Neural
  Networks}},  \href{https://arxiv.org/abs/1703.03507}{{\ttfamily 1703.03507}}.

\bibitem{Krohn:2009th}
D.~Krohn, J.~Thaler and L.-T. Wang, \emph{{Jet Trimming}},
  \href{http://dx.doi.org/10.1007/JHEP02(2010)084}{\emph{JHEP} {\bfseries 02}
  (2010) 084}, [\href{https://arxiv.org/abs/0912.1342}{{\ttfamily 0912.1342}}].

\bibitem{Ellis:2009me}
S.~D. Ellis, C.~K. Vermilion and J.~R. Walsh, \emph{{Recombination Algorithms
  and Jet Substructure: Pruning as a Tool for Heavy Particle Searches}},
  \href{http://dx.doi.org/10.1103/PhysRevD.81.094023}{\emph{Phys. Rev.}
  {\bfseries D81} (2010) 094023},
  [\href{https://arxiv.org/abs/0912.0033}{{\ttfamily 0912.0033}}].

\bibitem{Dasgupta:2013ihk}
M.~Dasgupta, A.~Fregoso, S.~Marzani and G.~P. Salam, \emph{{Towards an
  understanding of jet substructure}},
  \href{http://dx.doi.org/10.1007/JHEP09(2013)029}{\emph{JHEP} {\bfseries 09}
  (2013) 029}, [\href{https://arxiv.org/abs/1307.0007}{{\ttfamily 1307.0007}}].

\bibitem{Larkoski:2014wba}
A.~J. Larkoski, S.~Marzani, G.~Soyez and J.~Thaler, \emph{{Soft Drop}},
  \href{http://dx.doi.org/10.1007/JHEP05(2014)146}{\emph{JHEP} {\bfseries 05}
  (2014) 146}, [\href{https://arxiv.org/abs/1402.2657}{{\ttfamily 1402.2657}}].

\bibitem{Thaler:2010tr}
J.~Thaler and K.~Van~Tilburg, \emph{{Identifying Boosted Objects with
  N-subjettiness}},
  \href{http://dx.doi.org/10.1007/JHEP03(2011)015}{\emph{JHEP} {\bfseries 03}
  (2011) 015}, [\href{https://arxiv.org/abs/1011.2268}{{\ttfamily 1011.2268}}].

\bibitem{Gras:2017jty}
P.~Gras, S.~H{\"o}che, D.~Kar, A.~Larkoski, L.~L{\"o}nnblad, S.~Pl{\"a}tzer
  et~al., \emph{{Systematics of quark/gluon tagging}},
  \href{http://dx.doi.org/10.1007/JHEP07(2017)091}{\emph{JHEP} {\bfseries 07}
  (2017) 091}, [\href{https://arxiv.org/abs/1704.03878}{{\ttfamily
  1704.03878}}].

\bibitem{Thaler:2011gf}
J.~Thaler and K.~Van~Tilburg, \emph{{Maximizing Boosted Top Identification by
  Minimizing N-subjettiness}},
  \href{http://dx.doi.org/10.1007/JHEP02(2012)093}{\emph{JHEP} {\bfseries 02}
  (2012) 093}, [\href{https://arxiv.org/abs/1108.2701}{{\ttfamily 1108.2701}}].

\bibitem{Larkoski:2013eya}
A.~J. Larkoski, G.~P. Salam and J.~Thaler, \emph{{Energy Correlation Functions
  for Jet Substructure}},
  \href{http://dx.doi.org/10.1007/JHEP06(2013)108}{\emph{JHEP} {\bfseries 06}
  (2013) 108}, [\href{https://arxiv.org/abs/1305.0007}{{\ttfamily 1305.0007}}].

\bibitem{Larkoski:2015kga}
A.~J. Larkoski, I.~Moult and D.~Neill, \emph{{Analytic Boosted Boson
  Discrimination}},
  \href{http://dx.doi.org/10.1007/JHEP05(2016)117}{\emph{JHEP} {\bfseries 05}
  (2016) 117}, [\href{https://arxiv.org/abs/1507.03018}{{\ttfamily
  1507.03018}}].

\bibitem{Moult:2016cvt}
I.~Moult, L.~Necib and J.~Thaler, \emph{{New Angles on Energy Correlation
  Functions}}, \href{http://dx.doi.org/10.1007/JHEP12(2016)153}{\emph{JHEP}
  {\bfseries 12} (2016) 153},
  [\href{https://arxiv.org/abs/1609.07483}{{\ttfamily 1609.07483}}].

\bibitem{Pandolfi:1480598}
F.~Pandolfi and D.~Del~Re, \emph{{Search for the Standard Model Higgs Boson in
  the $H \to ZZ \to l^{+}l^{-}q\overline{-}q$ Decay Channel at CMS}},  2013.

\bibitem{Chatrchyan:2012sn}
{\scshape CMS} collaboration, S.~Chatrchyan et~al., \emph{{Search for a Higgs
  boson in the decay channel $H$ to ZZ(*) to $q$ qbar $\ell^-$ l+ in $pp$
  collisions at $\sqrt{s}=7$ TeV}},
  \href{http://dx.doi.org/10.1007/JHEP04(2012)036}{\emph{JHEP} {\bfseries 04}
  (2012) 036}, [\href{https://arxiv.org/abs/1202.1416}{{\ttfamily 1202.1416}}].

\bibitem{Larkoski:2014zma}
A.~J. Larkoski, I.~Moult and D.~Neill, \emph{{Building a Better Boosted Top
  Tagger}}, \href{http://dx.doi.org/10.1103/PhysRevD.91.034035}{\emph{Phys.
  Rev.} {\bfseries D91} (2015) 034035},
  [\href{https://arxiv.org/abs/1411.0665}{{\ttfamily 1411.0665}}].

\bibitem{Larkoski:2014pca}
A.~J. Larkoski, J.~Thaler and W.~J. Waalewijn, \emph{{Gaining (Mutual)
  Information about Quark/Gluon Discrimination}},
  \href{http://dx.doi.org/10.1007/JHEP11(2014)129}{\emph{JHEP} {\bfseries 11}
  (2014) 129}, [\href{https://arxiv.org/abs/1408.3122}{{\ttfamily 1408.3122}}].

\bibitem{Gallicchio:2011xq}
J.~Gallicchio and M.~D. Schwartz, \emph{{Quark and Gluon Tagging at the LHC}},
  \href{http://dx.doi.org/10.1103/PhysRevLett.107.172001}{\emph{Phys. Rev.
  Lett.} {\bfseries 107} (2011) 172001},
  [\href{https://arxiv.org/abs/1106.3076}{{\ttfamily 1106.3076}}].

\bibitem{Gallicchio:2012ez}
J.~Gallicchio and M.~D. Schwartz, \emph{{Quark and Gluon Jet Substructure}},
  \href{http://dx.doi.org/10.1007/JHEP04(2013)090}{\emph{JHEP} {\bfseries 04}
  (2013) 090}, [\href{https://arxiv.org/abs/1211.7038}{{\ttfamily 1211.7038}}].

\bibitem{Salam:2016yht}
G.~P. Salam, L.~Schunk and G.~Soyez, \emph{{Dichroic subjettiness ratios to
  distinguish colour flows in boosted boson tagging}},
  \href{http://dx.doi.org/10.1007/JHEP03(2017)022}{\emph{JHEP} {\bfseries 03}
  (2017) 022}, [\href{https://arxiv.org/abs/1612.03917}{{\ttfamily
  1612.03917}}].

\bibitem{Hocker:2007ht}
A.~Hoecker, P.~Speckmayer, J.~Stelzer, J.~Therhaag, E.~von Toerne and H.~Voss,
  \emph{{TMVA: Toolkit for Multivariate Data Analysis}}, {\emph{PoS} {\bfseries
  ACAT} (2007) 040}, [\href{https://arxiv.org/abs/physics/0703039}{{\ttfamily
  physics/0703039}}].

\bibitem{Friedman00greedyfunction}
J.~H. Friedman, \emph{Greedy function approximation: A gradient boosting
  machine}, {\emph{Annals of Statistics} {\bfseries 29} (2000) 1189--1232}.

\bibitem{deOliveira:2015xxd}
L.~de~Oliveira, M.~Kagan, L.~Mackey, B.~Nachman and A.~Schwartzman,
  \emph{{Jet-images ? deep learning edition}},
  \href{http://dx.doi.org/10.1007/JHEP07(2016)069}{\emph{JHEP} {\bfseries 07}
  (2016) 069}, [\href{https://arxiv.org/abs/1511.05190}{{\ttfamily
  1511.05190}}].

\bibitem{Baldi:2016fql}
P.~Baldi, K.~Bauer, C.~Eng, P.~Sadowski and D.~Whiteson, \emph{{Jet
  Substructure Classification in High-Energy Physics with Deep Neural
  Networks}}, \href{http://dx.doi.org/10.1103/PhysRevD.93.094034}{\emph{Phys.
  Rev.} {\bfseries D93} (2016) 094034},
  [\href{https://arxiv.org/abs/1603.09349}{{\ttfamily 1603.09349}}].

\bibitem{Datta:2017rhs}
K.~Datta and A.~Larkoski, \emph{{How Much Information is in a Jet?}},
  \href{http://dx.doi.org/10.1007/JHEP06(2017)073}{\emph{JHEP} {\bfseries 06}
  (2017) 073}, [\href{https://arxiv.org/abs/1704.08249}{{\ttfamily
  1704.08249}}].

\bibitem{Larkoski:2014gra}
A.~J. Larkoski, I.~Moult and D.~Neill, \emph{{Power Counting to Better Jet
  Observables}}, \href{http://dx.doi.org/10.1007/JHEP12(2014)009}{\emph{JHEP}
  {\bfseries 12} (2014) 009},
  [\href{https://arxiv.org/abs/1409.6298}{{\ttfamily 1409.6298}}].

\bibitem{CMS-PAS-JME-14-002}
{\scshape CMS} collaboration, \emph{{V Tagging Observables and Correlations}},
  Tech. Rep. CMS-PAS-JME-14-002, CERN, Geneva, 2014.

\bibitem{CMS-TDR-17-001}
{\scshape CMS} collaboration, \emph{{CMS Technical Design Report for HL-LHC}},
  Tech. Rep. CMS-TDR-17-001, CERN, 2017. Geneva, 2017.

\bibitem{CMS-PAS-JME-16-003}
{\scshape CMS} collaboration, \emph{{Jet algorithms performance in 13 TeV
  data}},  Tech. Rep. CMS-PAS-JME-16-003, CERN, Geneva, 2017.

\bibitem{Khachatryan:2014vla}
{\scshape CMS} collaboration, V.~Khachatryan et~al., \emph{{Identification
  techniques for highly boosted W bosons that decay into hadrons}},
  \href{http://dx.doi.org/10.1007/JHEP12(2014)017}{\emph{JHEP} {\bfseries 12}
  (2014) 017}, [\href{https://arxiv.org/abs/1410.4227}{{\ttfamily 1410.4227}}].

\bibitem{Sirunyan:2016cao}
{\scshape CMS} collaboration, A.~M. Sirunyan et~al., \emph{{Search for massive
  resonances decaying into WW, WZ or ZZ bosons in proton-proton collisions at
  $\sqrt{s} = $ 13 TeV}},
  \href{http://dx.doi.org/10.1007/JHEP03(2017)162}{\emph{JHEP} {\bfseries 03}
  (2017) 162}, [\href{https://arxiv.org/abs/1612.09159}{{\ttfamily
  1612.09159}}].

\bibitem{Chekanov:2016ppq}
S.~V. Chekanov, M.~Beydler, A.~V. Kotwal, L.~Gray, S.~Sen, N.~V. Tran et~al.,
  \emph{{Initial performance studies of a general-purpose detector for
  multi-TeV physics at a 100 TeV pp collider}},
  \href{http://dx.doi.org/10.1088/1748-0221/12/06/P06009}{\emph{JINST}
  {\bfseries 12} (2017) P06009},
  [\href{https://arxiv.org/abs/1612.07291}{{\ttfamily 1612.07291}}].

\bibitem{Krohn:2013lba}
D.~Krohn, M.~D. Schwartz, M.~Low and L.-T. Wang, \emph{{Jet Cleansing: Pileup
  Removal at High Luminosity}},
  \href{http://dx.doi.org/10.1103/PhysRevD.90.065020}{\emph{Phys. Rev.}
  {\bfseries D90} (2014) 065020},
  [\href{https://arxiv.org/abs/1309.4777}{{\ttfamily 1309.4777}}].

\bibitem{puppi}
D.~Bertolini, P.~Harris, M.~Low and N.~Tran, \emph{{Pileup per particle
  identification}},
  \href{http://dx.doi.org/10.1007/JHEP10(2014)059}{\emph{JHEP} {\bfseries 10}
  (2014) 059}, [\href{https://arxiv.org/abs/1407.6013}{{\ttfamily 1407.6013}}].

\bibitem{softkiller}
M.~Cacciari, G.~P. Salam and G.~Soyez, \emph{{SoftKiller, a particle-level
  pileup removal method}},
  \href{http://dx.doi.org/10.1140/epjc/s10052-015-3267-2}{\emph{Eur. Phys. J.}
  {\bfseries C75} (2015) 59},
  [\href{https://arxiv.org/abs/1407.0408}{{\ttfamily 1407.0408}}].

\bibitem{shapesub}
G.~Soyez, G.~P. Salam, J.~Kim, S.~Dutta and M.~Cacciari, \emph{{Pileup
  subtraction for jet shapes}},
  \href{http://dx.doi.org/10.1103/PhysRevLett.110.162001}{\emph{Phys. Rev.
  Lett.} {\bfseries 110} (2013) 162001},
  [\href{https://arxiv.org/abs/1211.2811}{{\ttfamily 1211.2811}}].

\bibitem{Komiske:2017ubm}
P.~T. Komiske, E.~M. Metodiev, B.~Nachman and M.~D. Schwartz, \emph{{Pileup
  Mitigation with Machine Learning (PUMML)}},
  \href{https://arxiv.org/abs/1707.08600}{{\ttfamily 1707.08600}}.

\bibitem{Aguilar-Saavedra:2017zuc}
J.~A. Aguilar-Saavedra, \emph{{Stealth multiboson signals}},
  \href{https://arxiv.org/abs/1705.07885}{{\ttfamily 1705.07885}}.

\bibitem{Sirunyan:2017dnz}
{\scshape CMS} collaboration, A.~M. Sirunyan et~al., \emph{{Search for low mass
  vector resonances decaying to quark-antiquark pairs in proton-proton
  collisions at sqrt(s) = 13 TeV}},
  \href{https://arxiv.org/abs/1705.10532}{{\ttfamily 1705.10532}}.

\bibitem{Aguilar-Saavedra:2017rzt}
J.~A. Aguilar-Saavedra, J.~H. Collins and R.~K. Mishra, \emph{{A generic
  anti-QCD jet tagger}},  \href{https://arxiv.org/abs/1709.01087}{{\ttfamily
  1709.01087}}.

\bibitem{Chekanov:talk1}
S.~Chekanov et~al., ``{A high granularity hadronic calorimeter for multi TeV
  jets}.'' {FCC Week 2017,
  https://indico.cern.ch/event/556692/contributions/2592529}, {2017}.

\bibitem{Chekanov:talk2}
S.~Chekanov et~al., ``{Simulations of detector response for multi-TeV physics
  at a 100 TeV pp collider}.'' {Boost 2017 Conference,
  https://indico.cern.ch/event/579660/contributions/2575089}, {2017}.

\end{thebibliography}\endgroup

\end{document}